\documentclass[preprint,prc,showpacs,showkeys,longbibliography]{revtex4-1}
\usepackage{epsfig,graphicx,float,color}
\usepackage{amssymb}
\usepackage[normalem]{ulem}
\begin{document}

\title{The subtle connection between shape coexistence and quantum phase transition. The Zr case}

\author{J.E.~Garc\'{\i}a-Ramos$^{1,2}$ and K.~Heyde$^3$}
\affiliation{
$^1$Departamento de  Ciencias Integradas y Centro de Estudios Avanzados en F\'isica, Matem\'atica y Computaci\'on, Universidad de Huelva,
21071 Huelva, Spain\\
$^2$Instituto Carlos I de F\'{\i}sica Te\'orica y Computacional,  Universidad de Granada, Fuentenueva s/n, 18071 Granada, Spain\\
$^3$Department of Physics and Astronomy, Ghent University  Proeftuinstraat, 86 B-9000 Gent, Belgium}
\begin{abstract} 
\begin{description}
\item [Background] Zr region is characterized by very rapid changes in the ground state structure of the nuclei. In particular, the onset of deformation when passing from $^{98}$Zr to $^{100}$Zr is one of the fastest ever observed in the nuclear chart. It has been probed both experimental and theoretically that certain low-lying excited states of Zr isotopes own different shapes than the ground state.   

\item [Purpose] We intend to disentangle the interplay between the sudden changes in the ground state shape, i.e., the existence of a quantum phase transition, and the presence in the spectra of coexisting states with very different deformation, i.e., the presence of shape coexistence.
 
\item [Method] We rely on a previous calculation using the Interacting Boson Model with Configuration Mixing (IBM-CM) which reproduces in detail the spectroscopic properties of $^{96-110}$Zr. This IBM-CM calculation allows to compute mean-field energy surfaces, wave functions and any other observable related with the presence of shape coexistence or with a quantum phase transition.
  
\item [Results] We obtain energy surfaces and the equilibrium value of the deformation parameter $\beta$, the U(5) decomposition of the wave functions and the density of states.
  
\item [Conclusions] We confirm that Zr is a clear example of quantum phase transition that originates from the crossing of two configurations with a very different degree of deformation. Moreover, we observe how the intruder configuration exhibits its own evolution which resembles a quantum phase transition too.
\end{description}
\end{abstract}
 
\pacs{21.10.-k, 21.60.-n, 21.60.Fw}

\keywords{Zr isotopes, shape coexistence, quantum phase transition, interacting boson model.}

\date{\today}
\maketitle

\section{Introduction}
\label{sec-intro}
Nuclei over the full nuclear mass (N,Z) surface are characterized by the presence of clear cut regularities as the appearance of nuclear shells at magic numbers. Symmetries, on the other hand, allow to establish hallmarks in the chart of nuclei defining spherical, rigidly deformed and gamma-unstable nuclei. Therefore, most of the nuclei can be grouped in one of the previous classes. The different patterns in the mass table are modulated by the interplay between the different components of the nuclear force, namely, monopole interaction, pairing interaction, quadrupole (low multipole in general) interaction. The monopole part explains the changing energy of the single-particle levels \cite{Tsuno14}; pairing tends to stabilize the atomic nuclei in a spherical shape; while the quadrupole interaction is at the origin of the appearance of a deformed mean-field  producing nuclear deformation. In short and playing a central role, the detailed balance between the different components of the nuclear force is the responsible of the final shape of the nucleus.

Another remarkable aspect of the nuclear chart is the existence of transitional regions, where the structure of the nuclei evolves from one of the limiting cases to another as a function of the neutron and/or the proton number. In certain regions the evolution can be very rapid and the nucleus changes from being spherical to strongly deformed in just a few mass units, e.g., in the rare-earth region around $N=90$ \cite{Cejn09,Cejn10} or in Zr-Sr \cite{Ansa17,Regi17} region around $N=60$. For such situations the term quantum phase transition (QPT) was coined \cite{Hert76,Sach11} due to the similarities with the very well known thermodynamic phase transitions. A QPT implies an abrupt change in the properties of the ground state that results in the rapid change of several observables, namely, two-neutron separation energies, E($4_1^+$)/E($2_1^+$), B(E2:$4_1^+\rightarrow 2_1^+$), among others. QPT can be of two types, either of first or of second order. In the first order case a discontinuity in the first derivative of the energy with respect to a  control parameter is present while in the second order case the discontinuity hold for the second order derivative. The first-order QPT's suppose a more abrupt change than the second order ones.
\begin{figure}[hbt]
  \centering
  \includegraphics[width=0.90\linewidth]{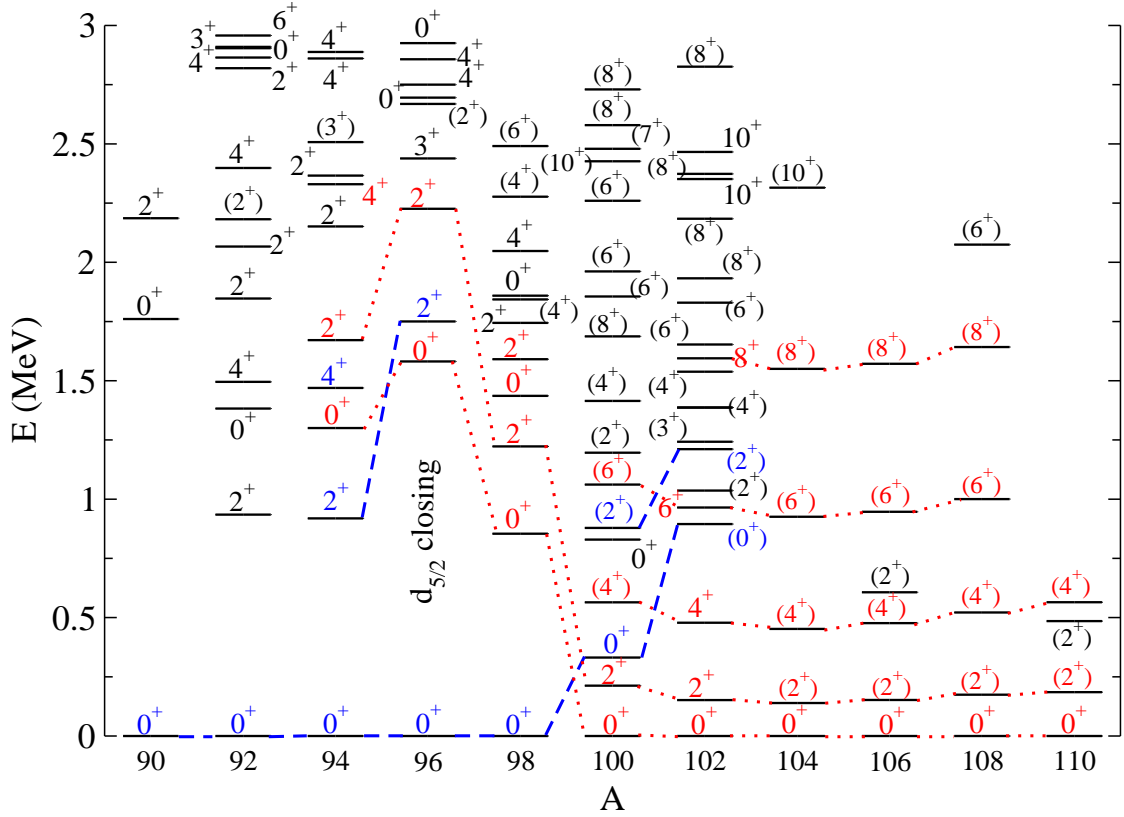}
  \caption{Energy systematics of even-even  Zr isotopes from $A=90$ to $A=110$. Levels identified as regular are marked in blue and connected with dashed blue lines, while those identified as intruder are marked in red and connected with dashed red lines.}
    \label{fig:energy-systematics}
  \end{figure}

  To complete the description of the mass table, one has to take into account that in many cases the change of shape not only occur when moving to neighbour nuclei, but also the properties in a given nucleus, as a function of the increasing excitation energy may be changing, giving evidences of deformed excited states. Hence, it is possible to find states with very different shapes in the spectrum of a single nucleus. The different shapes correspond to different n-particle n-hole excitations across a shell closure. These excitations give rise to an increasing number of active nucleons that can strongly affect the effective nuclear force. A notable example can be found in the doubly magic nucleus $^{40}$Ca which is spherical for the ground state but 2p-2h and 4p-4h excitations lead to deformed and superdeformed shapes, respectively \cite{Caur07}. The existence of states with different shapes in a narrow range of energies is known as shape coexistence and not only appears in nuclear physics, but also in molecules and other quantum systems \cite{heyde11}. Shape coexistence is present in many regions of the nuclear mass table, in general in those around shell or sub-shell closures of both protons and neutrons. Of particular interest is the region around the shell closure $Z=82$, i.e., the even-even Pt, Hg, Pb, and Po, where many of these isotopes have been studied both experimentally and theoretically, confirming the presence of two and even three families of states with a different degree of deformation. It is worth to note that in some cases the interaction among the different families is rather weak as in Hg and Pb while in the case of Pt and Po is rather strong. Moreover, in certain cases the intruder configuration, corresponding to 2p-2h excitations become the lowest-lying configurations, contrary to the 0p-0h configurations now describing highly-lying states, as is the case for Pt and Po. On the other hand, in the case of Hg or Pb the ground state always corresponds to a regular 0p-0h configuration.  

  QPT and shape coexistence share several aspects, namely, the rapid structural changes over a chain of isotopes or isotones and the presence of very low-lying $0^+$ states. As a consequence, the excitation energy of the $2_1^+$ can drop  or the value of the $B(E2:2_1^+\rightarrow 0_1^+)$ can increase suddenly in few mass units. Besides, in both cases, the value of the nuclear radius separates from the linear trend  or the two-nucleon transfer reaction intensities experience a rapid increase. However, some differences are in order, in particular, the existence in the case of the  shape coexistence of a set of states, intruder states, with a parabolic excitation energy systematics that is centered with respect to the mid shell. These states present also a different deformation than the ground state band. This behavior is not observed in the case of a QPT.  

  The region around $Z=40$, in particular, regarding the Zr isotopes, but also the Sr ones, is an ideal region to study the interplay between shape coexistence and the presence of a QPT, that is the reason for focusing on this mass region in this work. On one hand, this region shows all the hints pointing to the presence of a QPT but, at the same time, a set of states owns an energy systematics symmetric with respect to the mid shell and presents a different deformation than the rest of states, therefore one can say that shape coexistence is present in the spectrum.  

  The organization of the present paper is as follows: in Sec.~\ref{sec:evidences} some hints for the presence of shape coexistence are provided; in Sec.~\ref{sec-evidences-QPT} the corresponding evidences for the existence of QPT's are shown; in Sec.~\ref{sec-ibm-cm} a brief summary of the interacting boson model with configuration mixing (IBM-CM) is presented; Sec.~\ref{sec:results} is devoted to presenting the results of the IBM-CM calculations; finally in Secs.~\ref{sec:discussion} and \ref{sec-conclu}, discussion, conclusions and outlook are presented, respectively.

\section{Evidences of shape coexistence in Zr nuclei}
\label{sec:evidences}
Shape coexistence implies that in a narrow range of energy, states with different deformation coexist. Because deformation is not strictly an observable, it is needed to rely on observables that are tightly linked to deformation, such as excitation energy and B(E2) transition rate systematics, radii, etc. In addition, the existence of highly hindered E2 or E0 transitions points towards the presence of states with different degree of deformation.

Energy systematics is one of the first hints pointing to the presence of extra configurations in the spectra of an isotope chain \cite{heyde11}. In particular, in the case of even-even Hg and Pb  isotopes a parabolic-like trend in the energy levels centered at the mid shell, N=104, is clearly observed for a set of states, together with a rather flat behavior for another group of states. In both cases the flat systematics corresponds to regular states while the parabolic one to the intruder ones. There are other isotope chains where the presence of extra configurations is not so clearly observed. This is precisely the case of Pt isotopes, where due to the large mixing between regular and intruder states combined with the crossing of the bandheads of regular and intruders families makes the presence of intruder states somehow concealed (see \cite{Garc11,Garc12} for more details). The case of Po is also similar, though here, the mid shell is barely reached experimentally \cite{Garc15,Garc15c}. In both cases it is hard to separate a parabolic systematics from a flat one. The case of Zr is very much the same, as one can readily see in Fig.~\ref{fig:energy-systematics} where the presented energy systematics of the even-even Zr isotopes span the range from $A=90$ till 110. According to the experimental information accumulated during last decades (see references below in this section), it is possible to determine the character of certain states, hence the regular ones are labeled in blue while the intruder ones in red, moreover the states with the same internal structure are connected with lines. In this figure one can easily observe several features, the first one is the presence of shell and subshell closures for $A=90$ and $A=96$, respectively. The second is the increase in the density of levels below E$\approx 1.5$ MeV for $A\ge 100$ owing to the lowering of the intruder configuration that becomes the ground state for $A=100$ and onwards. However, the parabolic trend for the intruder energy systematics is not observed. Anyhow, the existence of low lying $0^+$ states is also a good indicator of the presence of additional particle-hole configurations. 
\label{sec-evidences-SC}
  \begin{figure}[hbt]
    \centering
    \includegraphics[width=1\linewidth]{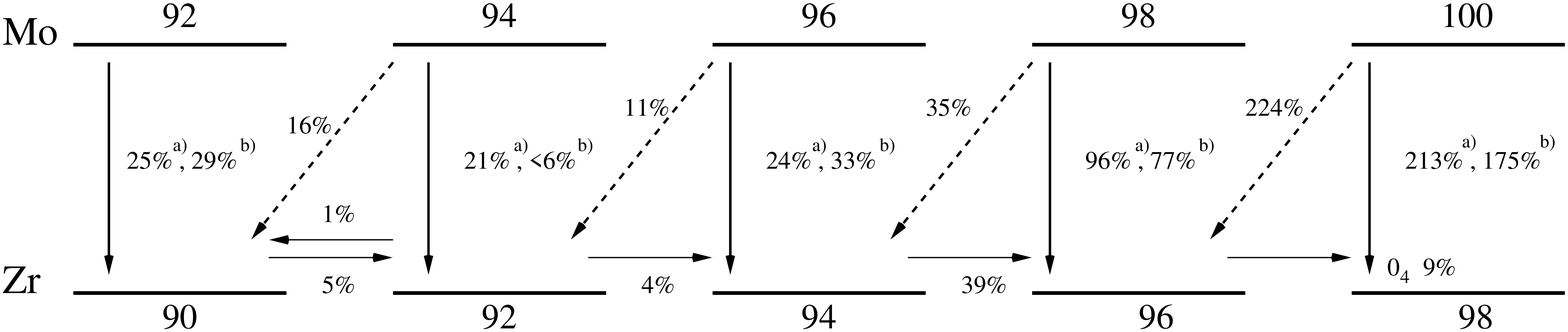}
    \caption{Two-nucleon or $\alpha$ transfer strengths to the $0_2^+$  state in the Zr isotopes, relative to the  $0_1^+$ state. (a) Stands for data from ($^6$Li, $^8$B) and (b) for data from  ($^{14}$C, $^{16}$O) reactions. Data taken from the corresponding Nuclear Data Sheets and from \cite{heyde11}.}
    \label{fig:transfer}
  \end{figure}

  The connection of the different shapes with different particle-hole configurations makes two-neutron, two-proton or $\alpha$ transfer reactions ideal tools to identity states with rather different shapes. In Fig.~\ref{fig:transfer}, the probability of populating a given state either with two-nucleon or with $\alpha$ transfer reaction, relative to the probability for populating the ground state is depicted.
  This figure shows strong evidences for pairing collectivity in both the $0^+_1$ and $0^+_2$ states, thereby being a clear evidence of the presence of ``coexisting'' characteristics in those two states. These data are pointing out to the underlying pairing collectivity in these nuclei and they are the result from two-neutron transfer in ($^6$Li, $^8$B) (labeled with (a)) and from the ($^{14}$C, $^{16}$O) (labeled with (b)) reactions, respectively. For more details on the Z$\approx 40$, N$\approx 60$ nuclei see also section III.A.3 and Fig.~29 and 30 of Ref.~\cite{heyde11}. All in all, these data are the hard evidence of the fact that the ground and first excited $0^+$ states exhibit a ``coexisting'' character, i.e., the transfer strength goes strongly to more than one ``pairing condensate''.  
  
In the last decades a huge experimental effort has been carried out to complete the picture of deformation in Zr area, mainly through the determination of excitation energies and lifetimes. For mass  $A=94$, besides the \cite{Abri06} compilation, recent information has been obtained from lifetime data using neutron scattering \cite{Chak13,Peters13} as well as the $B(E2)$ values extracted from inelastic electron scattering at low momentum transfer into the 2$^{+}_{1,2}$ states. This allowed a comparison with the data obtained at the Kentucky facility from neutron inelastic scattering such as branching ratios, multipole mixing ratios, and spin assignments  indicating very much the same results, however with an increased precision \cite{Scheik14,Elhalmi08}.  For mass $A=96$, the major part of data can be found in the compilation \cite{Abri08}, moreover, data on g-factors have been determined by Coulomb excitation of $^{96}$Zr beams in inverse kinematics. Besides, the lifetime of the 2$^+_1$ state has been redetermined using the Doppler Shift Attenuation (DSA) method \cite{Kumb03}. Interesting information has been extracted from single and double-$\beta$ decay Q values among the triplet of nuclei $^{96}$Zr, $^{96}$Nb and  $^{96}$Mo \cite{Alan16}; recent high-resolution electron scattering allowed to extract the $B(E2; 0^+_1\rightarrow 2^+_2)$ value as well as decay strengths originating from the 2$^+_2$ excited states by Kremer {\it et al.}\ at the S-DALINAC \cite{Krem16}.  The $E2$ collectivity deduced in this way was moreover discussed by Pietralla {\it et al.}\ \cite{Piet18} for the heavy Zr nuclei. Moving into the transitional nucleus at $A=98$ (compilation can be found in \cite{Sing03}), lifetime data have been measured using the EXILL-FATIMA array at the cold neutron beam line at the Institut Laue-Langevin (ILL-Grenoble) by Betterman \cite{Bett10} and Ansari {\it et al.}\ \cite{Ansa17}, the latter also extracting lifetimes for the $A=100$ and $A=102$ Zr isotopes.  These lifetime data are of essential use in obtaining a deeper understanding of the changing structure when moving from $A=96$ towards the isotopes beyond $A=100$.  Coming to the $^{98}$Zr nucleus, very recent data taken at the ATLAS/CARIBU facility at Argonne National Laboratory, through Coulomb excitation studies, have allowed to determine an upper limit of the  $B(E2; 2^+_1 \rightarrow 0^+_1)$ value in $^{98}$Zr. Moreover, new lifetime data by Singh {\it et al.}, using the recoil-distance Doppler shift method at GANIL, were published \cite{Singh18}. These data are of major importance to trace the transition from $A=96$ into the strongly deformed $A=100$ Zr isotope \cite{Wern18}. More details on that study  by Witt {\it et al.}\ have very recently been published \cite{Witt18}.  Data on $^{100}$Zr are presented in the compilation \cite{Sing08}, the ones on  $^{102}$Zr are in \cite{Defr09}, and $^{104}$Zr are in \cite{Blac07}.  Data for $^{106}$Zr and $^{108}$Zr are in \cite{Sing15} and \cite{Sing15b}, respectively. Regarding $^{106}$Zr, in Ref.~\cite{Navin14} the energy of the states $6_1^+$ and $8_1^+$ have been measured. This experiment was performed at GANIL using the large-acceptance spectrometer VAMOS++ trough the reaction ($^{238}$U, $^9$Be) in inverse kinematics with the aim of studying the spectra of $^{104-105-106}$Zr. A detailed study by Paul {\it et al.}\ \cite{Paul17} of the $^{110}$Zr nucleus and its lowest-lying states has been carried out very recently. The above information is the base to separate between regular and intruder states in Fig.~\ref{fig:energy-systematics}.
\begin{figure}[hbt]
\centering
\includegraphics[width=0.60\linewidth]{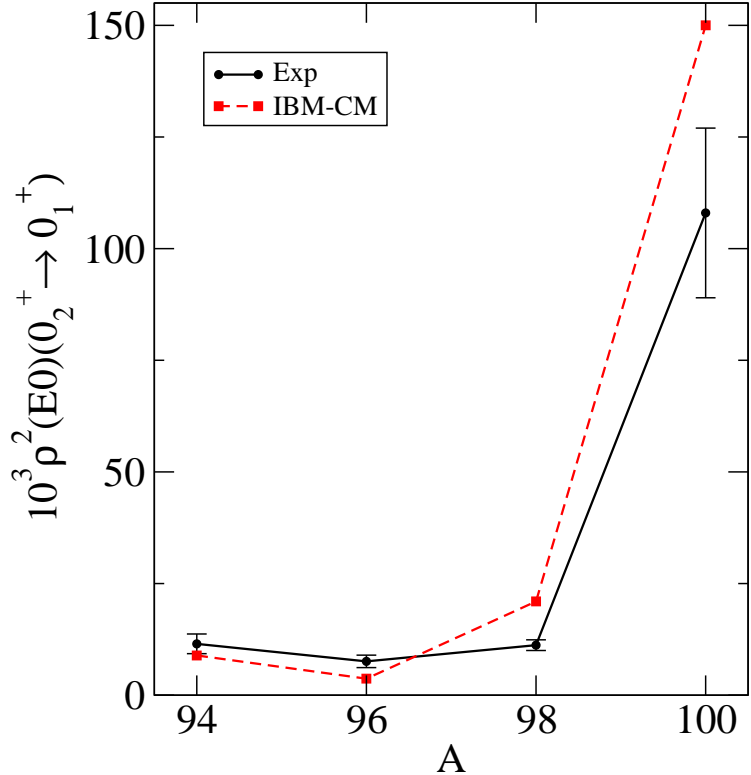}
\caption{Evolution of the value of $\rho^2(E0:0^+_2\rightarrow 0^+_1)$, from $^{94}$Zr till $^{100}$Zr. Black full line for experimental values while red dashed ones for IBM-CM calculations.}
\label{fig-e0}
\end{figure}

The analysis of the E0 transition rates has been proven to be an ideal tool to disentangle the presence of different configurations in the spectrum and its degree of mixing \cite{Wood99}. In particular, a small value of  $\rho^2(E0:0^+_i\rightarrow 0^+_j)$ suggests that states $0^+_i$ and $0^+_j$ have largely different degrees of deformation,
 implying no (or negligible) mixing between these two $0^+$ states. %
In Fig.~\ref{fig-e0} the $\rho^2(E0)$ experimental values for the known $0^+_2\rightarrow 0^+_1$ transitions together with the theoretical IBM-CM values (see Sec.~\ref{sec-ibm-cm}) are depicted. The values for the range $A=94-98$ are rather small, suggesting that the involved configurations are rather pure and owing a different structure. Indeed, the $0_1^+$ state is assumed to be regular while the $0_2^+$ one of intruder nature. However, the $\rho^2(E0)$ transition rate for $A=100$ is one of the largest ones observed over the full nuclear mass table, pointing to the existence of a large mixing between the intruder and the regular sectors concerning the two first $0^+$ states.

\section{Quantum phase transition indicators in Zr nuclei. Is there any hint of X(5) critical point symmetry?}
  \label{sec-evidences-QPT}
  
  A QPT develops in systems where the structure of the ground state changes abruptly for a specific value of the control parameter
and the temperature is equal to zero. %
In particular, a QPT is related to quantum systems which Hamiltonians, that express the change in between different symmetries called A and B, can be written down as,
\begin{equation}
  \label{eq:H_QPT}
  \hat H= (1-x) \hat {H}(\text{sym}_A)+ x\hat  H(\text{sym}_B),
\end{equation}
where $\hat {H}(\text{sym}_A)$ and $\hat {H}(\text{sym}_B)$ correspond to  Hamiltonians that own a given symmetry, either $A$ of $B$, that in most of cases can be considered as dynamical symmetries \cite{Iach98}.
In the familiar Ising model \cite{Isin25}, $\hat {H}(\text{sym}_A)$ can be identified with the term leading the system into a ferromagnetic phase, while $\hat {H}(\text{sym}_B)$ with that part leading into the paramagnetic phase \cite{Sach11}. 
The QPT happens for a critical value $x=x_c$ where the wave functions switch from having the symmetry $A$ to having the symmetry $B$. The existence of a QPT supposes also the abrupt change in the so called order parameter that presents a null value in the symmetric phase while different from zero in the broken phase \cite{Sach11}. Hence, the order parameter represents, albeit in a schematic way, the symmetry of a given phase.
\begin{figure}
\includegraphics[width=0.60\textwidth]{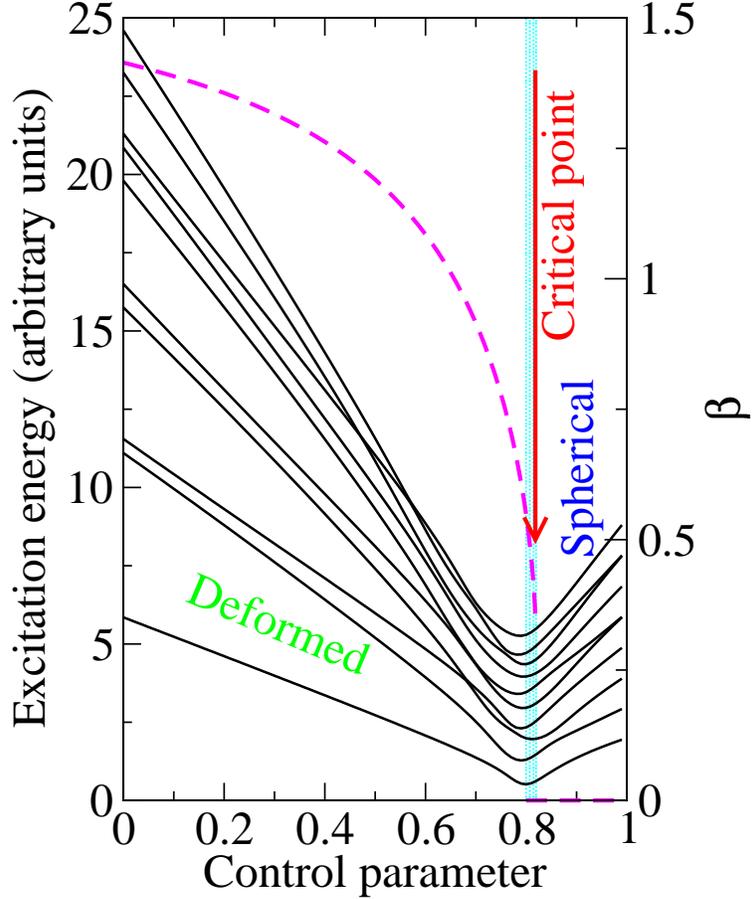}
\caption{Schematic spectra (full black lines and left y-axis) of few $0^+$ states for an IBM Hamiltonian undergoing a first order QPT at a value of the control parameter equal to $9/11$. Value of the order parameter $\beta$ (dashed magenta line and right y-axis).   Dashed cyan zone corresponds to the coexistence region. See Ref.~\cite{Garc01} for details on the calculation of $\beta$.}
\label{fig:gag-beta}
\end{figure}

Following the classical classification of QPT's, the transitions can be separated in first and second order (or continuum) QPT's \cite{Land69}. In the case of a first order QPT (according to the Erhenfest classification, the first derivative of the ground state energy with respect to the control parameter presents a discontinuity), there is a narrow region around $x_c$ where the two symmetries can coexist. However, in the case of a second order QPT (according to the Erhenfest classification, the second derivative of the ground state energy with respect to the control parameter presents a discontinuity), there is no coexistence of symmetries for any particular value of $x$. Another important feature is that in the case of a second order QPT, the excitation energy of the first excited state at the critical point vanishes, while for the first order QPT it takes a very small value. This is schematically presented for an IBM calculation (with a single configuration) in Fig.~\ref{fig:gag-beta} for the case of a first order QPT, where the spectra corresponding to few $0^+$ states and the value of the order parameter for a schematic IBM Hamiltonian are depicted.
The two limiting phases, deformed and spherical, and the critical point are marked. The aim of presenting this figure is two fold. First to illustrate how, in general, a QPT induces the lowering of a set of states and the compression of the spectra and, second, to see the evolution of the order parameter (calculated in the thermodynamic limit) \cite{Garc01}, that is connected with the deformation of the ground state, and that suffers a sudden change passing from zero to finite values when crossing the critical point. 
\begin{figure}
\includegraphics[width=0.60\textwidth]{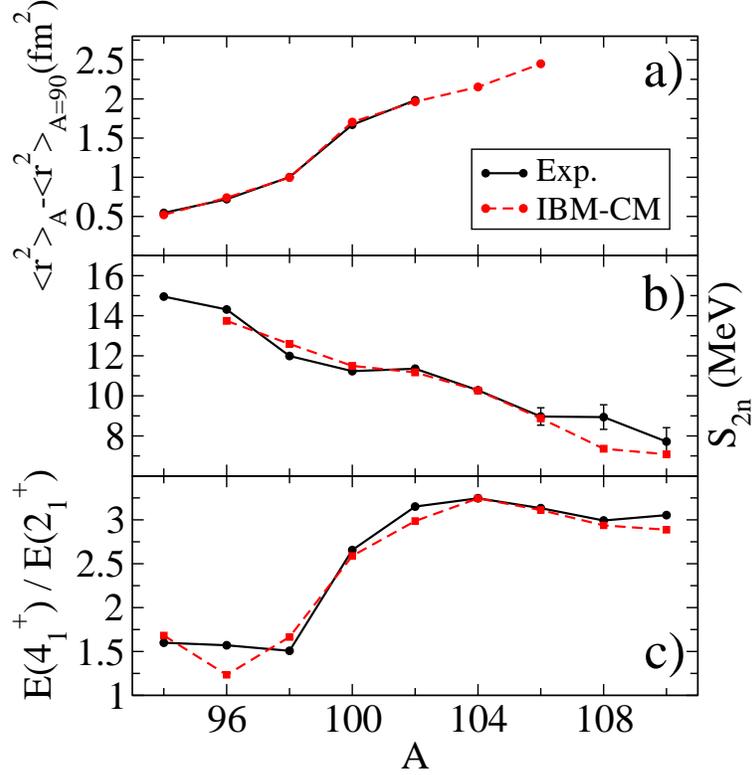}
\caption{Indicators of the presence of a QPT. Black full line for experimental values and red dashed one for IBM-CM calculations. (a) Radii square referred to $A=90$, (b) two-neutron separation energies, and (c) E($4_1^+$)/E($2_1^+$) ratio.}
\label{fig-QPT-evidences}
\end{figure}

The position of a QPT in a given model depends on the topology of its potential energy surface and the case of a first order QPT, that is the one of interest for this work, implies the existence of two degenerated minima which own two different shapes \cite{Cejn10}. In the case of the IBM with a single configuration, this degeneracy exists for the critical value of the control parameter, i.e., $9/11$, as shown in Fig.~\ref{fig:gag-beta}. Before the critical point, the order parameter is different from zero, while after, it is equal to zero (see magenta line in  Fig.~\ref{fig:gag-beta}). The two minima coexist only in a narrow region around the critical point. The use of the IBM with two configurations, i.e, the IBM-CM, introduces a net difference with respect to the case of a single configuration because both configurations coexist for the whole parameters space of the Hamiltonian. In this case, the first order QPT will appear where both configurations do cross, but, on the other hand, the lowest configuration could also experience a QPT.
Strictly speaking, a QPT is only well defined in the thermodynamic limit, i.e., for systems consisting of an infinite number of particles. Therefore, in nuclear systems, due to its finite size, all the abrupt changes will be somehow smoothed out \cite{Iach98}. On the other hand, there is an additional difficulty when dealing with QPT's in nuclei that is the absence of a real control parameter because the Hamiltonian of a given nucleus cannot be changed externally. This drawback is partially solved considering that the number of neutrons (protons), or A, in a chain of isotopes (isotones) acts as an effective control parameter. Note, that doing so, the order parameter is not a continuous but a discrete variable \cite{Cejn10}.
We mention that the variation of A may involve the change of several Hamiltonian parameters and not only one, as illustrated in the schematic case of Eq. (\ref{eq:H_QPT}). In fact, in this work 12 parameters should be determine in every Zr isotope. The way to do so has been extensively discussed in great detail in Ref.~\cite{Garc19} and in an abridged way in Section \ref{sec-ibm-cm}.
In  Fig.~\ref{fig-QPT-evidences} different observables that can be considered as indicators of the presence of a QPT are presented. In Fig.~\ref{fig-QPT-evidences}(a) the experimental and theoretical values (see Section \ref{sec-ibm-cm} and  \cite{Garc19}) of the nuclear radii are presented with $A=90$ taken as the reference nucleus. This observable is related with the evolution of the order parameter and it shows a sudden increase at $A=100$. In Fig.~\ref{fig-QPT-evidences}(b) the two-neutron separation energy is depicted and both the experimental and the IBM-CM theoretical values are shown. This observable is tightly connected with the derivative of the ground state energy and exhibits a sudden change in its trend at $A=100$. Finally, in Fig.~\ref{fig-QPT-evidences}(c) the ratio E($4_1^+$)/E($2_1^+$) (experimental and IBM-CM theoretical values) is presented. This ratio is connected with the evolution of the order parameter. Indeed the observed trend is similar to the behavior of the order parameter in Fig.~\ref{fig:gag-beta}. The value of this observable experiences a sudden increase at $A=100$. 

According to the previous figures, it is a logical step to assume that Zr undergoes a QPT for $A\approx 100$ being possible to define an order parameter (Fig.~\ref{fig-QPT-evidences}(c). Moreover, there exists a kind of discontinuity in the first derivative of the energy, i.e., as observed in the S$_{2n}$ (Fig.~\ref{fig-QPT-evidences}(b).

In the areas where a QPT develops it could be of application the concept of critical point symmetry. This concept was introduced twenty years ago by Iachello, first for the transition between spherical and gamma-unstable shapes \cite{Iach00}, named as E(5) critical point symmetry and later for a transition from sphericity to axial deformation, called X(5) \cite{Iach01}. In our case the relevant critical point symmetry is X(5) because the competition is between a spherical and an axially deformed shape. The interesting conclusion follows from the X(5) critical point because it provides the value of parameter free excitation energy and transition rate ratios that can be directly compared with the experimental information. In the case of Zr, the most promising nuclei to be compared with X(5) are $^{100}$Zr and $^{102}$Zr. In Table  \ref{tab:x5}, some X(5) key ratios are compared with $^{100}$Zr and $^{102}$Zr. The experimental data for $^{152}$Sm are also given for comparison purposes. Because of the scarcity of data, in some cases we have resorted to theoretical IBM-CM calculations (see Sec. \ref{sec-ibm-cm}) and in this case they appear marked with an asterisk.  Note that in the case of $^{100}$Zr all the considered states belong to the intruder band, hence in some cases the index for a given L is a unit higher than for the X(5) one, e.g., the 0$_{3}^+$ and 2$_3^+$ states. In the case of $^{102}$Zr, the  0$_2^+$ state is the one used to be compared with X(5) although theoretically has a regular character (76\%), however the 0$_3^+$ state is not known experimentally, but it is expected to presents a major intruder character (76\%). The $0_3^+$ state presents a theoretical value of E($0_{3}^+$)/E($2_1^+$) equal to $6.87$. The 2$_2^+$ state is mainly of intruder character and, therefore this is the one used in the comparison with X(5).
\begin{table}[htb]
  \caption{Comparison of key observables in the X(5) limit with $^{152}$Sm, $^{100}$Zr and $^{102}$Zr. $^*$ stands for theoretical IBM-CM values in absence of experimental data.}
  \label{tab:x5}
  \begin{center}
    \begin{tabular}{lcccc}
      \hline
      X(5) & ~~X(5)~~ & ~~$^{152}$Sm~~ & ~~$^{100}$Zr~~ & ~~$^{102}$Zr~~\\
      \hline
      E($4_1^+$)/E($2_1^+$) & 2.91 & 3.01 & 2.66 & 3.16\\
      E($6_1^+$)/E($2_1^+$) & 5.45 & 5.80 & 5.00 & 6.39\\
      E($0_{2,3}^+$)/E($2_1^+$) & 5.67 & 5.62 & 3.91 & 5.92\\
      $\frac{B(E2:4_1^+\rightarrow 2_1^+)}{B(E2:2_1^+\rightarrow 0_1^+)}$ &  1.58 & 1.50 & 1.39 & 1.59\\
      $\frac{B(E2:2_{2,3}^+\rightarrow 2_1^+)}{B(E2:2_1^+\rightarrow 0_1^+)}$ & 0.09 & 0.04 & 0.14$^*$ & 0.18$^*$\\
      $\frac{B(E2:2_{2,3}^+\rightarrow 4_1^+)}{B(E2:2_1^+\rightarrow 0_1^+)}$ & 0.36 & 0.13 & 0.006$^*$ & $6~ 10^{-7}$$^*$\\
      $\frac{B(E2:0_{2,3}^+\rightarrow 2_1^+)}{B(E2:2_1^+\rightarrow 0_1^+)}$ & 0.63 & 0.23 & 0.37$^*$ & 0.11$^*$\\
      $\frac{B(E2:2_{2_3}^+\rightarrow 0_1^+)}{B(E2:2_1^+\rightarrow 0_1^+)}$ & 0.02 & 0.007 & 0.017$^*$ & 0.027$^*$\\
           &  &  &  & \\
      \hline
    \end{tabular}
  \end{center}
\end{table}

Regarding $^{100}$Zr, the agreement with the X(5) limit is only acceptable (similar to the case of $^{152}$Sm) for  E($4_1^+$)/E($2_1^+$), E($6_1^+$)/E($2_1^+$), $\frac{B(E2:4_1^+\rightarrow 2_1^+)}{B(E2:2_1^+\rightarrow 0_1^+)}$, $\frac{B(E2:2_{2,3}^+\rightarrow 2_1^+)}{B(E2:2_1^+\rightarrow 0_1^+)}$, $\frac{B(E2:0_{2,3}^+\rightarrow 2_1^+)}{B(E2:2_1^+\rightarrow 0_1^+)}$ and $\frac{B(E2:2_{2_3}^+\rightarrow 0_1^+)}{B(E2:2_1^+\rightarrow 0_1^+)}$, however it fails for the most important ratio  E($0_{2,3}^+$)/E($2_1^+$) which is too low. In the case of $^{102}$Zr the comparison is similar with the difference that in this case the agreement for E($0_{2,3}^+$)/E($2_1^+$) is reasonable but not so for E($6_1^+$)/E($2_1^+$). In short, the excited $0^+$ states in $^{100}$Zr are too low in excitation energy while the yrast band in $^{102}$Zr is too rigid. Hence, it is obvious that there is no crystal-clear correspondence with the X(5) critical symmetry, however the agreement for certain ratios is similar when comparing with the corresponding ratios for the case of $^{152}$Sm and therefore, it is fair to claim that $^{100-102}$Zr presents certain resemblance with the X(5) critical symmetry.

\section{The Interacting Boson Model with configuration mixing formalism}
\label{sec-ibm-cm}
The selected formalism to perform theoretical calculations in the region of interest, i.e., $^{94-110}$Zr, is an enlarged version of the IBM \cite{iach87}. The original version of the IBM was proposed in the mid 1970's by \citet{iach87} for dealing with medium and heavy mass even-even nuclei and it supposes a strong reduction of the available Hilbert space for the nucleons because only the degrees of freedom corresponding to pairs of nucleons coupled either to angular momentum $L=0$ (S pairs) or to angular momentum $L=2$ (D pairs) are considered. Moreover, these pairs are bosonized and, hence, the building blocks of the model are s and d bosons. The number of effective bosons, N, corresponds to half the number of nucleon pairs, regardless of its proton or neutron nature (this is strictly true for the simplest version of the model, IBM-1, but in other versions it is possible to define proton and neutron bosons). The IBM-1 model was enlarged for treating nuclei where more than one particle-hole configuration was used \cite{duval81,duval82}.  In this model, the original boson space $[N]$ is enlarged to $[N]\oplus[N+2]$ where in the case of Zr isotopes $N$ is the boson number, corresponding to the number of active protons outside the $Z=40$, zero in the case of Zr, plus the number of active neutrons outside the shell closure $N=50$ divided my two. The $[N+2]$ space corresponds to considering two extra bosons that come from the promotion of a pair of protons across the shell closure $Z=40$, generating an extra boson made of proton holes and another made of proton particles.

The Hamiltonian in this formalism can be written as,
\begin{equation}
  \hat{H}=\hat{P}^{\dag}_{N}\hat{H}^N_{\rm ecqf}\hat{P}_{N}+
  \hat{P}^{\dag}_{N+2}\left(\hat{H}^{N+2}_{\rm ecqf}+
    \Delta^{N+2}\right)\hat{P}_{N+2}\
  +\hat{V}_{\rm mix}^{N,N+2}~,
\label{eq:ibmhamiltonian}
\end{equation}
where $\hat{P}_{N}$ and $\hat{P}_{N+2}$ are projection operators onto the $[N]$ and the $[N+2]$ boson spaces, respectively, $\hat{V}_{\rm mix}^{N,N+2}$  is describing the mixing between the $[N]$ and the $[N+2]$ boson subspaces, and
\begin{equation}
  \hat{H}^i_{\rm ecqf}=\varepsilon_i \hat{n}_d+\kappa'_i
  \hat{L}\cdot\hat{L}+
  \kappa_i
  \hat{Q}(\chi_i)\cdot\hat{Q}(\chi_i),
  \label{eq:cqfhamiltonian}
\end{equation}
is a simplified form of the general IBM Hamiltonian called extended consistent-Q Hamiltonian (ECQF) \cite{warner83,lipas85} with $i=N,N+2$, $\hat{n}_d$ being the $d$ boson number operator, 
\begin{equation}
  \hat{L}_\mu=[d^\dag\times\tilde{d}]^{(1)}_\mu ,
\label{eq:loperator}
\end{equation}
the angular momentum operator, and
\begin{equation}
  \hat{Q}_\mu(\chi_i)=[s^\dag\times\tilde{d}+ d^\dag\times
  s]^{(2)}_\mu+\chi_i[d^\dag\times\tilde{d}]^{(2)}_\mu~,
\label{eq:quadrupoleop}
\end{equation}
the quadrupole operator. The parameter $\Delta^{N+2}$ accounts for the  energy needed to promote a pair of protons across the shell closure $Z=40$, corrected for the pairing and the monopole interaction energy \cite{Hey85,Hey87}. The operator $\hat{V}_{\rm mix}^{N,N+2}$ describes the mixing between the $N$ and the $N+2$ configurations and it is defined as
\begin{equation}
  \hat{V}_{\rm mix}^{N,N+2}=\omega_0^{N,N+2}(s^\dag\times s^\dag + s\times
  s)+\omega_2^{N,N+2} (d^\dag\times d^\dag+\tilde{d}\times \tilde{d})^{(0)}.
\label{eq:vmix}
\end{equation}

This approach has been used successfully for describing the spectroscopic properties of Pt \cite{Garc09,Garc11,Garc12}, Hg \cite{Garc14b,Garc15b}, Po \cite{Garc15,Garc15c}, and Zr isotopes \cite{Garc18,Garc19}. 
The good reproduction of the available experimental data justify the use of only 0p-0h and 2p-2h excitations in the previous cited works. In the Zr isotopic series, a few data in the heavier Zr isotopes may hint to the need for including also 4p-4h excitations, not used in the present paper.
%


The parameters of the Hamiltonian and $\hat{T}(E2)$ operator need to be fixed either from microscopical considerations, e.g., through a mapping from a shell model Hamiltonian, or from a phenomenological procedure, i.e., performing a least-squares fit to energy levels and $B(E2)$ transition probabilities. In this work, the second procedure has been used to obtain the Hamiltonian's and the $\hat{T}(E2)$ parameters, as explained in detail in Section IV.B of Ref.~\cite{Garc19}. In Table III of Ref.~\cite{Garc19} the fitted parameters of the Hamiltonian and the $\hat{T}(E2)$ operators are given and explained. 

\section{Results}
\label{sec:results}  
\subsection{Mean-field energy surfaces and the evolution of deformation}
\label{sec-q-invariants}
One of the standard ways to study QPT's or shape coexistence is through the density functional theory (DFT) that allows to obtain a mean-field energy surface and provides a clear picture about the deformation of the system. In particular, in the IBM it is possible to calculate mean-field energy surfaces defining the so called intrinsic state formalism \cite{gino80,diep80a,diep80b,Gilm74} that is not more than a condensate of bosons (or coherent state), namely, a condensate of deformed bosons that depend on two deformation parameters $\beta_B$ and $\gamma_B$, where the sub-index B stands for the parameters corresponding to bosons. These deformation values should not be confused with the $\beta$ and $\gamma$ deformation parameters as describing a quadrupole deformed nuclear surface. The intrinsic state formalism is defined as, 
\begin{equation}
\label{GS}
|N; \beta_B,\gamma_B   \rangle = {1 \over \sqrt{N!}}
\left({1 \over \sqrt{1+\beta_B^2}} \left (s^\dagger + \beta_B
\cos     \gamma_B          \,d^\dagger_0          +{1\over\sqrt{2}}\beta_B
\sin\gamma_B\,(d^\dagger_2+d^\dagger_{-2}) \right) \right)^N | 0 \rangle
\end{equation}
and, therefore, the mean-field energy corresponds to the matrix element $E(N,\beta_B,\gamma_B)=\langle N; \beta_B,\gamma_B 
|\hat H| N; \beta_B,\gamma_B  \rangle$. The value of $\beta_B$ and $\gamma_B$ which minimize the expectation value of the energy provide the shape of the nucleus with a given number $N$ of bosons.

The latter formalism needs to be extended when dealing with IBM-CM, defining a matrix within coherent state bands,
\begin{equation}
H_{CM}=\left (
\begin{array}{cc}
E(N,\beta_B,\gamma_B)& \Omega(\beta_B)\\
\Omega(\beta_B)& E(N+2,\beta_B,\gamma_B)
\end{array}
\right ) .
\label{surf-cm}
\end{equation}
which lowest eigenvalue will correspond to the mean-field energy surface of the model (see \cite{Frank02,Frank04,Frank06,Mora08} for a detail description of the formalism). One can easily compute this energy surfaces once the parameters of the Hamiltonian are given. In our case we use the ones obtained in \cite{Garc19}.
\begin{figure}
  \centering
 \includegraphics[width=0.60\linewidth]{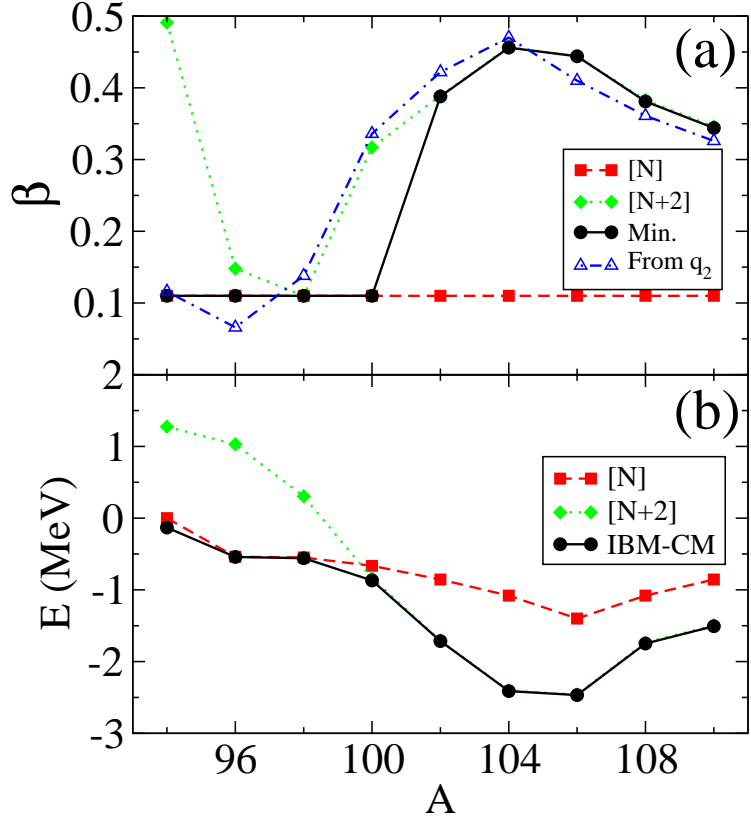}
 \caption{(a) Comparison of the scaled value of $\beta$ from the IBM-CM mean-field energy (``Min.'') and from Eqs.~(\ref{q2}) and (\ref{beta-q2}) (``From q$_2$''). The scaled value of $\beta$ corresponding to the unperturbed ``regular'', [N], and the unperturbed ``intruder'', [N+2], configurations are also depicted. (b) Energy of the unperturbed regular and intruder IBM bandheads, compared with the ground-state energy (full IBM-CM diagonalization).}
 \label{fig-beta-energy}
 \end{figure}

The comparison of the IBM-CM energy surface with the one from mean-field DFT using a realistic interaction needs some care because the IBM deformation parameters should  be rescaled. Therefore, one needs to establish a connection between $(\beta_B, \gamma_B)$ and  $(\beta, \gamma)$. To do so we closely follow  the technique presented in \cite{Garc14a}, using the {\it ansatz},
\begin{equation}
\beta= 1.18 \frac{2}{A} \beta_B (N+2~(1-\omega)) ~ \delta +\xi.
\label{beta2}
\end{equation} 
where $\delta$ and $\xi$ are fitting parameters to be obtained from the comparison of the IBM deformation parameters with the ones of the collective model. On the other hand, it is assumed that $\gamma_B=\gamma$.
\begin{figure}[hbt]
\centering
\includegraphics[width=0.80\linewidth]{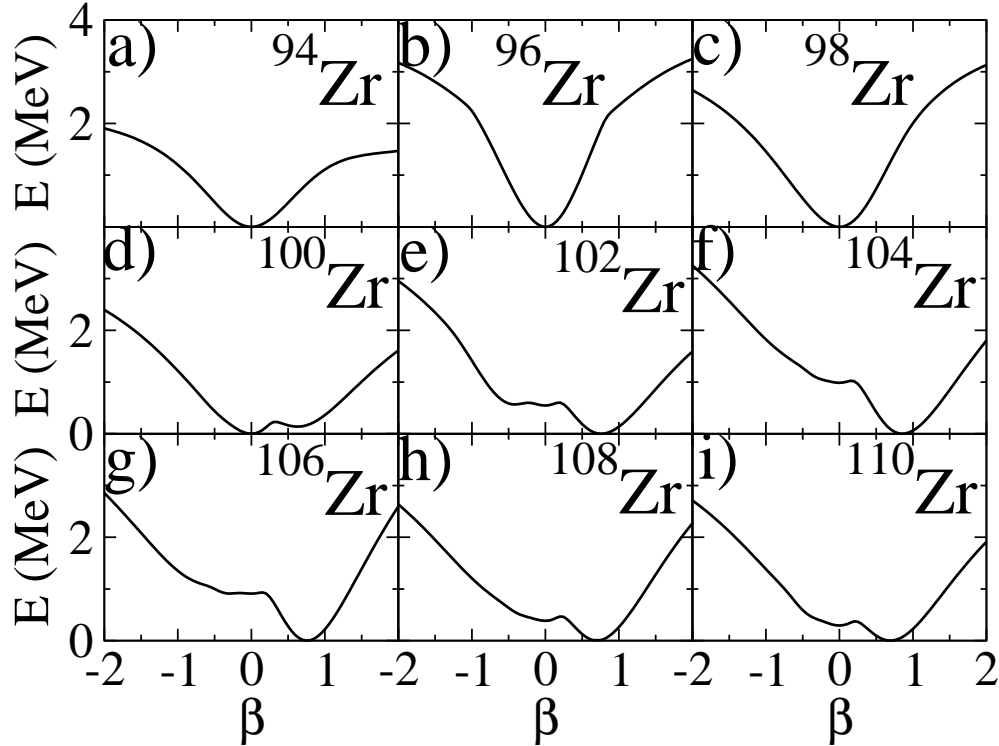}
\caption{Matrix coherent-state potential energy surface along the axial axis for even-even Zr isotopes, from $^{94}$Zr up to $^{110}$Zr. The deepest energy minimum is set to zero.}
\label{fig-ener-axial}
\end{figure}


The main novelty of this work with respect to \cite{Garc14a} is that in the present study, the value of $\beta$ is extracted from the quadrupole shape invariants using the wave function obtained from the IBM-CM calculation with the parameters given in \cite{Garc19} while in \cite{Garc14a} it was obtained from a mean-field energy surface using a Gogny-D1S interaction. 
The equations involved to extract $\beta$ from the quadrupole shape invariants are \cite{kumar72,Cline86},
\begin{equation}
q_{2}=\sqrt{5} \langle 0^+| [\hat{Q} \times \hat{Q} ]^{(0)}|0^+\rangle,
\label{q2}
\end{equation}
\begin{equation}
\beta=\frac{4\, \pi\, \sqrt{q_2}}{3\, Z\, e\, r_0^2\, A^{2/3}},
\label{beta-q2}
\end{equation}
where $r_0=1.2 A^{1/3}$ fm and $e$ the charge of the proton.
\begin{figure}
\includegraphics[width=0.32\textwidth]{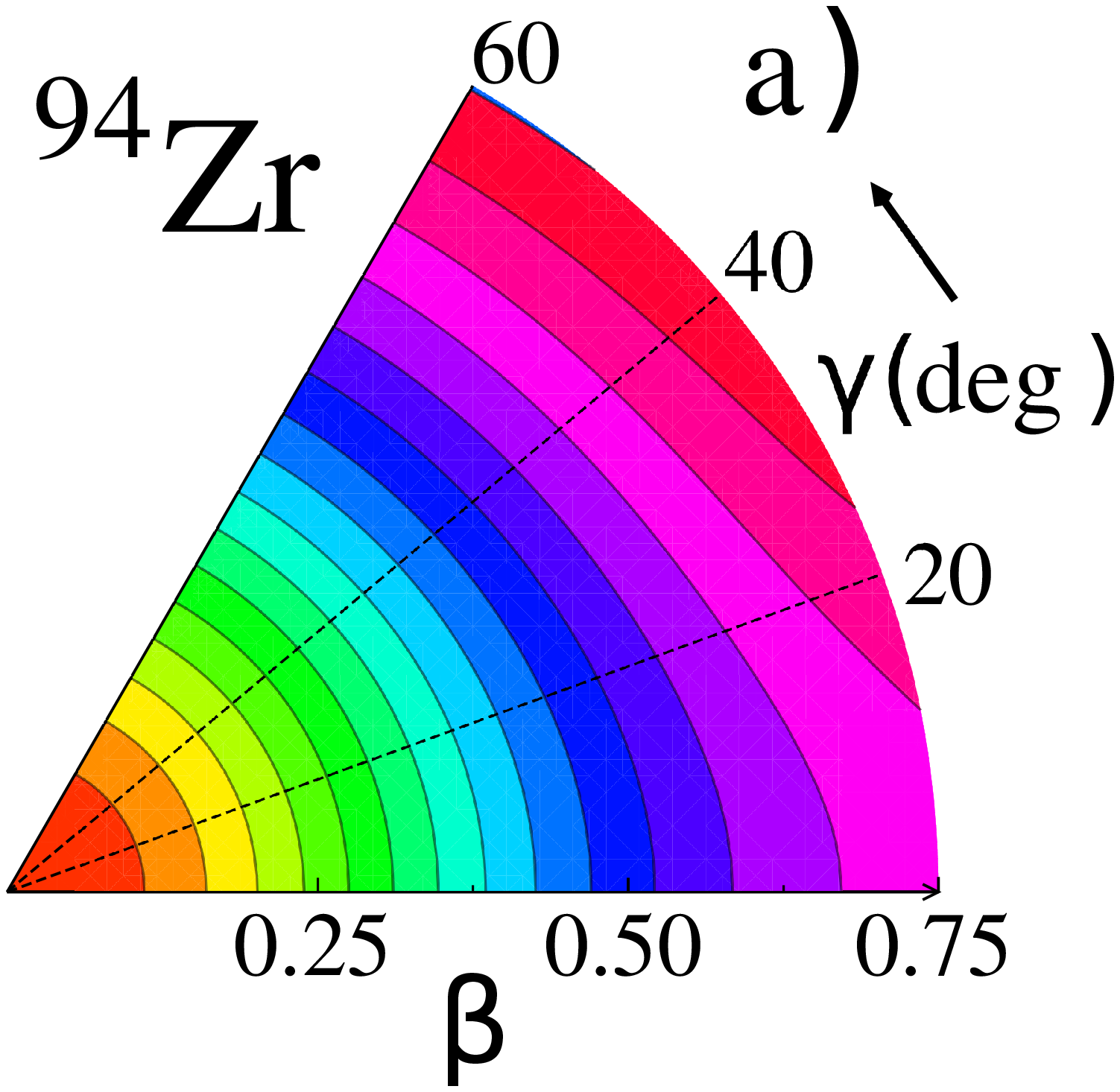}%
\includegraphics[width=0.32\textwidth]{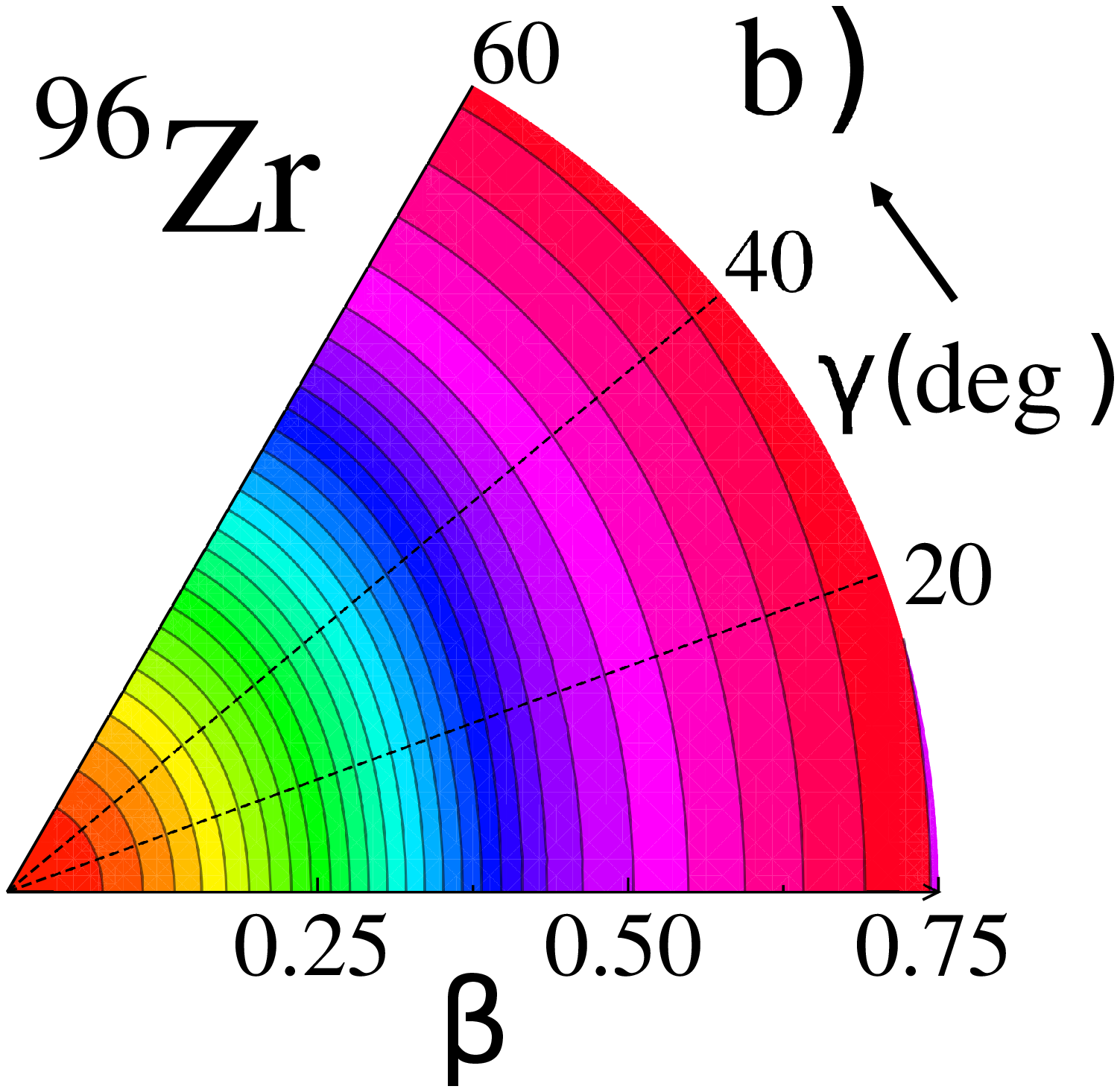}%
\includegraphics[width=0.32\textwidth]{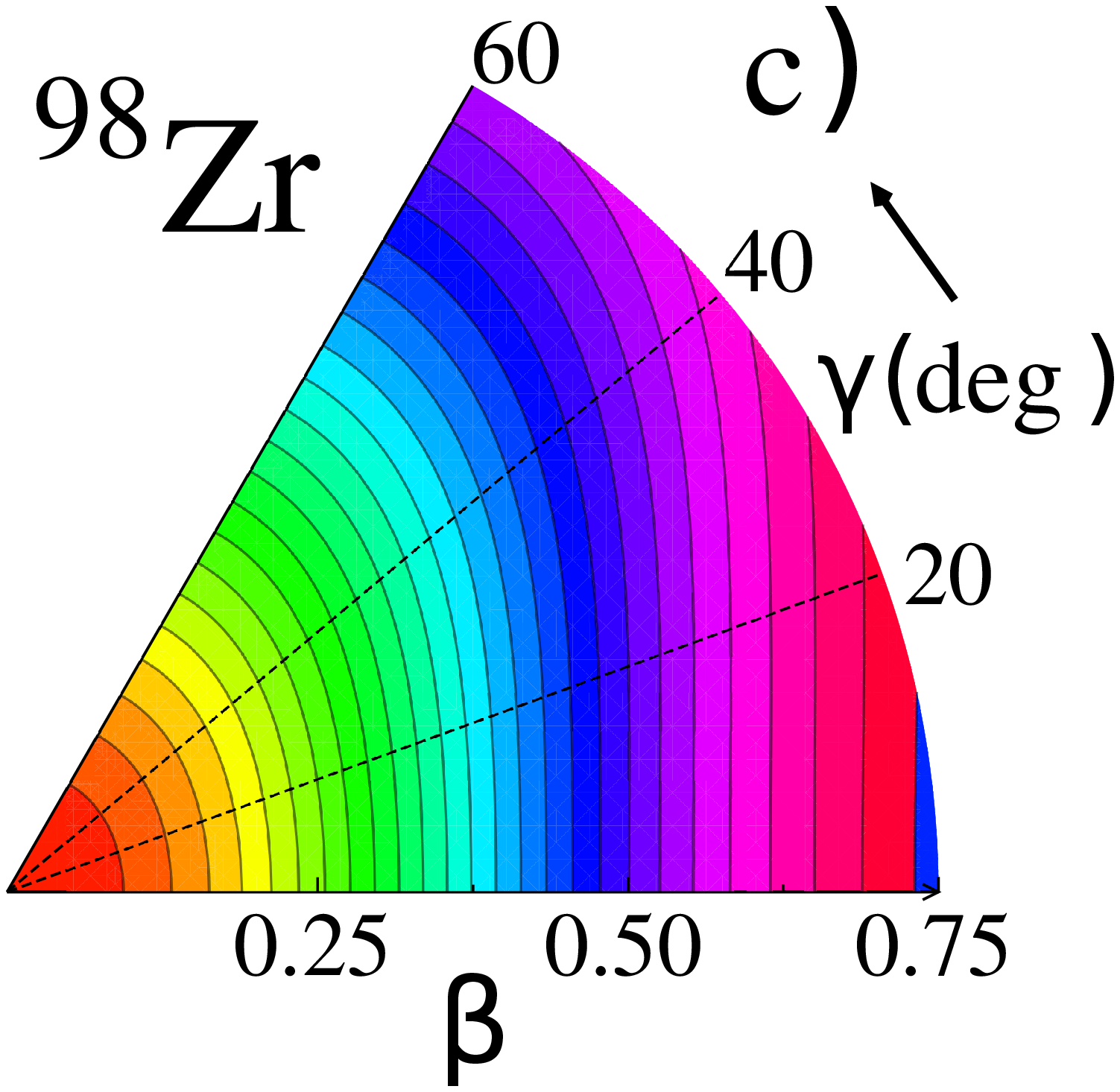}\\
\includegraphics[width=0.32\textwidth]{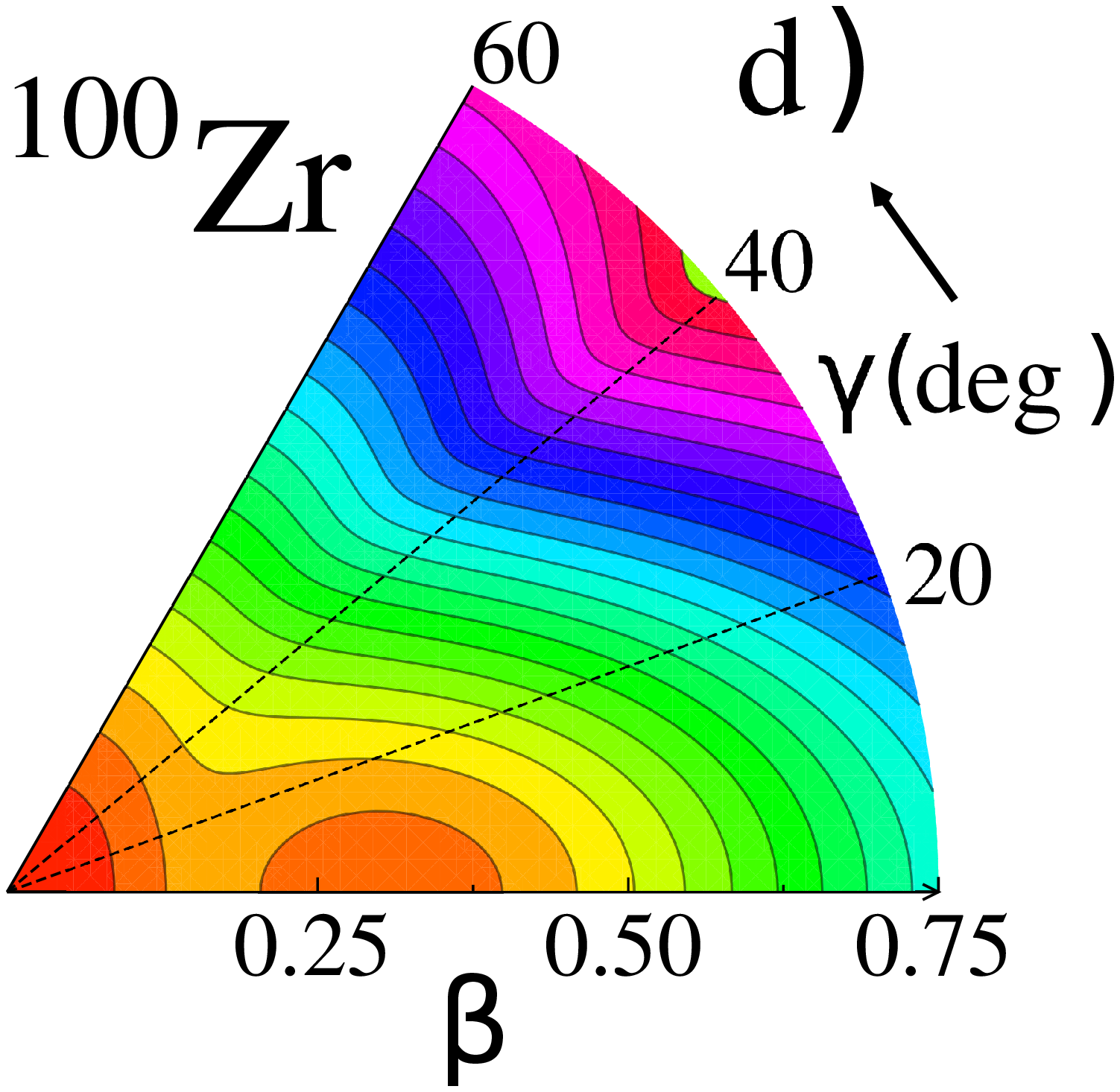}%
\includegraphics[width=0.32\textwidth]{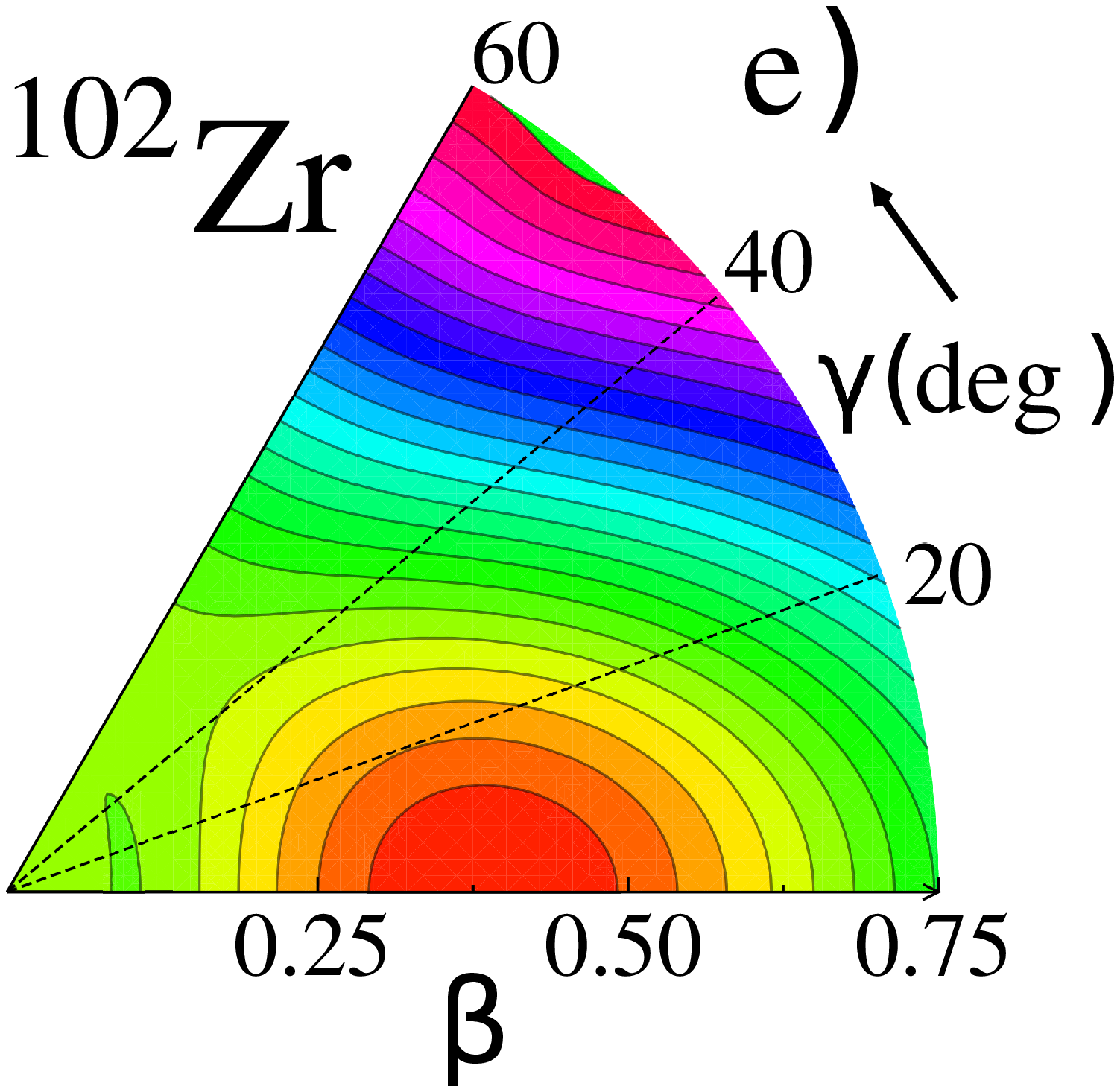}%
\includegraphics[width=0.32\textwidth]{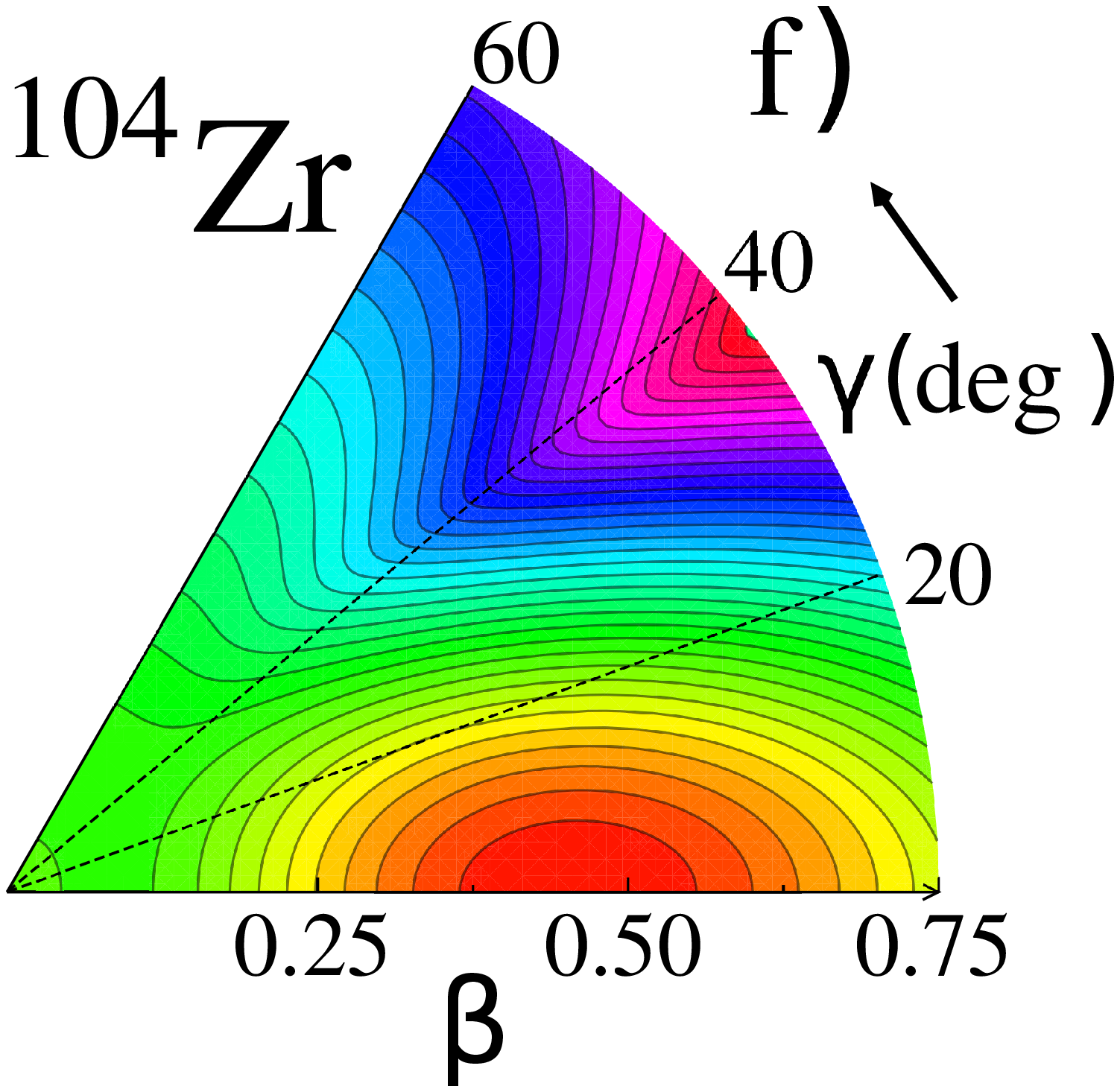}\\
\includegraphics[width=0.32\textwidth]{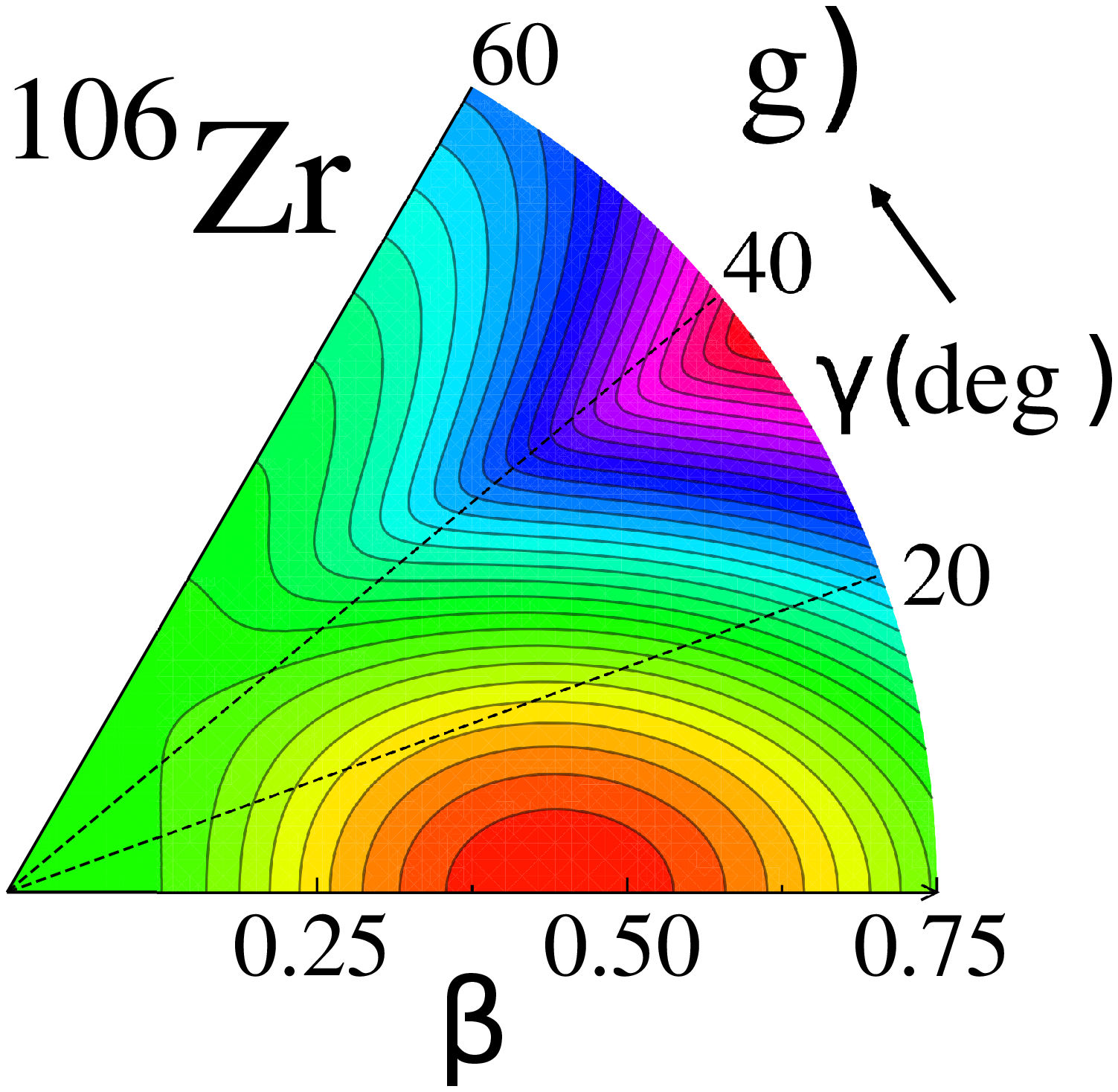}%
\includegraphics[width=0.32\textwidth]{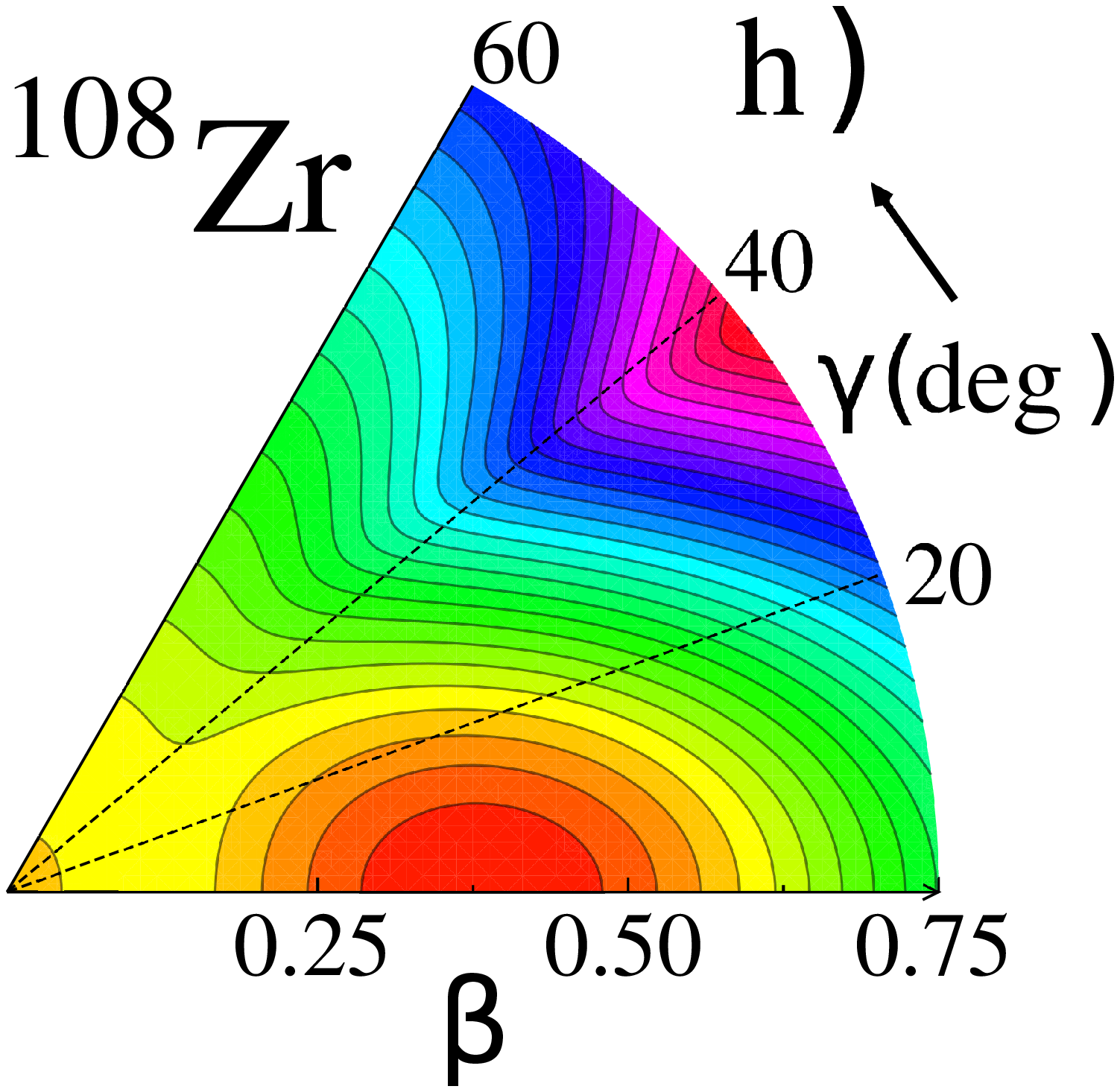}%
\includegraphics[width=0.32\textwidth]{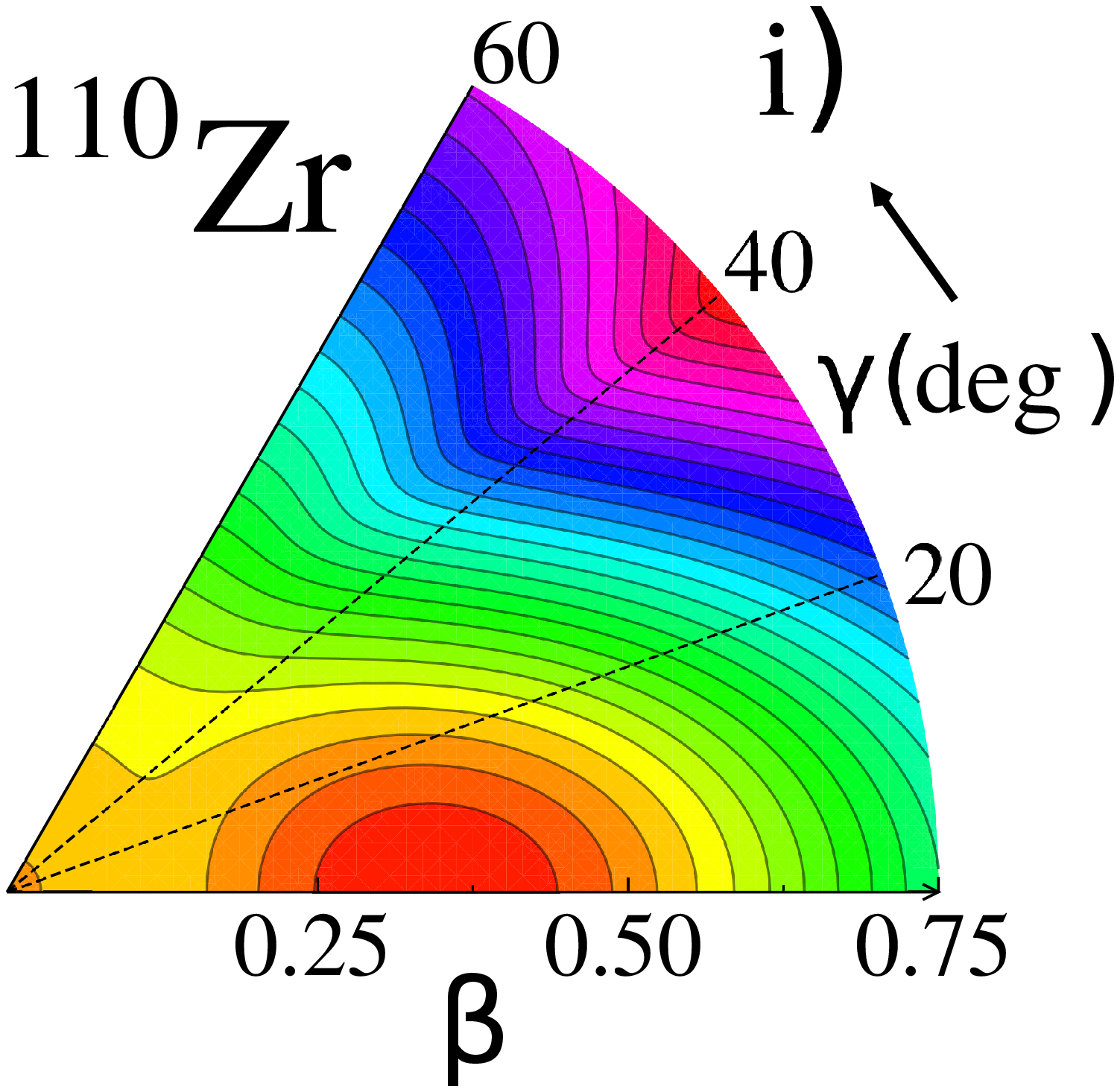}\\
\caption{Matrix coherent-state potential energy surface in the $\beta-\gamma$ plane for even-even Zr isotopes, from $^{94}$Zr up to $^{110}$Zr. The energy spacing between adjacent contour lines is $100$ keV and the deepest energy minimum is set to zero, corresponding to the red color.}
\label{fig_ibm_ener_surph}
\end{figure}

As next step we perform a least-squares fit using the value of $\beta$ extracted from the quadrupole shape invariants and the value of $\beta_B$ from IBM corresponding to the minimum of the mean-field energy surface. This procedure leads to the following values of the fitting coefficients including its respective errors $\delta=1.98(15)$, $\xi=0.110(17)$.

In Fig.~\ref{fig-beta-energy}(a) the scaled $\beta$ values extracted from Eq.~(\ref{beta2}) are compared  with the ones obtained from the quadrupole shape invariant (\ref{q2},\ref{beta-q2}). We also show the ones corresponding to the unperturbed [N] and [N+2] configurations. Note that the IBM-CM $\beta$ value  for $A=100$ corresponds to a spherical minimum, although the deformed one was used for performing the least-squares fit previously described, because it is known that for $A=100$ Zr is already strongly deformed. The spherical minima (for $A<100$) have values of $\beta\neq 0$ because the obtained intercept in the least-squares fit is also different from $0$. According to the presented results, the regular configuration always present a very small value of $\beta$ and therefore we identify them with spherical states. The intruder configuration shows a rather different trend. On one hand, it is strongly perturbed because of the vicinity of the $N=50$ and $N=56$ shell and subshell closures, respectively.  On the other hand, it presents the rapid change from a spherical shape towards a deformed one, around $A=100$. The agreement between the mean-field results and the ones extracted for the quadrupole shape invariants is remarkable. The switch in $\beta$ from the regular to the intruder configuration is perfectly understandable according to Fig.~\ref{fig-beta-energy}(b) where the unperturbed energies of the regular and intruder configurations cross at $A=100$. Note that the value of $\beta$ obtained from the quadrupole shape invariant already corresponds to a deformed shape for $A=100$ while not the one from mean field. This is due to the existence of two almost degenerate minima at the IBM-CM mean-field level for $A=100$, one spherical and the other deformed, being the spherical minimum a little deeper than the deformed one (see Figs.~\ref{fig-ener-axial}(d) and \ref{fig_ibm_ener_surph}(d)). However, the calculation in the laboratory frame leads to a deformed shape for $A=100$.

In Fig.~\ref{fig-ener-axial} the axial energy surface for the whole chain of isotopes is depicted, while in Fig.~\ref{fig_ibm_ener_surph} the energy surfaces in the $(\beta,\gamma)$ plane is presented. Note that the rescaled $\beta$ value has been used in all the figures. 
These two sets of figures confirm the existence of a single minimum of spherical nature for $^{94-98}$Zr; for $^{100}$Zr two minima coexist with almost the same energy, though the spherical is the deepest one; in $^{98-102}$Zr  a rapid transition from a spherical to well-deformed shapes is observed; from $^{102}$Zr onwards two minima coexist, although in this case the deformed is the deepest one and the spherical minimum is very shallow. This fact is more clearly observed in Fig.~\ref{fig-ener-axial} as compared with Fig.~\ref{fig_ibm_ener_surph}. Note that the deformed minimum for the heavier isotopes is always prolate. 

It can be of interest to compare the results presented in the present study with the mean-field energy surfaces obtained with realistic interactions. In the literature, we encountered several recent calculations in which the Zr isotopes are studied. In the Bruy\`eres-le-Ch\^atel compilation \cite{Hila08,Bruyere-surfaces}, Hartree-Fock-Bogoliubov  calculations based on the Gogny-D1S effective interaction are presented. The obtained values of $\beta$ in the deformed cases are quite similar to the ones obtained in the present study, although the minima of $^{98-100}$Zr are oblate (almost degenerate with the prolate one), and, in general, prolate and oblate minima are connected through a triaxial valley. The same holds for the other isotopes. In general, in the well deformed cases $\beta_{\text{oblate}}\approx 0.2$ while $\beta_{\text{prolate}}\approx 0.4$. The IBM-CM calculations for these nuclei present $\beta\approx 0.4$. In Ref.~\cite{Nomu16}, HFB energy surfaces using a Gogny-D1M interaction are used and mapped into an IBM-CM with three different particle-hole configurations. Here, also prolate and oblate almost degenerate minima coexist, but also a spherical minima, although at very high energy, exists. The main drawback of this approach is that it predicts a deformed shape for $^{96}$Zr and $^{98}$Zr. In Ref.~\cite{Togashi16}, Monte Carlo shell-model calculations are carried out for the Zr isotopes. Within this approach it is possible to generate total energy surfaces and to extract information on the deformation of the states. Here, the deformed nuclei corresponds to prolate shapes and the potential energy is flat in the $\gamma$ direction, with presence of a local minimum in the oblate side, but less deformed than the prolate one. The value of $\beta$ for $^{100-101}$Zr is approximately $0.35$ which is slightly smaller than the value obtained with a Gogny interaction. In Ref.~\cite{Rodr10}, the interaction Gogny-D1S has also been used and an oblate-prolate transition is observed, passing from $A=98$ to $A=100$, resulting in a prolate shape with $\beta\approx 0.44$ for $^{100}$Zr. In \cite{Abus17} a relativistic EDF is considered, obtaining a general prolate-oblate coexistence for the deformed part of the Zr chain and, in the case of $^{100}$Zr, also a spherical-prolate coexistence. Moreover, in the heaviest isotope the deepest minimum corresponds to the oblate one.

In summary, a comparison between the realistic mean-field calculations and the IBM-CM energy surfaces points to certain similarities, as for example, the spherical shape of the lightest isotopes and the presence of a local spherical minimum for $^{100}$Zr, although at the same time an oblate minimum is present in the realistic mean-field calculations together with a shallow valley connecting with the prolate one, which is almost degenerate with the first one.
However, in the IBM only the spherical and the prolate minima are present, which seems to be a limitation of the used approach.%
Most of the  realistic mean-field calculations  present an oblate shape (global minimum) for the heaviest isotopes while the IBM energy surface calculations always generate a prolate one.
\begin{figure}
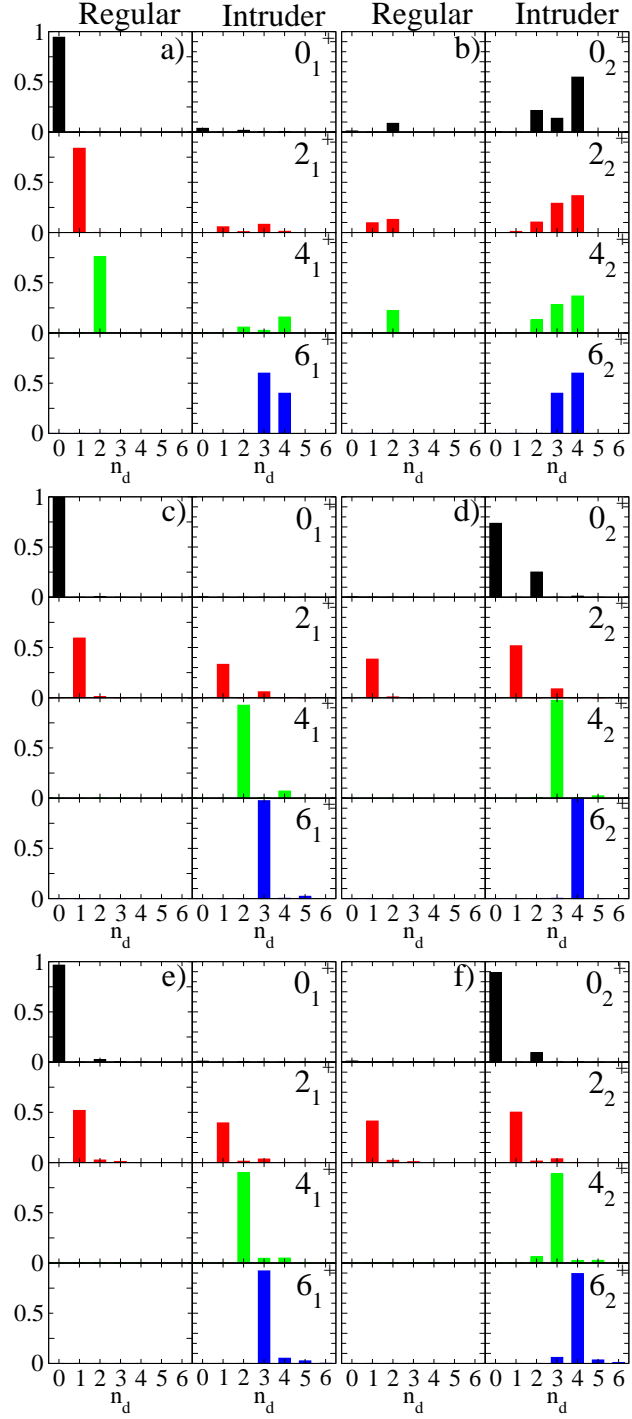

  \centering
  \includegraphics[width=.5\linewidth]{94ZR-U5.eps}\\
  \includegraphics[width=.5\linewidth]{96ZR-U5.eps}\\
  \includegraphics[width=.5\linewidth]{98ZR-U5.eps}
  \caption{U(5) decomposition ($n_d$ content, $NDC$, see text for details) of the yrast band and the states $0_2^+$, $2_2^+$, $4_2^+$, and $6_2^+$ of $^{94}$Zr (panels (a) and (b)), $^{96}$Zr (panels (c) and (d)), $^{98}$Zr (panels (e) and (f)). Columns labeled with ``Regular'' correspond to the decomposition in the regular sector, while labeled with ``Intruder'' to the decomposition in the intruder sector.}
  \label{fig-wf-u5-1}
\end{figure}

\subsection{Wave function structure}
\label{sec-evolution}


To know about the structure of the different eigenstates of Zr isotopes, it is very enlightening to express the wave function using the U(5) basis in both the regular and the intruder spaces. This basis allows to distinguish  between states with a spherical character, i.e., states described with wave functions having an almost single U(5) component as well as states with a broad distribution of U(5) components. To do so, we express  explicitly the eigenfunctions in the U(5) basis, making use of the corresponding U(5) quantum numbers, namely, $n_d$, $\tau$, and $n_\Delta$, for the N and N+2 subsystems, respectively,
\begin{eqnarray}
\Psi(k,JM) &=& \sum_{n_d,\tau,n_\Delta} a^{k}_{n_d,\tau,n_\Delta}(J;N) \psi((sd)^{N}_{n_d,\tau,n_\Delta};JM) 
\nonumber\\
&+& \
\sum_{n_d,\tau,n_\Delta} b^{k}_{n_d,\tau,n_\Delta}(J;N+2)\psi((sd)^{N+2}_{n_d,\tau,n_\Delta};JM)~.
\label{eq:wf:U5b}
\end{eqnarray}
Here, $n_d$ corresponds to the number of $d$ bosons, $\tau$ to the boson seniority,  $n_\Delta$ to the number of $d$ boson triplets coupled to zero, and  $k$ is a rank number to label the state. The $n_d$ content of the wave function, that we will call $NDC$, either in the regular or the intruder sector are defined as,
\begin{eqnarray}
  NDC(n_d,k,N)&=&\sum_{\tau,n_\Delta} \left(a^{k}_{n_d,\tau,n_\Delta}(J;N)\right)^2\\
  NDC(n_d,k,N+2)&=&\sum_{\tau,n_\Delta} \left(b^{k}_{n_d,\tau,n_\Delta}(J;N+2)\right)^2
\end{eqnarray}
fulfilling the normalization constraint,
\begin{equation}
  \label{eq:norm}
  \sum_{n_d} NDC(n_d,k,N)+ \sum_{n_d} NDC(n_d,k,N+2)=1.
\end{equation}
Here, the sum extends over all allowed $n_d$ values in the Hilbert spaces. $NDC$ provides precise information on the structure of the considered states. On one hand, if only one, or very few, of the $n_d$ components have a sizeable value, it implies that the state has a vibrational or spherical character, either in the regular or in the intruder sector. On the other hand, if the $NDC$ values are broadly distributed over several $n_d$ components, the state can be defined to exhibit a rotational or deformed character.  This is a possible way to express the underlying structure of the wave functions derived from the IBM-CM, but the $NDC$ values do not correspond in any direct way with observables as measured in the various experiments.
\begin{figure}
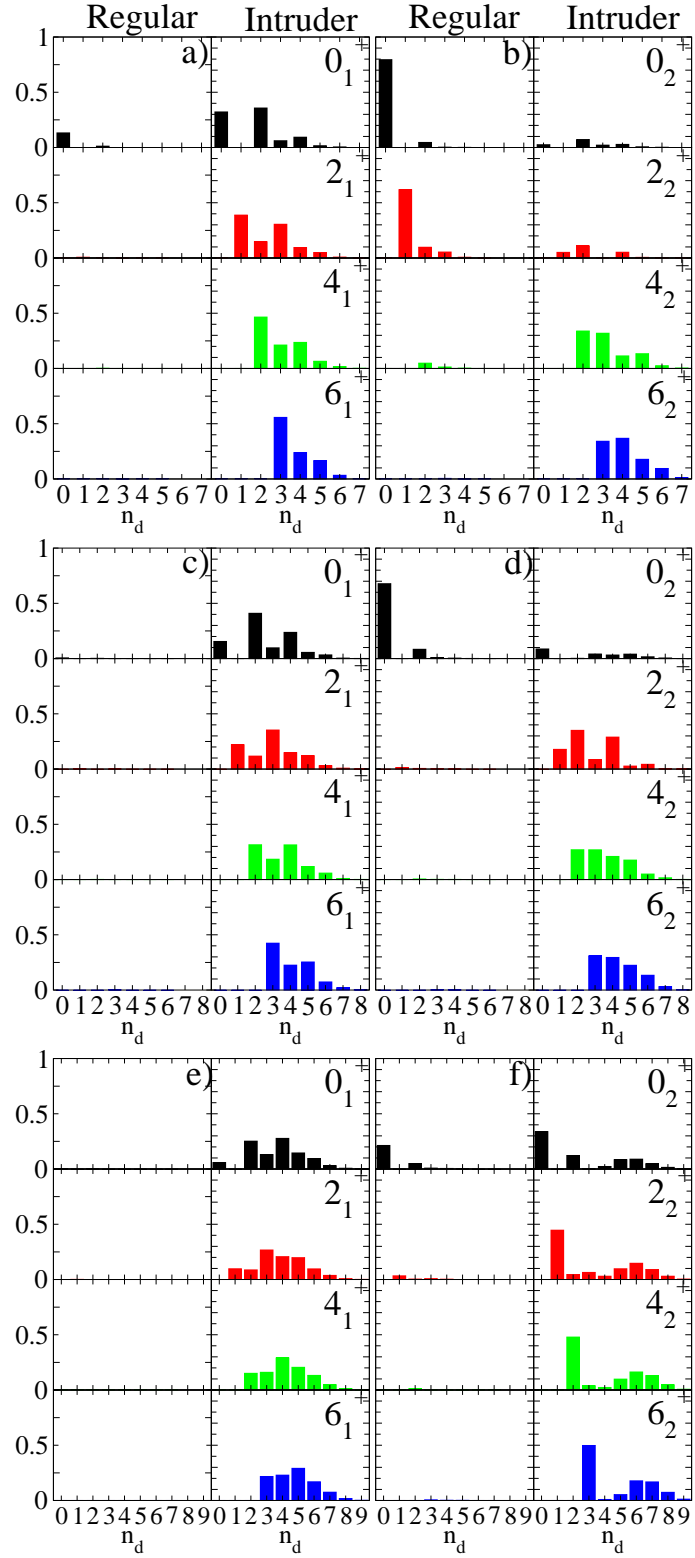

  \centering
  \includegraphics[width=.55\linewidth]{100ZR-U5.eps}\\
  \includegraphics[width=.55\linewidth]{102ZR-U5.eps}\\
  \includegraphics[width=.55\linewidth]{104ZR-U5.eps}
  \caption{Same caption than Fig.~\ref{fig-wf-u5-1}, but for $^{100}$Zr (panels (a) and (b)), $^{102}$Zr (panels (c) and (d)), $^{104}$Zr(panels (e) and (f)).}
  \label{fig-wf-u5-2}
\end{figure}

In Figs.\ \ref{fig-wf-u5-1} and \ref{fig-wf-u5-2} we depict the value of $NDC$ for the first and the second member of each angular momentum $0^+$, $2^+$, $4^+$,  $6^+$. We present the results for the isotopes where the most rapid changes in structure are expected. In panels (a) and (b) of Fig.~\ref{fig-wf-u5-1}, $^{94}$Zr is presented. It is easily appreciated how the states $0_1^+$, $2_1^+$, $4_1^+$ own a regular and spherical character, while  $6_1^+$ is also of spherical nature but situated mainly in the intruder sector. On the other hand, the states  $0_2^+$, $2_2^+$, $4_2^+$ are indicative of a deformed behavior (the distribution is spread over several components) in the intruder sector, although  containing a small contribution in the regular side. Finally, the structure of the state $6_2^+$ is intruder and spherical. The nuclei  $^{96}$Zr and  $^{98}$Zr (panels (c), (d), (e), and (f)) exhibit a similar structure, namely,  $0_1^+$, $4_1^+$, and $6_1^+$ present a regular and spherical behavior; $2_1^+$ and $2_2^+$ also present a spherical behavior but with components in the regular and in the intruder sectors; finally,  $0_2^+$, $4_2^+$, and $6_2^+$ owns an intruder, but spherical character. In Fig.~\ref{fig-wf-u5-2} we enter into the region where deformation quickly appears and consequently, the yrast band of the three nuclei (panels (a), (c), and (e)) exhibit an almost $100\%$ intruder component for all analyzed angular momenta, with $NDC$ values spread over many components, again acting as a strong hint for the presence of a deformed shape. For the non-yrast states, the situation is slightly different in every isotope. In the case of $^{100}$Zr (panel (b)), the states $0_2^+$ and $2_2^+$ present a spherical and regular character, while  $4_2^+$ and $6_2^+$ an intruder and deformed one. In the case of $^{102}$Zr (panel (d)), only the $0_2^+$ state presents a regular and spherical character, while the rest of the band exhibits an intruder and rather deformed character, with the $NDC$  distributed over many $n_d$ components. Finally, in $^{104}$Zr (panel (f)), all the states show an intruder character with a rather spread $NDC$ distribution, but also with a very large component at the lowest possible value of $n_d$. The corresponding figures for $^{104-110}$Zr (not shown) present the same trend as in $^{102}$Zr.

In summary, one observes how the structure of the regular states evolves from being spherical for  $^{94,96,98,100,102}$Zr to being slightly deformed in $^{104}$Zr and onwards (not shown in figures). Concerning the structure of the intruder states, they start with a deformed structure in $^{94}$Zr, but then the internal structure changes into a spherical one for $^{96-98}$Zr, in $^{100}$Zr it presents a mixed structure, finally, becoming deformed in $^{102-104-106}$Zr and onwards.

\subsection{Density of states}
\label{sec-density}
The density of low lying states is a property that is expected to be strongly perturbed by the presence of intruder states or by the existence of a QPT. Hence, it is worth to be analyzed. As a matter of fact, \citet{Buca18} propose the density of states as an indicator to distinguish between shape coexistence and the presence of a QPT.
\begin{figure}[htb]
\centering
\includegraphics[width=.45\linewidth]{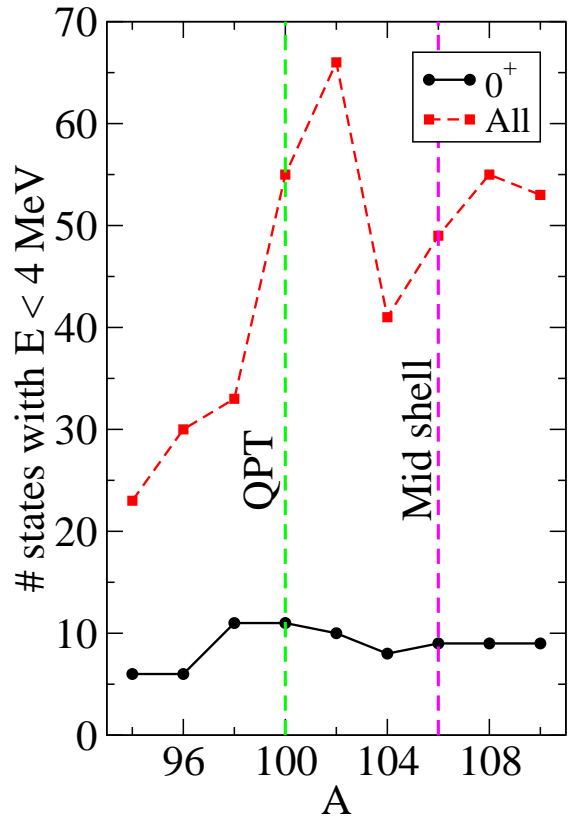}
\caption{Number of states below an excitation energy E=$4$ MeV as a function of A, computed using IBM-CM calculations \cite{Garc18}. Black full line corresponds to the number of $0^+$ states and red dashed one to the total number of states, up to and including L$=8$.}
\label{fig:density}
\end{figure}

In this section we present a quantity that is related with the density of states as it is the number of states below a certain energy, namely $4$ MeV. To such an end we rely on the IBM-CM results from \cite{Garc18}. Owing to that the presence of low lying $0^+$ states is a hint for the existence of shape coexistence we compute the number of $0^+$ states together with the total number of states up to and including $L=8$ below $4$ MeV in Fig.~\ref{fig:density}.

In Fig.~\ref{fig:density} the theoretical number of $0^+$ and the total number of states are presented. Both present a sudden increase when reaching the critical point, $A=100$. The number of $0^+$ states remains rather constant after the critical point although with a smooth decrease, but it is barely affected by the presence of the mid shell. Concerning the total number of states below 4 MeV, it also suffers a rapid increase at the QPT point, but in this case there are also abrupt changes after this point.

\citet{Buca18} suggest that the behavior of the density of states can provide some hints to distinguish between the presence of shape coexistence from the presence of a QPT. The key point is that a QPT implies the existence of a maximum, with a peak, in the density of states, while the existence of shape coexistence supposes a rapid increase in the density of states but rather with a plateau than with a peak. 

According to our results it seems that the number of $0^+$ states points towards the presence of shape coexistence. However, the peak in the total number of states rather suggests the existence of a QPT. Hence, no clear conclusions can be obtained from this analysis.

\section{Discussion}
\label{sec:discussion}
\begin{figure}[hbt]
  \centering
  \includegraphics[width=0.60\linewidth]{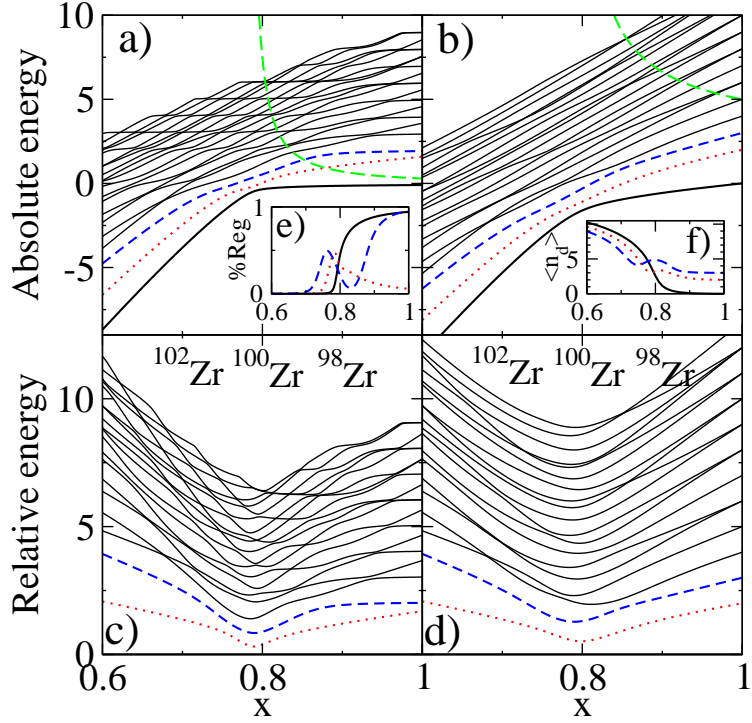}
  \caption{ (a) Absolute energy systematics of $J^\pi=0^+$ states for an IBM Hamiltonian with two configurations, one with constant parameters and another one that experiences a QPT (see text); (b) absolute energy systematics of $J^\pi=0^+$ states for an IBM Hamiltonian with a single configuration that experiences a QPT (see text); (c) the same as in panel (a) but for the relative energy;  (d) the same as in panel (b) but for the relative energy; (e) regular component of the  first three low-lying states corresponding to panel (a); in panel (f), the $\langle n_d\rangle$ value for the first three  low-lying states corresponding to panel (b). The green curve stands for the derivative of the ground state energy.  All energies are given in arbitrary units.}
  \label{fig:CM-QPT}
 \end{figure}
 So far we have accumulated evidences pointing either to the presence of shape coexistence or to the existence of a QPT and most probably the reader got the impression that both phenomena are, indeed, difficult to be separated because both are tightly connected. To shed some light on this issue it is useful to perform a set of schematic calculations comparing both situations but at the same time being close to what happens experimentally for the Zr isotopes. This is illustrated in Fig.~\ref{fig:CM-QPT} where the spectrum of an IBM-CM mixing calculation is compared with an IBM single configuration one. In panels (a), (c), and (e) we used a configuration mixing Hamiltonian with $\varepsilon_N=1$ (in arbitrary units) and the remaining of the regular Hamiltonian parameters are put equal to zero, while $\varepsilon_{N+2}=x$, $\kappa_{N+2}=\frac{x-1}{N+2}$, $\chi=-\sqrt{7}/2$, and the remaining of the intruder parameters are put equal to zero. The number of bosons is set to $N=18 \;(N+2=20)$ and the mixing term to $\omega_0^{N,N+2}=\omega_2^{N,N+2}=0.02$. This small value will ensure a tiny mixing between the regular and the intruder sectors in the wave function. The shift parameter is fixed to $\Delta^{N+2}=0.75$ to guaranty a crossing of the regular and the intruder $0^+$ bandheads at $x=9/11 \approx 0.8181$ for $\omega_0^{N,N+2}=\omega_2^{N,N+2}=0.0$. Moreover, in panels (b), (d), and (f) we present the calculation for a single IBM configuration with  $\varepsilon=x$, $\kappa=\frac{x-1}{N}$, $\chi=-\sqrt{7}/2$, with the remaining parameters equal to zero and using $N=20$. Note that $x$ acts as a control parameter in both Hamiltonians.

 The schematic IBM-CM Hamiltonian was built in order to mimic the Zr systematics, with a regular configuration which Hamiltonian remains essentially constant and an intruder configuration that undergoes a QPT. The mass number $A=100$ qualitatively should correspond to the control parameter $x=9/11$, simply because, as pointed out in Section \ref{sec-evidences-QPT}, $^{100}$Zr corresponds to a critical point. Moreover, increasing (decreasing) values of $x$ qualitatively correspond to decreasing (increasing) values of A because the nuclei become more spherical (deformed) as $x$ increases. On the other hand, the IBM single configuration calculation only aims at reproducing what happens in the Zr ground states. Once more, there is a QPT at $x=9/11$. To help the understanding of Fig.~\ref{fig:CM-QPT}, we have added the labels $^{98}$Zr, $^{100}$Zr,  and $^{102}$Zr as an extra reference to guide the eye inspecting the content of this figure.%

 What strikes in the comparison is that for $x<9/11$ and $E<0$ both spectra are almost identical. The reason is that for this area the leading Hamiltonians in left and right panels are approximately the same. However, above $E=0$ clear differences are present because the almost harmonic spectrum of the regular Hamiltonian, i.e., E=$0, 2,3,4, \ldots$, is present in panels (a) and (c), but not in (b) and (d). These vibrational states are clearly observed because of the small mixing that has been used. Besides, in panels (a) and (c) one can note how the spherical states repel the deformed ones and avoided crossings are present, happening in particular for the ground state. Regarding panels (b) and (d), the global spectrum undergoes an evolution from deformed to spherical shapes because of the dependence of the Hamiltonian on $x$. To better appreciate the analogies and differences between both calculations, the lowest three states are marked. Concerning these three states, for $x<9/11$, the left panel and the right panel show essentially the same trend because for this zone the spherical states (regular) are not yet influencing the lower part of the spectrum. However, for $x=9/11$ the slope of the energy curves changes more rapidly in the IBM-CM case as compared with the single configuration one. The reason is due to the crossing and repulsion between regular and intruder families of states. The most evident consequence is that the change in the first derivative (green long dashed curves) of the ground state energy is much faster in the left than in the right panels. Second consequence refers to the phonon structure of the states for $x>9/11$. In panel (a) and (c) the first and the second states have $\langle n_d\rangle\approx 0$, while the third has $\langle n_d\rangle\approx 2$ (not shown in the figure). This follows from the fact that the first and the third states have intruder nature while the second one is mainly of regular character, as shown in panel (e).  In panel (b) and (d), however, the ground state has $\langle n_d\rangle\approx 0$, the second  $\langle n_d\rangle\approx 2$, while the third  $\langle n_d\rangle\approx 3$, as illustrated as a function of x in panel (f). As a consequence, the presence in the spectrum of two low lying $0^+$ states with $\langle n_d\rangle\approx 0$ is a direct consequence of the existence of two families of states. The existence of two families of states is responsible of the more rapid lowering of the first excited state as compared with the case of a single configuration [see panels (c) and (d)].
 
 The quantity presented in panels (e) and (f), the regular component and  $\langle n_d \rangle$, respectively, are both ideal candidates to serve as an order parameter. The observed changes are much more rapid in the case of two configurations than in the case of a single one, as already observed for the ground state energy. Therefore, an extra consequence of the presence of two configurations is that the induced QPT happens abrupt as compared with the case when considering only a single configuration. This is also observed in the green long dashed line of Fig.~\ref{fig:CM-QPT}(a) and Fig.~\ref{fig:CM-QPT}(b).  

 In left panels, one observes that the QPT that the intruder configuration undergoes favours the more rapid lowering of the intruder states. It is evident that the crossing of two configurations with a constant Hamiltonian induces a certain discontinuity in the derivative of the ground state energy with respect to a control parameter, but this discontinuity will increase if on top of the crossing, the intruder ground state evolves from being spherical to becoming deformed. This fact can be explained inspecting Fig.~\ref{fig-beta-energy}(b) in which one observes a large energy gain in binding energy being due to the onset of deformation.
\begin{figure}[hbt]
  \centering \includegraphics[width=0.60\linewidth]{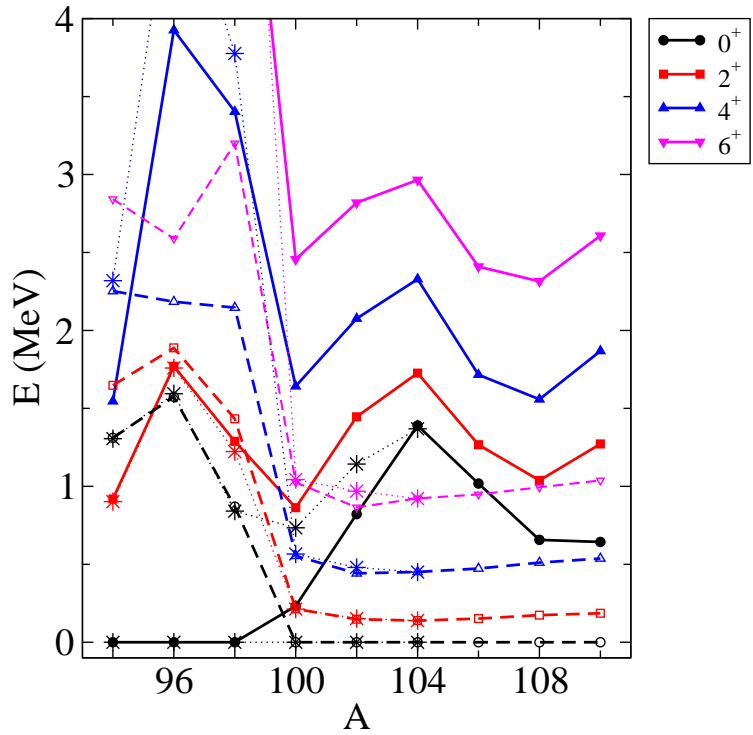}
  \caption{Energy systematics for Zr isotopes from the IBM-CM calculation, separating in states which are mainly regular (full lines with full symbol) and mainly of intruder nature (dashed lines with open symbols), compared with an IBM calculation using a single configuration \cite{Garc05} (dotted line and starts).}
  \label{fig:energ-separation}
 \end{figure}

 What has been presented so far in the schematic calculations turns out to be equivalent to the comparison of the IBM-CM results used in this work and extracted from Ref.~\cite{Garc19} with the results of an IBM calculation for Zr using a single configuration, e.g, the one used in Ref.~\cite{Garc05} (only calculated up to $A=104$). In the latter paper, the IBM parameters were obtained through a least-squares fit to the available information (available in 2005), with special emphasis in the two-neutron separation energy, that was considered as the main indicator hinting towards the existence of a QPT. In Ref.~\cite{Garc05} it was proven that the even-even Zr isotopes undergo a QPT with $^{100}$Zr being the critical nucleus. To study how both approaches are compared, in Fig.~\ref{fig:energ-separation} we depict part of the spectra of both Hamiltonians. The regular and the intruder states in the IBM-CM calculation are separated taking into account the value of its regular component, i.e., if larger than $0.5$ are marked as regular while if smaller than $0.5$ are marked as intruder. A first relevant fact is that the single configuration calculation gives rise to too high excitation energies for $L>2$ in the lightest isotopes. Moreover, it follows reasonably well that the excitation energy of the $2_1^+$ state all the way and the one of the $4_1^+$ and $6_1^+$ states from $A=100$ and onwards. A second important result corresponds to the good reproduction of the $0_2^+$ excitation energy for $A=94$, $96$, $98$ and $104$, but not for $A=100$ and $102$. Therefore, some states present in the spectrum, namely, the $0_2^+$ state for $^{100}$Zr, cannot be correctly reproduced using a  single configuration. Consequently, in Ref.~\cite{Garc05} it was explained that the experimental $0_2^+$ state of $^{100}$Zr, which is of approximate spherical character, was excluded from the least-squares fit. This is, therefore, the main drawback of the single-configuration calculation, i.e., some states are beyond the model space and cannot be correctly reproduced, particularly the very low-lying excited $0^+$ states. In the case of the IBM-CM calculation, very low-lying $0^+$ states are correctly explained because the crossing of two different configurations, intruder and regular. 

 Therefore, one can conclude that the differences between the energy systematics of both approaches can be understood considering the level repulsion and quantum integrability as explained in \cite{Aria03}. For a single configuration, the existence of a sizeable interaction matrix elements between the first two $0^+$ states prevents its approaching, and instead, a large repulsion results. However, in the case of regular and intruder configurations, the existence of a quantum number with different value for both, namely N and N+2, allows the crossing of both configurations in the case of no mixing term, or the presence of  a small repulsion for sufficiently small values of the mixing term, leading to a much more abrupt change in the slope of the ground state energy than in the case of a QPT with a single configuration.

\section{Conclusions and outlook}
\label{sec-conclu}
In the present study, a connection between shape coexistence and QPT has been systematically studied from different points of view and exemplified for the case of Zr nuclei. First, we have reviewed the many experimental evidences for shape coexistence and for QPT. Next, the possible fulfillment of the X(5) critical point symmetry has been considered concluding that the evidences are rather mixed and that $^{100}$Zr and $^{102}$Zr present certain characteristics of the X(5) symmetry but not all of them. The IBM-CM mean-field energy surfaces have been presented, both in the axial direction and  in the $\beta-\gamma$ plane. The scale that connects the IBM deformation parameters and the one from the collective model has been obtained for every isotope. With the goal of analyzing in depth the structure of the wave function, a decomposition in the U(5) basis was carried out, observing how the structure of the intruder states changes from being spherical to being well deformed. Also, the evolution of the density of states as a function of mass number has been explored, without obtaining a clear conclusion. Finally, a schematic calculation has been conducted to obtain some hints to distinguish between shape coexistence and QPT.

In short, in the even-even Zr isotopes, ample experimental evidence exists for a QPT showing  up , with $^{100}$Zr as the ``critical nucleus''. This conclusion is based on the experimental information [S$_{2n}$, $E(4_1^+)/E(2_1^+)$, or B(E2:$4_1^+\rightarrow 2_1^+$) evolution] and supported by several theoretical models, including the IBM used in the present work. It is also self-evident, in view of the available experimental information, that states with a very different structure coexist in a narrow region of energy and, consequently, shape coexistence shows up in the Zr isotopic chain of nuclei
and is responsible for the evolution of the nuclear structure in this region of the nuclear mass table. %
It is fair to say that
the existence of a %
QPT is \textbf{triggered} by shape coexistence and is the result of the crossing in the $0^+$ ground state of two configurations corresponding  with  different structural properties, i.e., spherical and deformed character.
The reduce mixing between regular and intruder configurations and the evolution of the intruder configuration from a spherical shape to a deformed one when passing from $^{98}$Zr to $^{100}$Zr is at the origin of the unique properties of the shape evolution observed in this mass area. Similar conclusions have been formulated in a recent conference contribution by \citet{Poves18} and in \cite{Caur07} with a discussion for the doubly-magic $^{40}$Ca nucleus. A possible conclusion is that if the shell gap at $Z, N=20$ single-particle levels would have been a little smaller, this doubly magic nucleus would have turned up as deformed in its ground state band \cite{heyde11}.

The study of the possible relationship between shape coexistence and QPT in Zr isotopes has been also carried out in Refs.~\cite{Gavr19,Gavr19b}, obtaining similar conclusions, although in these works the authors claimed that there is an addition evolution into gamma-unstable shapes in $^{106}$Zr an onwards.

The presented analysis can encourage further studies in this mass region, for example in the Sr isotopes, but also at the border of this deformation area, i.e., for Kr and Mo isotopes.

\section{Acknowledgment}
We are very grateful to A.\ Leviatan and J.L.\ Wood for enlightening discussions. This work was supported (KH) by the InterUniversity Attraction Poles Program of the Belgian State-Federal Office for Scientific, Technical and Cultural Affairs (IAP Grant  P7/12) and it has also been partially supported (JEGR) by the Ministerio de Ciencia, Innovaci\'on y Universidades (Spain) under project number PID2019-104002GB-C21, by the Consejer\'{\i}a de Econom\'{\i}a, Conocimiento, Empresas y Universidad de la Junta de Andaluc\'{\i}a (Spain) under Group FQM-370 and by European Regional Development Fund (ERDF), ref.\ SOMM17/6105/UGR. Resources supporting this work were provided by the CEAFMC and Universidad de Huelva High Performance Computer (HPC@UHU) funded by ERDF/MINECO project UNHU-15CE-2848. 

\bibliography{references,references-QPT}

\begin{thebibliography}{77}%
\makeatletter
\providecommand \@ifxundefined [1]{%
 \@ifx{#1\undefined}
}%
\providecommand \@ifnum [1]{%
 \ifnum #1\expandafter \@firstoftwo
 \else \expandafter \@secondoftwo
 \fi
}%
\providecommand \@ifx [1]{%
 \ifx #1\expandafter \@firstoftwo
 \else \expandafter \@secondoftwo
 \fi
}%
\providecommand \natexlab [1]{#1}%
\providecommand \enquote  [1]{``#1''}%
\providecommand \bibnamefont  [1]{#1}%
\providecommand \bibfnamefont [1]{#1}%
\providecommand \citenamefont [1]{#1}%
\providecommand \href@noop [0]{\@secondoftwo}%
\providecommand \href [0]{\begingroup \@sanitize@url \@href}%
\providecommand \@href[1]{\@@startlink{#1}\@@href}%
\providecommand \@@href[1]{\endgroup#1\@@endlink}%
\providecommand \@sanitize@url [0]{\catcode `\\12\catcode `\$12\catcode
  `\&12\catcode `\#12\catcode `\^12\catcode `\_12\catcode `\%12\relax}%
\providecommand \@@startlink[1]{}%
\providecommand \@@endlink[0]{}%
\providecommand \url  [0]{\begingroup\@sanitize@url \@url }%
\providecommand \@url [1]{\endgroup\@href {#1}{\urlprefix }}%
\providecommand \urlprefix  [0]{URL }%
\providecommand \Eprint [0]{\href }%
\providecommand \doibase [0]{http://dx.doi.org/}%
\providecommand \selectlanguage [0]{\@gobble}%
\providecommand \bibinfo  [0]{\@secondoftwo}%
\providecommand \bibfield  [0]{\@secondoftwo}%
\providecommand \translation [1]{[#1]}%
\providecommand \BibitemOpen [0]{}%
\providecommand \bibitemStop [0]{}%
\providecommand \bibitemNoStop [0]{.\EOS\space}%
\providecommand \EOS [0]{\spacefactor3000\relax}%
\providecommand \BibitemShut  [1]{\csname bibitem#1\endcsname}%
\let\auto@bib@innerbib\@empty
\bibitem [{\citenamefont {Tsunoda}\ \emph {et~al.}(2014)\citenamefont
  {Tsunoda}, \citenamefont {Otsuka}, \citenamefont {Shimizu}, \citenamefont
  {Honma},\ and\ \citenamefont {Utsuno}}]{Tsuno14}%
  \BibitemOpen
  \bibfield  {author} {\bibinfo {author} {\bibfnamefont {Yusuke}\ \bibnamefont
  {Tsunoda}}, \bibinfo {author} {\bibfnamefont {Takaharu}\ \bibnamefont
  {Otsuka}}, \bibinfo {author} {\bibfnamefont {Noritaka}\ \bibnamefont
  {Shimizu}}, \bibinfo {author} {\bibfnamefont {Michio}\ \bibnamefont {Honma}},
  \ and\ \bibinfo {author} {\bibfnamefont {Yutaka}\ \bibnamefont {Utsuno}},\
  }\bibfield  {title} {\enquote {\bibinfo {title} {{Novel shape evolution in
  exotic Ni isotopes and configuration-dependent shell structure}},}\ }\href
  {\doibase 10.1103/PhysRevC.89.031301} {\bibfield  {journal} {\bibinfo
  {journal} {Phys. Rev. C}\ }\textbf {\bibinfo {volume} {89}},\ \bibinfo
  {pages} {031301} (\bibinfo {year} {2014})}\BibitemShut {NoStop}%
\bibitem [{\citenamefont {Cejnar}\ and\ \citenamefont {Jolie}(2009)}]{Cejn09}%
  \BibitemOpen
  \bibfield  {author} {\bibinfo {author} {\bibfnamefont {Pavel}\ \bibnamefont
  {Cejnar}}\ and\ \bibinfo {author} {\bibfnamefont {Jan}\ \bibnamefont
  {Jolie}},\ }\bibfield  {title} {\enquote {\bibinfo {title} {{Quantum phase
  transitions in the interacting boson model}},}\ }\href {\doibase
  https://doi.org/10.1016/j.ppnp.2008.08.001} {\bibfield  {journal} {\bibinfo
  {journal} {Progress in Particle and Nuclear Physics}\ }\textbf {\bibinfo
  {volume} {62}},\ \bibinfo {pages} {210 -- 256} (\bibinfo {year}
  {2009})}\BibitemShut {NoStop}%
\bibitem [{\citenamefont {Cejnar}\ \emph {et~al.}(2010)\citenamefont {Cejnar},
  \citenamefont {Jolie},\ and\ \citenamefont {Casten}}]{Cejn10}%
  \BibitemOpen
  \bibfield  {author} {\bibinfo {author} {\bibfnamefont {Pavel}\ \bibnamefont
  {Cejnar}}, \bibinfo {author} {\bibfnamefont {Jan}\ \bibnamefont {Jolie}}, \
  and\ \bibinfo {author} {\bibfnamefont {Richard~F.}\ \bibnamefont {Casten}},\
  }\bibfield  {title} {\enquote {\bibinfo {title} {{Quantum phase transitions
  in the shapes of atomic nuclei}},}\ }\href {\doibase
  10.1103/RevModPhys.82.2155} {\bibfield  {journal} {\bibinfo  {journal} {Rev.
  Mod. Phys.}\ }\textbf {\bibinfo {volume} {82}},\ \bibinfo {pages}
  {2155--2212} (\bibinfo {year} {2010})}\BibitemShut {NoStop}%
\bibitem [{\citenamefont {Ansari}\ \emph {et~al.}(2017)\citenamefont {Ansari},
  \citenamefont {R\'egis}, \citenamefont {Jolie}, \citenamefont {Saed-Samii},
  \citenamefont {Warr}, \citenamefont {Korten}, \citenamefont
  {Zieli\ifmmode~\acute{n}\else \'{n}\fi{}ska}, \citenamefont {Salsac},
  \citenamefont {Blanc}, \citenamefont {Jentschel}, \citenamefont {K\"oster},
  \citenamefont {Mutti}, \citenamefont {Soldner}, \citenamefont {Simpson},
  \citenamefont {Drouet}, \citenamefont {Vancraeyenest}, \citenamefont
  {de~France}, \citenamefont {Cl\'ement}, \citenamefont {Stezowski},
  \citenamefont {Ur}, \citenamefont {Urban}, \citenamefont {Regan},
  \citenamefont {Podoly\'ak}, \citenamefont {Larijani}, \citenamefont
  {Townsley}, \citenamefont {Carroll}, \citenamefont {Wilson}, \citenamefont
  {Mach}, \citenamefont {Fraile}, \citenamefont {Paziy}, \citenamefont
  {Olaizola}, \citenamefont {Vedia}, \citenamefont {Bruce}, \citenamefont
  {Roberts}, \citenamefont {Smith}, \citenamefont {Scheck}, \citenamefont
  {Kr\"oll}, \citenamefont {Hartig}, \citenamefont {Ignatov}, \citenamefont
  {Ilieva}, \citenamefont {Lalkovski}, \citenamefont {M\ifmmode~\u{a}\else
  \u{a}\fi{}rginean}, \citenamefont {Otsuka}, \citenamefont {Shimizu},
  \citenamefont {Togashi},\ and\ \citenamefont {Tsunoda}}]{Ansa17}%
  \BibitemOpen
  \bibfield  {author} {\bibinfo {author} {\bibfnamefont {S.}~\bibnamefont
  {Ansari}}, \bibinfo {author} {\bibfnamefont {J.-M.}\ \bibnamefont {R\'egis}},
  \bibinfo {author} {\bibfnamefont {J.}~\bibnamefont {Jolie}}, \bibinfo
  {author} {\bibfnamefont {N.}~\bibnamefont {Saed-Samii}}, \bibinfo {author}
  {\bibfnamefont {N.}~\bibnamefont {Warr}}, \bibinfo {author} {\bibfnamefont
  {W.}~\bibnamefont {Korten}}, \bibinfo {author} {\bibfnamefont
  {M.}~\bibnamefont {Zieli\ifmmode~\acute{n}\else \'{n}\fi{}ska}}, \bibinfo
  {author} {\bibfnamefont {M.-D.}\ \bibnamefont {Salsac}}, \bibinfo {author}
  {\bibfnamefont {A.}~\bibnamefont {Blanc}}, \bibinfo {author} {\bibfnamefont
  {M.}~\bibnamefont {Jentschel}}, \bibinfo {author} {\bibfnamefont
  {U.}~\bibnamefont {K\"oster}}, \bibinfo {author} {\bibfnamefont
  {P.}~\bibnamefont {Mutti}}, \bibinfo {author} {\bibfnamefont
  {T.}~\bibnamefont {Soldner}}, \bibinfo {author} {\bibfnamefont {G.~S.}\
  \bibnamefont {Simpson}}, \bibinfo {author} {\bibfnamefont {F.}~\bibnamefont
  {Drouet}}, \bibinfo {author} {\bibfnamefont {A.}~\bibnamefont
  {Vancraeyenest}}, \bibinfo {author} {\bibfnamefont {G.}~\bibnamefont
  {de~France}}, \bibinfo {author} {\bibfnamefont {E.}~\bibnamefont
  {Cl\'ement}}, \bibinfo {author} {\bibfnamefont {O.}~\bibnamefont
  {Stezowski}}, \bibinfo {author} {\bibfnamefont {C.~A.}\ \bibnamefont {Ur}},
  \bibinfo {author} {\bibfnamefont {W.}~\bibnamefont {Urban}}, \bibinfo
  {author} {\bibfnamefont {P.~H.}\ \bibnamefont {Regan}}, \bibinfo {author}
  {\bibfnamefont {Zs.}\ \bibnamefont {Podoly\'ak}}, \bibinfo {author}
  {\bibfnamefont {C.}~\bibnamefont {Larijani}}, \bibinfo {author}
  {\bibfnamefont {C.}~\bibnamefont {Townsley}}, \bibinfo {author}
  {\bibfnamefont {R.}~\bibnamefont {Carroll}}, \bibinfo {author} {\bibfnamefont
  {E.}~\bibnamefont {Wilson}}, \bibinfo {author} {\bibfnamefont
  {H.}~\bibnamefont {Mach}}, \bibinfo {author} {\bibfnamefont {L.~M.}\
  \bibnamefont {Fraile}}, \bibinfo {author} {\bibfnamefont {V.}~\bibnamefont
  {Paziy}}, \bibinfo {author} {\bibfnamefont {B.}~\bibnamefont {Olaizola}},
  \bibinfo {author} {\bibfnamefont {V.}~\bibnamefont {Vedia}}, \bibinfo
  {author} {\bibfnamefont {A.~M.}\ \bibnamefont {Bruce}}, \bibinfo {author}
  {\bibfnamefont {O.~J.}\ \bibnamefont {Roberts}}, \bibinfo {author}
  {\bibfnamefont {J.~F.}\ \bibnamefont {Smith}}, \bibinfo {author}
  {\bibfnamefont {M.}~\bibnamefont {Scheck}}, \bibinfo {author} {\bibfnamefont
  {T.}~\bibnamefont {Kr\"oll}}, \bibinfo {author} {\bibfnamefont {A.-L.}\
  \bibnamefont {Hartig}}, \bibinfo {author} {\bibfnamefont {A.}~\bibnamefont
  {Ignatov}}, \bibinfo {author} {\bibfnamefont {S.}~\bibnamefont {Ilieva}},
  \bibinfo {author} {\bibfnamefont {S.}~\bibnamefont {Lalkovski}}, \bibinfo
  {author} {\bibfnamefont {N.}~\bibnamefont {M\ifmmode~\u{a}\else
  \u{a}\fi{}rginean}}, \bibinfo {author} {\bibfnamefont {T.}~\bibnamefont
  {Otsuka}}, \bibinfo {author} {\bibfnamefont {N.}~\bibnamefont {Shimizu}},
  \bibinfo {author} {\bibfnamefont {T.}~\bibnamefont {Togashi}}, \ and\
  \bibinfo {author} {\bibfnamefont {Y.}~\bibnamefont {Tsunoda}},\ }\bibfield
  {title} {\enquote {\bibinfo {title} {{Experimental study of the lifetime and
  phase transition in neutron-rich $^{98,100,102}\mathrm{Zr}$}},}\ }\href
  {\doibase 10.1103/PhysRevC.96.054323} {\bibfield  {journal} {\bibinfo
  {journal} {Phys. Rev. C}\ }\textbf {\bibinfo {volume} {96}},\ \bibinfo
  {pages} {054323} (\bibinfo {year} {2017})}\BibitemShut {NoStop}%
\bibitem [{\citenamefont {R\'egis}\ \emph {et~al.}(2017)\citenamefont
  {R\'egis}, \citenamefont {Jolie}, \citenamefont {Saed-Samii}, \citenamefont
  {Warr}, \citenamefont {Pfeiffer}, \citenamefont {Blanc}, \citenamefont
  {Jentschel}, \citenamefont {K\"oster}, \citenamefont {Mutti}, \citenamefont
  {Soldner}, \citenamefont {Simpson}, \citenamefont {Drouet}, \citenamefont
  {Vancraeyenest}, \citenamefont {de~France}, \citenamefont {Cl\'ement},
  \citenamefont {Stezowski}, \citenamefont {Ur}, \citenamefont {Urban},
  \citenamefont {Regan}, \citenamefont {Podoly\'ak}, \citenamefont {Larijani},
  \citenamefont {Townsley}, \citenamefont {Carroll}, \citenamefont {Wilson},
  \citenamefont {Fraile}, \citenamefont {Mach}, \citenamefont {Paziy},
  \citenamefont {Olaizola}, \citenamefont {Vedia}, \citenamefont {Bruce},
  \citenamefont {Roberts}, \citenamefont {Smith}, \citenamefont {Scheck},
  \citenamefont {Kr\"oll}, \citenamefont {Hartig}, \citenamefont {Ignatov},
  \citenamefont {Ilieva}, \citenamefont {Lalkovski}, \citenamefont {Korten},
  \citenamefont {M\ifmmode~\u{a}\else \u{a}\fi{}rginean}, \citenamefont
  {Otsuka}, \citenamefont {Shimizu}, \citenamefont {Togashi},\ and\
  \citenamefont {Tsunoda}}]{Regi17}%
  \BibitemOpen
  \bibfield  {author} {\bibinfo {author} {\bibfnamefont {J.-M.}\ \bibnamefont
  {R\'egis}}, \bibinfo {author} {\bibfnamefont {J.}~\bibnamefont {Jolie}},
  \bibinfo {author} {\bibfnamefont {N.}~\bibnamefont {Saed-Samii}}, \bibinfo
  {author} {\bibfnamefont {N.}~\bibnamefont {Warr}}, \bibinfo {author}
  {\bibfnamefont {M.}~\bibnamefont {Pfeiffer}}, \bibinfo {author}
  {\bibfnamefont {A.}~\bibnamefont {Blanc}}, \bibinfo {author} {\bibfnamefont
  {M.}~\bibnamefont {Jentschel}}, \bibinfo {author} {\bibfnamefont
  {U.}~\bibnamefont {K\"oster}}, \bibinfo {author} {\bibfnamefont
  {P.}~\bibnamefont {Mutti}}, \bibinfo {author} {\bibfnamefont
  {T.}~\bibnamefont {Soldner}}, \bibinfo {author} {\bibfnamefont {G.~S.}\
  \bibnamefont {Simpson}}, \bibinfo {author} {\bibfnamefont {F.}~\bibnamefont
  {Drouet}}, \bibinfo {author} {\bibfnamefont {A.}~\bibnamefont
  {Vancraeyenest}}, \bibinfo {author} {\bibfnamefont {G.}~\bibnamefont
  {de~France}}, \bibinfo {author} {\bibfnamefont {E.}~\bibnamefont
  {Cl\'ement}}, \bibinfo {author} {\bibfnamefont {O.}~\bibnamefont
  {Stezowski}}, \bibinfo {author} {\bibfnamefont {C.~A.}\ \bibnamefont {Ur}},
  \bibinfo {author} {\bibfnamefont {W.}~\bibnamefont {Urban}}, \bibinfo
  {author} {\bibfnamefont {P.~H.}\ \bibnamefont {Regan}}, \bibinfo {author}
  {\bibfnamefont {Zs.}\ \bibnamefont {Podoly\'ak}}, \bibinfo {author}
  {\bibfnamefont {C.}~\bibnamefont {Larijani}}, \bibinfo {author}
  {\bibfnamefont {C.}~\bibnamefont {Townsley}}, \bibinfo {author}
  {\bibfnamefont {R.}~\bibnamefont {Carroll}}, \bibinfo {author} {\bibfnamefont
  {E.}~\bibnamefont {Wilson}}, \bibinfo {author} {\bibfnamefont {L.~M.}\
  \bibnamefont {Fraile}}, \bibinfo {author} {\bibfnamefont {H.}~\bibnamefont
  {Mach}}, \bibinfo {author} {\bibfnamefont {V.}~\bibnamefont {Paziy}},
  \bibinfo {author} {\bibfnamefont {B.}~\bibnamefont {Olaizola}}, \bibinfo
  {author} {\bibfnamefont {V.}~\bibnamefont {Vedia}}, \bibinfo {author}
  {\bibfnamefont {A.~M.}\ \bibnamefont {Bruce}}, \bibinfo {author}
  {\bibfnamefont {O.~J.}\ \bibnamefont {Roberts}}, \bibinfo {author}
  {\bibfnamefont {J.~F.}\ \bibnamefont {Smith}}, \bibinfo {author}
  {\bibfnamefont {M.}~\bibnamefont {Scheck}}, \bibinfo {author} {\bibfnamefont
  {T.}~\bibnamefont {Kr\"oll}}, \bibinfo {author} {\bibfnamefont {A.-L.}\
  \bibnamefont {Hartig}}, \bibinfo {author} {\bibfnamefont {A.}~\bibnamefont
  {Ignatov}}, \bibinfo {author} {\bibfnamefont {S.}~\bibnamefont {Ilieva}},
  \bibinfo {author} {\bibfnamefont {S.}~\bibnamefont {Lalkovski}}, \bibinfo
  {author} {\bibfnamefont {W.}~\bibnamefont {Korten}}, \bibinfo {author}
  {\bibfnamefont {N.}~\bibnamefont {M\ifmmode~\u{a}\else \u{a}\fi{}rginean}},
  \bibinfo {author} {\bibfnamefont {T.}~\bibnamefont {Otsuka}}, \bibinfo
  {author} {\bibfnamefont {N.}~\bibnamefont {Shimizu}}, \bibinfo {author}
  {\bibfnamefont {T.}~\bibnamefont {Togashi}}, \ and\ \bibinfo {author}
  {\bibfnamefont {Y.}~\bibnamefont {Tsunoda}},\ }\bibfield  {title} {\enquote
  {\bibinfo {title} {{Abrupt shape transition at neutron number $N=60$: $B(E2)$
  values in $^{94,96,98}$Sr from fast $\gamma-\gamma$ timing}},}\ }\href
  {\doibase 10.1103/PhysRevC.95.054319} {\bibfield  {journal} {\bibinfo
  {journal} {Phys. Rev. C}\ }\textbf {\bibinfo {volume} {95}},\ \bibinfo
  {pages} {054319} (\bibinfo {year} {2017})}\BibitemShut {NoStop}%
\bibitem [{\citenamefont {Hertz}(1976)}]{Hert76}%
  \BibitemOpen
  \bibfield  {author} {\bibinfo {author} {\bibfnamefont {John~A.}\ \bibnamefont
  {Hertz}},\ }\bibfield  {title} {\enquote {\bibinfo {title} {Quantum critical
  phenomena},}\ }\href {\doibase 10.1103/PhysRevB.14.1165} {\bibfield
  {journal} {\bibinfo  {journal} {Phys. Rev. B}\ }\textbf {\bibinfo {volume}
  {14}},\ \bibinfo {pages} {1165--1184} (\bibinfo {year} {1976})}\BibitemShut
  {NoStop}%
\bibitem [{\citenamefont {Sachdev}(2011)}]{Sach11}%
  \BibitemOpen
  \bibfield  {author} {\bibinfo {author} {\bibfnamefont {S.}~\bibnamefont
  {Sachdev}},\ }\href@noop {} {\emph {\bibinfo {title} {Quantum Phase
  Transitions}}}\ (\bibinfo  {publisher} {Cambridge University Press,
  Cambridge, UK},\ \bibinfo {year} {2011})\BibitemShut {NoStop}%
\bibitem [{\citenamefont {Caurier}\ \emph {et~al.}(2007)\citenamefont
  {Caurier}, \citenamefont {Men\'endez}, \citenamefont {Nowacki},\ and\
  \citenamefont {Poves}}]{Caur07}%
  \BibitemOpen
  \bibfield  {author} {\bibinfo {author} {\bibfnamefont {E.}~\bibnamefont
  {Caurier}}, \bibinfo {author} {\bibfnamefont {J.}~\bibnamefont {Men\'endez}},
  \bibinfo {author} {\bibfnamefont {F.}~\bibnamefont {Nowacki}}, \ and\
  \bibinfo {author} {\bibfnamefont {A.}~\bibnamefont {Poves}},\ }\bibfield
  {title} {\enquote {\bibinfo {title} {{Coexistence of spherical states with
  deformed and superdeformed bands in doubly magic $^{40}\mathrm{Ca}$: A
  shell-model challenge}},}\ }\href {\doibase 10.1103/PhysRevC.75.054317}
  {\bibfield  {journal} {\bibinfo  {journal} {Phys. Rev. C}\ }\textbf {\bibinfo
  {volume} {75}},\ \bibinfo {pages} {054317} (\bibinfo {year}
  {2007})}\BibitemShut {NoStop}%
\bibitem [{\citenamefont {Heyde}\ and\ \citenamefont {Wood}(2011)}]{heyde11}%
  \BibitemOpen
  \bibfield  {author} {\bibinfo {author} {\bibfnamefont {Kris}\ \bibnamefont
  {Heyde}}\ and\ \bibinfo {author} {\bibfnamefont {John~L.}\ \bibnamefont
  {Wood}},\ }\bibfield  {title} {\enquote {\bibinfo {title} {{Shape coexistence
  in atomic nuclei}},}\ }\href {\doibase 10.1103/RevModPhys.83.1467} {\bibfield
   {journal} {\bibinfo  {journal} {Rev. Mod. Phys.}\ }\textbf {\bibinfo
  {volume} {83}},\ \bibinfo {pages} {1467--1521} (\bibinfo {year}
  {2011})}\BibitemShut {NoStop}%
\bibitem [{\citenamefont {Garc\'{\i}a-Ramos}\ \emph {et~al.}(2011)\citenamefont
  {Garc\'{\i}a-Ramos}, \citenamefont {Hellemans},\ and\ \citenamefont
  {Heyde}}]{Garc11}%
  \BibitemOpen
  \bibfield  {author} {\bibinfo {author} {\bibfnamefont {J.~E.}\ \bibnamefont
  {Garc\'{\i}a-Ramos}}, \bibinfo {author} {\bibfnamefont {V.}~\bibnamefont
  {Hellemans}}, \ and\ \bibinfo {author} {\bibfnamefont {K.}~\bibnamefont
  {Heyde}},\ }\bibfield  {title} {\enquote {\bibinfo {title} {{Platinum nuclei:
  Concealed configuration mixing and shape coexistence}},}\ }\href {\doibase
  10.1103/PhysRevC.84.014331} {\bibfield  {journal} {\bibinfo  {journal} {Phys.
  Rev. C}\ }\textbf {\bibinfo {volume} {84}},\ \bibinfo {pages} {014331}
  (\bibinfo {year} {2011})}\BibitemShut {NoStop}%
\bibitem [{\citenamefont {Garc\'{\i}a-Ramos}\ \emph {et~al.}(2012)\citenamefont
  {Garc\'{\i}a-Ramos}, \citenamefont {Hellemans},\ and\ \citenamefont
  {Heyde}}]{Garc12}%
  \BibitemOpen
  \bibfield  {author} {\bibinfo {author} {\bibfnamefont {J.~E.}\ \bibnamefont
  {Garc\'{\i}a-Ramos}}, \bibinfo {author} {\bibfnamefont {V.}~\bibnamefont
  {Hellemans}}, \ and\ \bibinfo {author} {\bibfnamefont {K.}~\bibnamefont
  {Heyde}},\ }\bibfield  {title} {\enquote {\bibinfo {title} {Concealed
  configuration mixing and shape coexistence in the platinum nuclei},}\ }\href
  {\doibase 10.1063/1.4764214} {\bibfield  {journal} {\bibinfo  {journal} {AIP
  Conference Proceedings}\ }\textbf {\bibinfo {volume} {1491}},\ \bibinfo
  {pages} {109--112} (\bibinfo {year} {2012})},\ \Eprint
  {http://arxiv.org/abs/https://aip.scitation.org/doi/pdf/10.1063/1.4764214}
  {https://aip.scitation.org/doi/pdf/10.1063/1.4764214} \BibitemShut {NoStop}%
\bibitem [{\citenamefont {Garc\'{\i}a-Ramos}\ and\ \citenamefont
  {Heyde}(2015{\natexlab{a}})}]{Garc15}%
  \BibitemOpen
  \bibfield  {author} {\bibinfo {author} {\bibfnamefont {J.~E.}\ \bibnamefont
  {Garc\'{\i}a-Ramos}}\ and\ \bibinfo {author} {\bibfnamefont {K.}~\bibnamefont
  {Heyde}},\ }\bibfield  {title} {\enquote {\bibinfo {title} {{Nuclear shape
  coexistence in Po isotopes: An interacting boson model study}},}\ }\href
  {\doibase 10.1103/PhysRevC.92.034309} {\bibfield  {journal} {\bibinfo
  {journal} {Phys. Rev. C}\ }\textbf {\bibinfo {volume} {92}},\ \bibinfo
  {pages} {034309} (\bibinfo {year} {2015}{\natexlab{a}})}\BibitemShut
  {NoStop}%
\bibitem [{\citenamefont {Garc\'{\i}a-Ramos}\ and\ \citenamefont
  {Heyde}(2015{\natexlab{b}})}]{Garc15c}%
  \BibitemOpen
  \bibfield  {author} {\bibinfo {author} {\bibfnamefont {J.~E.}\ \bibnamefont
  {Garc\'{\i}a-Ramos}}\ and\ \bibinfo {author} {\bibfnamefont {K.}~\bibnamefont
  {Heyde}},\ }\bibfield  {title} {\enquote {\bibinfo {title} {The influence of
  intruder states in even-even po isotopes},}\ }\href {\doibase
  10.1063/1.4932269} {\bibfield  {journal} {\bibinfo  {journal} {AIP Conference
  Proceedings}\ }\textbf {\bibinfo {volume} {1681}},\ \bibinfo {pages} {040008}
  (\bibinfo {year} {2015}{\natexlab{b}})},\ \Eprint
  {http://arxiv.org/abs/https://aip.scitation.org/doi/pdf/10.1063/1.4932269}
  {https://aip.scitation.org/doi/pdf/10.1063/1.4932269} \BibitemShut {NoStop}%
\bibitem [{\citenamefont {Abriola}\ and\ \citenamefont
  {Sonzogni}(2006)}]{Abri06}%
  \BibitemOpen
  \bibfield  {author} {\bibinfo {author} {\bibfnamefont {D.}~\bibnamefont
  {Abriola}}\ and\ \bibinfo {author} {\bibfnamefont {A.A.}\ \bibnamefont
  {Sonzogni}},\ }\bibfield  {title} {\enquote {\bibinfo {title} {{Nuclear Data
  Sheets for A = 94}},}\ }\href {\doibase
  https://doi.org/10.1016/j.nds.2006.08.001} {\bibfield  {journal} {\bibinfo
  {journal} {{Nuclear Data Sheets}}\ }\textbf {\bibinfo {volume} {107}},\
  \bibinfo {pages} {2423 -- 2578} (\bibinfo {year} {2006})}\BibitemShut
  {NoStop}%
\bibitem [{\citenamefont {Chakraborty}\ \emph {et~al.}(2013)\citenamefont
  {Chakraborty}, \citenamefont {Peters}, \citenamefont {Crider}, \citenamefont
  {Andreoiu}, \citenamefont {Bender}, \citenamefont {Cross}, \citenamefont
  {Demand}, \citenamefont {Garnsworthy}, \citenamefont {Garrett}, \citenamefont
  {Hackman}, \citenamefont {Hadinia}, \citenamefont {Ketelhut}, \citenamefont
  {Kumar}, \citenamefont {Leach}, \citenamefont {McEllistrem}, \citenamefont
  {Pore}, \citenamefont {Prados-Est\'evez}, \citenamefont {Rand}, \citenamefont
  {Singh}, \citenamefont {Tardiff}, \citenamefont {Wang}, \citenamefont
  {Wood},\ and\ \citenamefont {Yates}}]{Chak13}%
  \BibitemOpen
  \bibfield  {author} {\bibinfo {author} {\bibfnamefont {A.}~\bibnamefont
  {Chakraborty}}, \bibinfo {author} {\bibfnamefont {E.~E.}\ \bibnamefont
  {Peters}}, \bibinfo {author} {\bibfnamefont {B.~P.}\ \bibnamefont {Crider}},
  \bibinfo {author} {\bibfnamefont {C.}~\bibnamefont {Andreoiu}}, \bibinfo
  {author} {\bibfnamefont {P.~C.}\ \bibnamefont {Bender}}, \bibinfo {author}
  {\bibfnamefont {D.~S.}\ \bibnamefont {Cross}}, \bibinfo {author}
  {\bibfnamefont {G.~A.}\ \bibnamefont {Demand}}, \bibinfo {author}
  {\bibfnamefont {A.~B.}\ \bibnamefont {Garnsworthy}}, \bibinfo {author}
  {\bibfnamefont {P.~E.}\ \bibnamefont {Garrett}}, \bibinfo {author}
  {\bibfnamefont {G.}~\bibnamefont {Hackman}}, \bibinfo {author} {\bibfnamefont
  {B.}~\bibnamefont {Hadinia}}, \bibinfo {author} {\bibfnamefont
  {S.}~\bibnamefont {Ketelhut}}, \bibinfo {author} {\bibfnamefont {Ajay}\
  \bibnamefont {Kumar}}, \bibinfo {author} {\bibfnamefont {K.~G.}\ \bibnamefont
  {Leach}}, \bibinfo {author} {\bibfnamefont {M.~T.}\ \bibnamefont
  {McEllistrem}}, \bibinfo {author} {\bibfnamefont {J.}~\bibnamefont {Pore}},
  \bibinfo {author} {\bibfnamefont {F.~M.}\ \bibnamefont {Prados-Est\'evez}},
  \bibinfo {author} {\bibfnamefont {E.~T.}\ \bibnamefont {Rand}}, \bibinfo
  {author} {\bibfnamefont {B.}~\bibnamefont {Singh}}, \bibinfo {author}
  {\bibfnamefont {E.~R.}\ \bibnamefont {Tardiff}}, \bibinfo {author}
  {\bibfnamefont {Z.-M.}\ \bibnamefont {Wang}}, \bibinfo {author}
  {\bibfnamefont {J.~L.}\ \bibnamefont {Wood}}, \ and\ \bibinfo {author}
  {\bibfnamefont {S.~W.}\ \bibnamefont {Yates}},\ }\bibfield  {title} {\enquote
  {\bibinfo {title} {{Collective Structure in $^{94}\mathrm{Zr}$ and Subshell
  Effects in Shape Coexistence}},}\ }\href {\doibase
  10.1103/PhysRevLett.110.022504} {\bibfield  {journal} {\bibinfo  {journal}
  {Phys. Rev. Lett.}\ }\textbf {\bibinfo {volume} {110}},\ \bibinfo {pages}
  {022504} (\bibinfo {year} {2013})}\BibitemShut {NoStop}%
\bibitem [{\citenamefont {Peters}\ \emph {et~al.}(2013)\citenamefont {Peters},
  \citenamefont {Chakraborty}, \citenamefont {Crider}, \citenamefont {Davis},
  \citenamefont {Gnanamani}, \citenamefont {McEllistrem}, \citenamefont
  {Prados-Est\'evez}, \citenamefont {Vanhoy},\ and\ \citenamefont
  {Yates}}]{Peters13}%
  \BibitemOpen
  \bibfield  {author} {\bibinfo {author} {\bibfnamefont {E.~E.}\ \bibnamefont
  {Peters}}, \bibinfo {author} {\bibfnamefont {A.}~\bibnamefont {Chakraborty}},
  \bibinfo {author} {\bibfnamefont {B.~P.}\ \bibnamefont {Crider}}, \bibinfo
  {author} {\bibfnamefont {B.~H.}\ \bibnamefont {Davis}}, \bibinfo {author}
  {\bibfnamefont {M.~K.}\ \bibnamefont {Gnanamani}}, \bibinfo {author}
  {\bibfnamefont {M.~T.}\ \bibnamefont {McEllistrem}}, \bibinfo {author}
  {\bibfnamefont {F.~M.}\ \bibnamefont {Prados-Est\'evez}}, \bibinfo {author}
  {\bibfnamefont {J.~R.}\ \bibnamefont {Vanhoy}}, \ and\ \bibinfo {author}
  {\bibfnamefont {S.~W.}\ \bibnamefont {Yates}},\ }\bibfield  {title} {\enquote
  {\bibinfo {title} {{Level lifetimes in the stable Zr nuclei: Effects of
  chemical properties in Doppler-shift measurements}},}\ }\href {\doibase
  10.1103/PhysRevC.88.024317} {\bibfield  {journal} {\bibinfo  {journal} {Phys.
  Rev. C}\ }\textbf {\bibinfo {volume} {88}},\ \bibinfo {pages} {024317}
  (\bibinfo {year} {2013})}\BibitemShut {NoStop}%
\bibitem [{\citenamefont {Scheikh~Obeid}\ \emph {et~al.}(2014)\citenamefont
  {Scheikh~Obeid}, \citenamefont {Aslanidou}, \citenamefont {Birkhan},
  \citenamefont {Krugmann}, \citenamefont {von Neumann-Cosel}, \citenamefont
  {Pietralla}, \citenamefont {Poltoratska},\ and\ \citenamefont
  {Ponomarev}}]{Scheik14}%
  \BibitemOpen
  \bibfield  {author} {\bibinfo {author} {\bibfnamefont {A.}~\bibnamefont
  {Scheikh~Obeid}}, \bibinfo {author} {\bibfnamefont {S.}~\bibnamefont
  {Aslanidou}}, \bibinfo {author} {\bibfnamefont {J.}~\bibnamefont {Birkhan}},
  \bibinfo {author} {\bibfnamefont {A.}~\bibnamefont {Krugmann}}, \bibinfo
  {author} {\bibfnamefont {P.}~\bibnamefont {von Neumann-Cosel}}, \bibinfo
  {author} {\bibfnamefont {N.}~\bibnamefont {Pietralla}}, \bibinfo {author}
  {\bibfnamefont {I.}~\bibnamefont {Poltoratska}}, \ and\ \bibinfo {author}
  {\bibfnamefont {V.~Yu.}\ \bibnamefont {Ponomarev}},\ }\bibfield  {title}
  {\enquote {\bibinfo {title} {{$B(E2)$ strength ratio of one-phonon ${2}^{+}$
  states of ${}^{94}$Zr from electron scattering at low momentum transfer}},}\
  }\href {\doibase 10.1103/PhysRevC.89.037301} {\bibfield  {journal} {\bibinfo
  {journal} {Phys. Rev. C}\ }\textbf {\bibinfo {volume} {89}},\ \bibinfo
  {pages} {037301} (\bibinfo {year} {2014})}\BibitemShut {NoStop}%
\bibitem [{\citenamefont {Elhami}\ \emph {et~al.}(2008)\citenamefont {Elhami},
  \citenamefont {Orce}, \citenamefont {Scheck}, \citenamefont {Mukhopadhyay},
  \citenamefont {Choudry}, \citenamefont {McEllistrem}, \citenamefont {Yates},
  \citenamefont {Angell}, \citenamefont {Boswell}, \citenamefont {Fallin},
  \citenamefont {Howell}, \citenamefont {Hutcheson}, \citenamefont {Karwowski},
  \citenamefont {Kelley}, \citenamefont {Parpottas}, \citenamefont {Tonchev},\
  and\ \citenamefont {Tornow}}]{Elhalmi08}%
  \BibitemOpen
  \bibfield  {author} {\bibinfo {author} {\bibfnamefont {E.}~\bibnamefont
  {Elhami}}, \bibinfo {author} {\bibfnamefont {J.~N.}\ \bibnamefont {Orce}},
  \bibinfo {author} {\bibfnamefont {M.}~\bibnamefont {Scheck}}, \bibinfo
  {author} {\bibfnamefont {S.}~\bibnamefont {Mukhopadhyay}}, \bibinfo {author}
  {\bibfnamefont {S.~N.}\ \bibnamefont {Choudry}}, \bibinfo {author}
  {\bibfnamefont {M.~T.}\ \bibnamefont {McEllistrem}}, \bibinfo {author}
  {\bibfnamefont {S.~W.}\ \bibnamefont {Yates}}, \bibinfo {author}
  {\bibfnamefont {C.}~\bibnamefont {Angell}}, \bibinfo {author} {\bibfnamefont
  {M.}~\bibnamefont {Boswell}}, \bibinfo {author} {\bibfnamefont
  {B.}~\bibnamefont {Fallin}}, \bibinfo {author} {\bibfnamefont {C.~R.}\
  \bibnamefont {Howell}}, \bibinfo {author} {\bibfnamefont {A.}~\bibnamefont
  {Hutcheson}}, \bibinfo {author} {\bibfnamefont {H.~J.}\ \bibnamefont
  {Karwowski}}, \bibinfo {author} {\bibfnamefont {J.~H.}\ \bibnamefont
  {Kelley}}, \bibinfo {author} {\bibfnamefont {Y.}~\bibnamefont {Parpottas}},
  \bibinfo {author} {\bibfnamefont {A.~P.}\ \bibnamefont {Tonchev}}, \ and\
  \bibinfo {author} {\bibfnamefont {W.}~\bibnamefont {Tornow}},\ }\bibfield
  {title} {\enquote {\bibinfo {title} {{Experimental study of the low-lying
  structure of $^{94}\mathrm{Zr}$ with the $(n,{n}^{'}\ensuremath{\gamma})$
  reaction}},}\ }\href {\doibase 10.1103/PhysRevC.78.064303} {\bibfield
  {journal} {\bibinfo  {journal} {Phys. Rev. C}\ }\textbf {\bibinfo {volume}
  {78}},\ \bibinfo {pages} {064303} (\bibinfo {year} {2008})}\BibitemShut
  {NoStop}%
\bibitem [{\citenamefont {Abriola}\ and\ \citenamefont
  {Sonzogni}(2008)}]{Abri08}%
  \BibitemOpen
  \bibfield  {author} {\bibinfo {author} {\bibfnamefont {D.}~\bibnamefont
  {Abriola}}\ and\ \bibinfo {author} {\bibfnamefont {A.A.}\ \bibnamefont
  {Sonzogni}},\ }\bibfield  {title} {\enquote {\bibinfo {title} {{Nuclear Data
  Sheets for A = 96}},}\ }\href {\doibase
  https://doi.org/10.1016/j.nds.2008.10.002} {\bibfield  {journal} {\bibinfo
  {journal} {Nuclear Data Sheets}\ }\textbf {\bibinfo {volume} {109}},\
  \bibinfo {pages} {2501 -- 2655} (\bibinfo {year} {2008})}\BibitemShut
  {NoStop}%
\bibitem [{\citenamefont {Kumbartzki}\ \emph {et~al.}(2003)\citenamefont
  {Kumbartzki}, \citenamefont {Benczer-Koller}, \citenamefont {Holden},
  \citenamefont {Jakob}, \citenamefont {Mertzimekis}, \citenamefont {Taylor},
  \citenamefont {Speidel}, \citenamefont {Ernst}, \citenamefont {Stuchbery},
  \citenamefont {Beausang},\ and\ \citenamefont {Krücken}}]{Kumb03}%
  \BibitemOpen
  \bibfield  {author} {\bibinfo {author} {\bibfnamefont {G.}~\bibnamefont
  {Kumbartzki}}, \bibinfo {author} {\bibfnamefont {N.}~\bibnamefont
  {Benczer-Koller}}, \bibinfo {author} {\bibfnamefont {J.}~\bibnamefont
  {Holden}}, \bibinfo {author} {\bibfnamefont {G.}~\bibnamefont {Jakob}},
  \bibinfo {author} {\bibfnamefont {T.J.}\ \bibnamefont {Mertzimekis}},
  \bibinfo {author} {\bibfnamefont {M.J.}\ \bibnamefont {Taylor}}, \bibinfo
  {author} {\bibfnamefont {K.-H.}\ \bibnamefont {Speidel}}, \bibinfo {author}
  {\bibfnamefont {R.}~\bibnamefont {Ernst}}, \bibinfo {author} {\bibfnamefont
  {A.E.}\ \bibnamefont {Stuchbery}}, \bibinfo {author} {\bibfnamefont {C.W.}\
  \bibnamefont {Beausang}}, \ and\ \bibinfo {author} {\bibfnamefont
  {R.}~\bibnamefont {Krücken}},\ }\bibfield  {title} {\enquote {\bibinfo
  {title} {{Competition between proton and neutron excitations in
  $^{96}$Zr}},}\ }\href {\doibase
  https://doi.org/10.1016/S0370-2693(03)00608-7} {\bibfield  {journal}
  {\bibinfo  {journal} {Phys. Lett. B}\ }\textbf {\bibinfo {volume} {562}},\
  \bibinfo {pages} {193 -- 200} (\bibinfo {year} {2003})}\BibitemShut {NoStop}%
\bibitem [{\citenamefont {Alanssari}\ \emph {et~al.}(2016)\citenamefont
  {Alanssari}, \citenamefont {Frekers}, \citenamefont {Eronen}, \citenamefont
  {Canete}, \citenamefont {Dilling}, \citenamefont {Haaranen}, \citenamefont
  {Hakala}, \citenamefont {Holl}, \citenamefont {Je\ifmmode~\check{s}\else
  \v{s}\fi{}kovsk\'y}, \citenamefont {Jokinen}, \citenamefont {Kankainen},
  \citenamefont {Koponen}, \citenamefont {Mayer}, \citenamefont {Moore},
  \citenamefont {Nesterenko}, \citenamefont {Pohjalainen}, \citenamefont
  {Povinec}, \citenamefont {Reinikainen}, \citenamefont {Rinta-Antila},
  \citenamefont {Srivastava}, \citenamefont {Suhonen}, \citenamefont
  {Thompson}, \citenamefont {Voss},\ and\ \citenamefont {Wieser}}]{Alan16}%
  \BibitemOpen
  \bibfield  {author} {\bibinfo {author} {\bibfnamefont {M.}~\bibnamefont
  {Alanssari}}, \bibinfo {author} {\bibfnamefont {D.}~\bibnamefont {Frekers}},
  \bibinfo {author} {\bibfnamefont {T.}~\bibnamefont {Eronen}}, \bibinfo
  {author} {\bibfnamefont {L.}~\bibnamefont {Canete}}, \bibinfo {author}
  {\bibfnamefont {J.}~\bibnamefont {Dilling}}, \bibinfo {author} {\bibfnamefont
  {M.}~\bibnamefont {Haaranen}}, \bibinfo {author} {\bibfnamefont
  {J.}~\bibnamefont {Hakala}}, \bibinfo {author} {\bibfnamefont
  {M.}~\bibnamefont {Holl}}, \bibinfo {author} {\bibfnamefont {M.}~\bibnamefont
  {Je\ifmmode~\check{s}\else \v{s}\fi{}kovsk\'y}}, \bibinfo {author}
  {\bibfnamefont {A.}~\bibnamefont {Jokinen}}, \bibinfo {author} {\bibfnamefont
  {A.}~\bibnamefont {Kankainen}}, \bibinfo {author} {\bibfnamefont
  {J.}~\bibnamefont {Koponen}}, \bibinfo {author} {\bibfnamefont {A.~J.}\
  \bibnamefont {Mayer}}, \bibinfo {author} {\bibfnamefont {I.~D.}\ \bibnamefont
  {Moore}}, \bibinfo {author} {\bibfnamefont {D.~A.}\ \bibnamefont
  {Nesterenko}}, \bibinfo {author} {\bibfnamefont {I.}~\bibnamefont
  {Pohjalainen}}, \bibinfo {author} {\bibfnamefont {P.}~\bibnamefont
  {Povinec}}, \bibinfo {author} {\bibfnamefont {J.}~\bibnamefont
  {Reinikainen}}, \bibinfo {author} {\bibfnamefont {S.}~\bibnamefont
  {Rinta-Antila}}, \bibinfo {author} {\bibfnamefont {P.~C.}\ \bibnamefont
  {Srivastava}}, \bibinfo {author} {\bibfnamefont {J.}~\bibnamefont {Suhonen}},
  \bibinfo {author} {\bibfnamefont {R.~I.}\ \bibnamefont {Thompson}}, \bibinfo
  {author} {\bibfnamefont {A.}~\bibnamefont {Voss}}, \ and\ \bibinfo {author}
  {\bibfnamefont {M.~E.}\ \bibnamefont {Wieser}},\ }\bibfield  {title}
  {\enquote {\bibinfo {title} {{Single and Double Beta-Decay $Q$ Values among
  the Triplet $^{96}\mathrm{Zr}$, $^{96}\mathrm{Nb}$, and
  $^{96}\mathrm{Mo}$}},}\ }\href {\doibase 10.1103/PhysRevLett.116.072501}
  {\bibfield  {journal} {\bibinfo  {journal} {Phys. Rev. Lett.}\ }\textbf
  {\bibinfo {volume} {116}},\ \bibinfo {pages} {072501} (\bibinfo {year}
  {2016})}\BibitemShut {NoStop}%
\bibitem [{\citenamefont {Kremer}\ \emph {et~al.}(2016)\citenamefont {Kremer},
  \citenamefont {Aslanidou}, \citenamefont {Bassauer}, \citenamefont {Hilcker},
  \citenamefont {Krugmann}, \citenamefont {von Neumann-Cosel}, \citenamefont
  {Otsuka}, \citenamefont {Pietralla}, \citenamefont {Ponomarev}, \citenamefont
  {Shimizu}, \citenamefont {Singer}, \citenamefont {Steinhilber}, \citenamefont
  {Togashi}, \citenamefont {Tsunoda}, \citenamefont {Werner},\ and\
  \citenamefont {Zweidinger}}]{Krem16}%
  \BibitemOpen
  \bibfield  {author} {\bibinfo {author} {\bibfnamefont {C.}~\bibnamefont
  {Kremer}}, \bibinfo {author} {\bibfnamefont {S.}~\bibnamefont {Aslanidou}},
  \bibinfo {author} {\bibfnamefont {S.}~\bibnamefont {Bassauer}}, \bibinfo
  {author} {\bibfnamefont {M.}~\bibnamefont {Hilcker}}, \bibinfo {author}
  {\bibfnamefont {A.}~\bibnamefont {Krugmann}}, \bibinfo {author}
  {\bibfnamefont {P.}~\bibnamefont {von Neumann-Cosel}}, \bibinfo {author}
  {\bibfnamefont {T.}~\bibnamefont {Otsuka}}, \bibinfo {author} {\bibfnamefont
  {N.}~\bibnamefont {Pietralla}}, \bibinfo {author} {\bibfnamefont {V.~Yu.}\
  \bibnamefont {Ponomarev}}, \bibinfo {author} {\bibfnamefont {N.}~\bibnamefont
  {Shimizu}}, \bibinfo {author} {\bibfnamefont {M.}~\bibnamefont {Singer}},
  \bibinfo {author} {\bibfnamefont {G.}~\bibnamefont {Steinhilber}}, \bibinfo
  {author} {\bibfnamefont {T.}~\bibnamefont {Togashi}}, \bibinfo {author}
  {\bibfnamefont {Y.}~\bibnamefont {Tsunoda}}, \bibinfo {author} {\bibfnamefont
  {V.}~\bibnamefont {Werner}}, \ and\ \bibinfo {author} {\bibfnamefont
  {M.}~\bibnamefont {Zweidinger}},\ }\bibfield  {title} {\enquote {\bibinfo
  {title} {{First Measurement of Collectivity of Coexisting Shapes Based on
  Type II Shell Evolution: The Case of $^{96}\mathrm{Zr}$}},}\ }\href {\doibase
  10.1103/PhysRevLett.117.172503} {\bibfield  {journal} {\bibinfo  {journal}
  {Phys. Rev. Lett.}\ }\textbf {\bibinfo {volume} {117}},\ \bibinfo {pages}
  {172503} (\bibinfo {year} {2016})}\BibitemShut {NoStop}%
\bibitem [{\citenamefont {Pietralla}\ \emph {et~al.}(2018)\citenamefont
  {Pietralla}, \citenamefont {Kremer}, \citenamefont {Beck}, \citenamefont
  {Witt}, \citenamefont {Gayer}, \citenamefont {von Neumann-Cosel},\ and\
  \citenamefont {Werner}}]{Piet18}%
  \BibitemOpen
  \bibfield  {author} {\bibinfo {author} {\bibfnamefont {N.}~\bibnamefont
  {Pietralla}}, \bibinfo {author} {\bibfnamefont {C.}~\bibnamefont {Kremer}},
  \bibinfo {author} {\bibfnamefont {T.}~\bibnamefont {Beck}}, \bibinfo {author}
  {\bibfnamefont {W.}~\bibnamefont {Witt}}, \bibinfo {author} {\bibfnamefont
  {U.}~\bibnamefont {Gayer}}, \bibinfo {author} {\bibfnamefont
  {P.}~\bibnamefont {von Neumann-Cosel}}, \ and\ \bibinfo {author}
  {\bibfnamefont {V.}~\bibnamefont {Werner}},\ }\bibfield  {title} {\enquote
  {\bibinfo {title} {{Shell Evolution and E2 Collectivity: New Spectroscopic
  Information}},}\ }\bibfield  {booktitle} {\emph {\bibinfo {booktitle}
  {{Proceedings, 16th International Symposium on Capture Gamma-Ray Spectroscopy
  and Related Topics (CGS16): Shanghai, China, September 18-22, 2017}}},\
  }\href {\doibase 10.1051/epjconf/201817802007} {\bibfield  {journal}
  {\bibinfo  {journal} {EPJ Web Conf.}\ }\textbf {\bibinfo {volume} {178}},\
  \bibinfo {pages} {02007} (\bibinfo {year} {2018})}\BibitemShut {NoStop}%
\bibitem [{\citenamefont {Singh}\ and\ \citenamefont {Hu}(2003)}]{Sing03}%
  \BibitemOpen
  \bibfield  {author} {\bibinfo {author} {\bibfnamefont {Balraj}\ \bibnamefont
  {Singh}}\ and\ \bibinfo {author} {\bibfnamefont {Zhiqiang}\ \bibnamefont
  {Hu}},\ }\bibfield  {title} {\enquote {\bibinfo {title} {{Nuclear Data Sheets
  for A = 98}},}\ }\href {\doibase https://doi.org/10.1006/ndsh.2003.0003}
  {\bibfield  {journal} {\bibinfo  {journal} {Nuclear Data Sheets}\ }\textbf
  {\bibinfo {volume} {98}},\ \bibinfo {pages} {335 -- 513} (\bibinfo {year}
  {2003})}\BibitemShut {NoStop}%
\bibitem [{\citenamefont {Bettermann}\ \emph {et~al.}(2010)\citenamefont
  {Bettermann}, \citenamefont {Regis}, \citenamefont {Materna}, \citenamefont
  {Jolie}, \citenamefont {Koster}, \citenamefont {Moschner},\ and\
  \citenamefont {Radeck}}]{Bett10}%
  \BibitemOpen
  \bibfield  {author} {\bibinfo {author} {\bibfnamefont {L.}~\bibnamefont
  {Bettermann}}, \bibinfo {author} {\bibfnamefont {J.~M.}\ \bibnamefont
  {Regis}}, \bibinfo {author} {\bibfnamefont {T.}~\bibnamefont {Materna}},
  \bibinfo {author} {\bibfnamefont {J.}~\bibnamefont {Jolie}}, \bibinfo
  {author} {\bibfnamefont {U.}~\bibnamefont {Koster}}, \bibinfo {author}
  {\bibfnamefont {K.}~\bibnamefont {Moschner}}, \ and\ \bibinfo {author}
  {\bibfnamefont {D.}~\bibnamefont {Radeck}},\ }\bibfield  {title} {\enquote
  {\bibinfo {title} {{Lifetime measurement of excited states in the
  shape-phase-transitional nucleus Zr-98}},}\ }\href {\doibase
  10.1103/PhysRevC.82.044310} {\bibfield  {journal} {\bibinfo  {journal} {Phys.
  Rev.}\ }\textbf {\bibinfo {volume} {C82}},\ \bibinfo {pages} {044310}
  (\bibinfo {year} {2010})}\BibitemShut {NoStop}%
\bibitem [{\citenamefont {Singh}\ \emph {et~al.}(2018)\citenamefont {Singh},
  \citenamefont {Korten}, \citenamefont {Hagen}, \citenamefont {G\"orgen},
  \citenamefont {Grente}, \citenamefont {Salsac}, \citenamefont {Farget},
  \citenamefont {Cl\'ement}, \citenamefont {de~France}, \citenamefont
  {Braunroth}, \citenamefont {Bruyneel}, \citenamefont {Celikovic},
  \citenamefont {Delaune}, \citenamefont {Dewald}, \citenamefont {Dijon},
  \citenamefont {Delaroche}, \citenamefont {Girod}, \citenamefont {Hackstein},
  \citenamefont {Jacquot}, \citenamefont {Libert}, \citenamefont {Litzinger},
  \citenamefont {Ljungvall}, \citenamefont {Louchart}, \citenamefont
  {Gottardo}, \citenamefont {Michelagnoli}, \citenamefont {M\"uller-Gatermann},
  \citenamefont {Napoli}, \citenamefont {Otsuka}, \citenamefont {Pillet},
  \citenamefont {Recchia}, \citenamefont {Rother}, \citenamefont {Sahin},
  \citenamefont {Siem}, \citenamefont {Sulignano}, \citenamefont {Togashi},
  \citenamefont {Tsunoda}, \citenamefont {Theisen},\ and\ \citenamefont
  {Valiente-Dobon}}]{Singh18}%
  \BibitemOpen
  \bibfield  {author} {\bibinfo {author} {\bibfnamefont {Purnima}\ \bibnamefont
  {Singh}}, \bibinfo {author} {\bibfnamefont {W.}~\bibnamefont {Korten}},
  \bibinfo {author} {\bibfnamefont {T.~W.}\ \bibnamefont {Hagen}}, \bibinfo
  {author} {\bibfnamefont {A.}~\bibnamefont {G\"orgen}}, \bibinfo {author}
  {\bibfnamefont {L.}~\bibnamefont {Grente}}, \bibinfo {author} {\bibfnamefont
  {M.-D.}\ \bibnamefont {Salsac}}, \bibinfo {author} {\bibfnamefont
  {F.}~\bibnamefont {Farget}}, \bibinfo {author} {\bibfnamefont
  {E.}~\bibnamefont {Cl\'ement}}, \bibinfo {author} {\bibfnamefont
  {G.}~\bibnamefont {de~France}}, \bibinfo {author} {\bibfnamefont
  {T.}~\bibnamefont {Braunroth}}, \bibinfo {author} {\bibfnamefont
  {B.}~\bibnamefont {Bruyneel}}, \bibinfo {author} {\bibfnamefont
  {I.}~\bibnamefont {Celikovic}}, \bibinfo {author} {\bibfnamefont
  {O.}~\bibnamefont {Delaune}}, \bibinfo {author} {\bibfnamefont
  {A.}~\bibnamefont {Dewald}}, \bibinfo {author} {\bibfnamefont
  {A.}~\bibnamefont {Dijon}}, \bibinfo {author} {\bibfnamefont {J.-P.}\
  \bibnamefont {Delaroche}}, \bibinfo {author} {\bibfnamefont {M.}~\bibnamefont
  {Girod}}, \bibinfo {author} {\bibfnamefont {M.}~\bibnamefont {Hackstein}},
  \bibinfo {author} {\bibfnamefont {B.}~\bibnamefont {Jacquot}}, \bibinfo
  {author} {\bibfnamefont {J.}~\bibnamefont {Libert}}, \bibinfo {author}
  {\bibfnamefont {J.}~\bibnamefont {Litzinger}}, \bibinfo {author}
  {\bibfnamefont {J.}~\bibnamefont {Ljungvall}}, \bibinfo {author}
  {\bibfnamefont {C.}~\bibnamefont {Louchart}}, \bibinfo {author}
  {\bibfnamefont {A.}~\bibnamefont {Gottardo}}, \bibinfo {author}
  {\bibfnamefont {C.}~\bibnamefont {Michelagnoli}}, \bibinfo {author}
  {\bibfnamefont {C.}~\bibnamefont {M\"uller-Gatermann}}, \bibinfo {author}
  {\bibfnamefont {D.~R.}\ \bibnamefont {Napoli}}, \bibinfo {author}
  {\bibfnamefont {T.}~\bibnamefont {Otsuka}}, \bibinfo {author} {\bibfnamefont
  {N.}~\bibnamefont {Pillet}}, \bibinfo {author} {\bibfnamefont
  {F.}~\bibnamefont {Recchia}}, \bibinfo {author} {\bibfnamefont
  {W.}~\bibnamefont {Rother}}, \bibinfo {author} {\bibfnamefont
  {E.}~\bibnamefont {Sahin}}, \bibinfo {author} {\bibfnamefont
  {S.}~\bibnamefont {Siem}}, \bibinfo {author} {\bibfnamefont {B.}~\bibnamefont
  {Sulignano}}, \bibinfo {author} {\bibfnamefont {T.}~\bibnamefont {Togashi}},
  \bibinfo {author} {\bibfnamefont {Y.}~\bibnamefont {Tsunoda}}, \bibinfo
  {author} {\bibfnamefont {Ch.}\ \bibnamefont {Theisen}}, \ and\ \bibinfo
  {author} {\bibfnamefont {J.~J.}\ \bibnamefont {Valiente-Dobon}},\ }\bibfield
  {title} {\enquote {\bibinfo {title} {{Evidence for Coexisting Shapes through
  Lifetime Measurements in $^{98}\mathrm{Zr}$}},}\ }\href {\doibase
  10.1103/PhysRevLett.121.192501} {\bibfield  {journal} {\bibinfo  {journal}
  {Phys. Rev. Lett.}\ }\textbf {\bibinfo {volume} {121}},\ \bibinfo {pages}
  {192501} (\bibinfo {year} {2018})}\BibitemShut {NoStop}%
\bibitem [{\citenamefont {Werner}\ \emph {et~al.}(2018)\citenamefont {Werner}
  \emph {et~al.}}]{Wern18}%
  \BibitemOpen
  \bibfield  {author} {\bibinfo {author} {\bibfnamefont {V.}~\bibnamefont
  {Werner}} \emph {et~al.},\ }\bibfield  {title} {\enquote {\bibinfo {title}
  {{Nuclear shapes: Quest for triaxiality in $^{86}$Ge and the shape of
  $^{98}$Zr}},}\ }\bibfield  {booktitle} {\emph {\bibinfo {booktitle}
  {{Proceedings, 16th International Symposium on Capture Gamma-Ray Spectroscopy
  and Related Topics (CGS16): Shanghai, China, September 18-22, 2017}}},\
  }\href {\doibase 10.1051/epjconf/201817802013} {\bibfield  {journal}
  {\bibinfo  {journal} {EPJ Web Conf.}\ }\textbf {\bibinfo {volume} {178}},\
  \bibinfo {pages} {02013} (\bibinfo {year} {2018})}\BibitemShut {NoStop}%
\bibitem [{\citenamefont {Witt}\ \emph {et~al.}(2018)\citenamefont {Witt},
  \citenamefont {Werner}, \citenamefont {Pietralla}, \citenamefont {Albers},
  \citenamefont {Ayangeakaa}, \citenamefont {Bucher}, \citenamefont
  {Carpenter}, \citenamefont {Cline}, \citenamefont {David}, \citenamefont
  {Hayes}, \citenamefont {Hoffman}, \citenamefont {Janssens}, \citenamefont
  {Kay}, \citenamefont {Kondev}, \citenamefont {Korten}, \citenamefont
  {Lauritsen}, \citenamefont {M\"oller}, \citenamefont {Rainovski},
  \citenamefont {Savard}, \citenamefont {Seweryniak}, \citenamefont {Smith},
  \citenamefont {Stegmann}, \citenamefont {Zhu},\ and\ \citenamefont
  {Wu}}]{Witt18}%
  \BibitemOpen
  \bibfield  {author} {\bibinfo {author} {\bibfnamefont {W.}~\bibnamefont
  {Witt}}, \bibinfo {author} {\bibfnamefont {V.}~\bibnamefont {Werner}},
  \bibinfo {author} {\bibfnamefont {N.}~\bibnamefont {Pietralla}}, \bibinfo
  {author} {\bibfnamefont {M.}~\bibnamefont {Albers}}, \bibinfo {author}
  {\bibfnamefont {A.~D.}\ \bibnamefont {Ayangeakaa}}, \bibinfo {author}
  {\bibfnamefont {B.}~\bibnamefont {Bucher}}, \bibinfo {author} {\bibfnamefont
  {M.~P.}\ \bibnamefont {Carpenter}}, \bibinfo {author} {\bibfnamefont
  {D.}~\bibnamefont {Cline}}, \bibinfo {author} {\bibfnamefont {H.~M.}\
  \bibnamefont {David}}, \bibinfo {author} {\bibfnamefont {A.}~\bibnamefont
  {Hayes}}, \bibinfo {author} {\bibfnamefont {C.}~\bibnamefont {Hoffman}},
  \bibinfo {author} {\bibfnamefont {R.~V.~F.}\ \bibnamefont {Janssens}},
  \bibinfo {author} {\bibfnamefont {B.~P.}\ \bibnamefont {Kay}}, \bibinfo
  {author} {\bibfnamefont {F.~G.}\ \bibnamefont {Kondev}}, \bibinfo {author}
  {\bibfnamefont {W.}~\bibnamefont {Korten}}, \bibinfo {author} {\bibfnamefont
  {T.}~\bibnamefont {Lauritsen}}, \bibinfo {author} {\bibfnamefont
  {O.}~\bibnamefont {M\"oller}}, \bibinfo {author} {\bibfnamefont
  {G.}~\bibnamefont {Rainovski}}, \bibinfo {author} {\bibfnamefont
  {G.}~\bibnamefont {Savard}}, \bibinfo {author} {\bibfnamefont
  {D.}~\bibnamefont {Seweryniak}}, \bibinfo {author} {\bibfnamefont
  {J.}~\bibnamefont {Smith}}, \bibinfo {author} {\bibfnamefont
  {R.}~\bibnamefont {Stegmann}}, \bibinfo {author} {\bibfnamefont
  {S.}~\bibnamefont {Zhu}}, \ and\ \bibinfo {author} {\bibfnamefont {C.~Y.}\
  \bibnamefont {Wu}},\ }\bibfield  {title} {\enquote {\bibinfo {title}
  {{Sub-shell closure and shape coexistence in the transitional nucleus
  $^{98}\mathrm{Zr}$}},}\ }\href {\doibase 10.1103/PhysRevC.98.041302}
  {\bibfield  {journal} {\bibinfo  {journal} {Phys. Rev. C}\ }\textbf {\bibinfo
  {volume} {98}},\ \bibinfo {pages} {041302} (\bibinfo {year}
  {2018})}\BibitemShut {NoStop}%
\bibitem [{\citenamefont {Singh}(2008)}]{Sing08}%
  \BibitemOpen
  \bibfield  {author} {\bibinfo {author} {\bibfnamefont {Balraj}\ \bibnamefont
  {Singh}},\ }\bibfield  {title} {\enquote {\bibinfo {title} {{Nuclear Data
  Sheets for A = 100}},}\ }\href {\doibase
  https://doi.org/10.1016/j.nds.2008.01.001} {\bibfield  {journal} {\bibinfo
  {journal} {Nuclear Data Sheets}\ }\textbf {\bibinfo {volume} {109}},\
  \bibinfo {pages} {297 -- 516} (\bibinfo {year} {2008})}\BibitemShut {NoStop}%
\bibitem [{\citenamefont {DeFrenne}(2009)}]{Defr09}%
  \BibitemOpen
  \bibfield  {author} {\bibinfo {author} {\bibfnamefont {D.}~\bibnamefont
  {DeFrenne}},\ }\bibfield  {title} {\enquote {\bibinfo {title} {{Nuclear Data
  Sheets for A = 102}},}\ }\href {\doibase
  https://doi.org/10.1016/j.nds.2009.06.002} {\bibfield  {journal} {\bibinfo
  {journal} {Nuclear Data Sheets}\ }\textbf {\bibinfo {volume} {110}},\
  \bibinfo {pages} {1745 -- 1915} (\bibinfo {year} {2009})}\BibitemShut
  {NoStop}%
\bibitem [{\citenamefont {Blachot}(2007)}]{Blac07}%
  \BibitemOpen
  \bibfield  {author} {\bibinfo {author} {\bibfnamefont {Jean}\ \bibnamefont
  {Blachot}},\ }\bibfield  {title} {\enquote {\bibinfo {title} {{Nuclear Data
  Sheets for A = 104}},}\ }\href {\doibase
  https://doi.org/10.1016/j.nds.2007.09.001} {\bibfield  {journal} {\bibinfo
  {journal} {Nuclear Data Sheets}\ }\textbf {\bibinfo {volume} {108}},\
  \bibinfo {pages} {2035 -- 2172} (\bibinfo {year} {2007})}\BibitemShut
  {NoStop}%
\bibitem [{\citenamefont {Singh}(2015{\natexlab{a}})}]{Sing15}%
  \BibitemOpen
  \bibfield  {author} {\bibinfo {author} {\bibfnamefont {B.}~\bibnamefont
  {Singh}},\ }\href@noop {} {\enquote {\bibinfo {title} {{Nuclear Data Sheets
  for $^{106}$Zr}},}\ }\bibinfo {howpublished} {http://www.nndc.bnl.gov/ensdf}
  (\bibinfo {year} {2015}{\natexlab{a}})\BibitemShut {NoStop}%
\bibitem [{\citenamefont {Singh}(2015{\natexlab{b}})}]{Sing15b}%
  \BibitemOpen
  \bibfield  {author} {\bibinfo {author} {\bibfnamefont {B.}~\bibnamefont
  {Singh}},\ }\href@noop {} {\enquote {\bibinfo {title} {{Nuclear Data Sheets
  for $^{108}$Zr}},}\ }\bibinfo {howpublished} {http://www.nndc.bnl.gov/ensdf}
  (\bibinfo {year} {2015}{\natexlab{b}})\BibitemShut {NoStop}%
\bibitem [{\citenamefont {Navin}\ \emph {et~al.}(2014)\citenamefont {Navin},
  \citenamefont {Rejmund}, \citenamefont {Schmitt}, \citenamefont
  {Bhattacharyya}, \citenamefont {Lhersonneau}, \citenamefont {Isacker},
  \citenamefont {Caamaño}, \citenamefont {Cl\'ement}, \citenamefont {Delaune},
  \citenamefont {Farget}, \citenamefont {de~France},\ and\ \citenamefont
  {Jacquot}}]{Navin14}%
  \BibitemOpen
  \bibfield  {author} {\bibinfo {author} {\bibfnamefont {A.}~\bibnamefont
  {Navin}}, \bibinfo {author} {\bibfnamefont {M.}~\bibnamefont {Rejmund}},
  \bibinfo {author} {\bibfnamefont {C.}~\bibnamefont {Schmitt}}, \bibinfo
  {author} {\bibfnamefont {S.}~\bibnamefont {Bhattacharyya}}, \bibinfo {author}
  {\bibfnamefont {G.}~\bibnamefont {Lhersonneau}}, \bibinfo {author}
  {\bibfnamefont {P.~Van}\ \bibnamefont {Isacker}}, \bibinfo {author}
  {\bibfnamefont {M.}~\bibnamefont {Caamaño}}, \bibinfo {author}
  {\bibfnamefont {E.}~\bibnamefont {Cl\'ement}}, \bibinfo {author}
  {\bibfnamefont {O.}~\bibnamefont {Delaune}}, \bibinfo {author} {\bibfnamefont
  {F.}~\bibnamefont {Farget}}, \bibinfo {author} {\bibfnamefont
  {G.}~\bibnamefont {de~France}}, \ and\ \bibinfo {author} {\bibfnamefont
  {B.}~\bibnamefont {Jacquot}},\ }\bibfield  {title} {\enquote {\bibinfo
  {title} {{Towards the high spin–isospin frontier using
  isotopically-identified fission fragments}},}\ }\href {\doibase
  https://doi.org/10.1016/j.physletb.2013.11.024} {\bibfield  {journal}
  {\bibinfo  {journal} {Phys. Lett. B}\ }\textbf {\bibinfo {volume} {728}},\
  \bibinfo {pages} {136 -- 140} (\bibinfo {year} {2014})}\BibitemShut {NoStop}%
\bibitem [{\citenamefont {Paul}\ \emph {et~al.}(2017)\citenamefont {Paul},
  \citenamefont {Corsi}, \citenamefont {Obertelli}, \citenamefont {Doornenbal},
  \citenamefont {Authelet}, \citenamefont {Baba}, \citenamefont {Bally},
  \citenamefont {Bender}, \citenamefont {Calvet}, \citenamefont {Ch\^ateau},
  \citenamefont {Chen}, \citenamefont {Delaroche}, \citenamefont {Delbart},
  \citenamefont {Gheller}, \citenamefont {Giganon}, \citenamefont {Gillibert},
  \citenamefont {Girod}, \citenamefont {Heenen}, \citenamefont {Lapoux},
  \citenamefont {Libert}, \citenamefont {Motobayashi}, \citenamefont {Niikura},
  \citenamefont {Otsuka}, \citenamefont {Rodr\'{\i}guez}, \citenamefont
  {Rouss\'e}, \citenamefont {Sakurai}, \citenamefont {Santamaria},
  \citenamefont {Shimizu}, \citenamefont {Steppenbeck}, \citenamefont
  {Taniuchi}, \citenamefont {Togashi}, \citenamefont {Tsunoda}, \citenamefont
  {Uesaka}, \citenamefont {Ando}, \citenamefont {Arici}, \citenamefont
  {Blazhev}, \citenamefont {Browne}, \citenamefont {Bruce}, \citenamefont
  {Carroll}, \citenamefont {Chung}, \citenamefont {Cort\'es}, \citenamefont
  {Dewald}, \citenamefont {Ding}, \citenamefont {Flavigny}, \citenamefont
  {Franchoo}, \citenamefont {G\'orska}, \citenamefont {Gottardo}, \citenamefont
  {Jungclaus}, \citenamefont {Lee}, \citenamefont {Lettmann}, \citenamefont
  {Linh}, \citenamefont {Liu}, \citenamefont {Liu}, \citenamefont {Lizarazo},
  \citenamefont {Momiyama}, \citenamefont {Moschner}, \citenamefont {Nagamine},
  \citenamefont {Nakatsuka}, \citenamefont {Nita}, \citenamefont {Nobs},
  \citenamefont {Olivier}, \citenamefont {Patel}, \citenamefont {Podoly\'ak},
  \citenamefont {Rudigier}, \citenamefont {Saito}, \citenamefont {Shand},
  \citenamefont {S\"oderstr\"om}, \citenamefont {Stefan}, \citenamefont
  {Orlandi}, \citenamefont {Vaquero}, \citenamefont {Werner}, \citenamefont
  {Wimmer},\ and\ \citenamefont {Xu}}]{Paul17}%
  \BibitemOpen
  \bibfield  {author} {\bibinfo {author} {\bibfnamefont {N.}~\bibnamefont
  {Paul}}, \bibinfo {author} {\bibfnamefont {A.}~\bibnamefont {Corsi}},
  \bibinfo {author} {\bibfnamefont {A.}~\bibnamefont {Obertelli}}, \bibinfo
  {author} {\bibfnamefont {P.}~\bibnamefont {Doornenbal}}, \bibinfo {author}
  {\bibfnamefont {G.}~\bibnamefont {Authelet}}, \bibinfo {author}
  {\bibfnamefont {H.}~\bibnamefont {Baba}}, \bibinfo {author} {\bibfnamefont
  {B.}~\bibnamefont {Bally}}, \bibinfo {author} {\bibfnamefont
  {M.}~\bibnamefont {Bender}}, \bibinfo {author} {\bibfnamefont
  {D.}~\bibnamefont {Calvet}}, \bibinfo {author} {\bibfnamefont
  {F.}~\bibnamefont {Ch\^ateau}}, \bibinfo {author} {\bibfnamefont
  {S.}~\bibnamefont {Chen}}, \bibinfo {author} {\bibfnamefont {J.-P.}\
  \bibnamefont {Delaroche}}, \bibinfo {author} {\bibfnamefont {A.}~\bibnamefont
  {Delbart}}, \bibinfo {author} {\bibfnamefont {J.-M.}\ \bibnamefont
  {Gheller}}, \bibinfo {author} {\bibfnamefont {A.}~\bibnamefont {Giganon}},
  \bibinfo {author} {\bibfnamefont {A.}~\bibnamefont {Gillibert}}, \bibinfo
  {author} {\bibfnamefont {M.}~\bibnamefont {Girod}}, \bibinfo {author}
  {\bibfnamefont {P.-H.}\ \bibnamefont {Heenen}}, \bibinfo {author}
  {\bibfnamefont {V.}~\bibnamefont {Lapoux}}, \bibinfo {author} {\bibfnamefont
  {J.}~\bibnamefont {Libert}}, \bibinfo {author} {\bibfnamefont
  {T.}~\bibnamefont {Motobayashi}}, \bibinfo {author} {\bibfnamefont
  {M.}~\bibnamefont {Niikura}}, \bibinfo {author} {\bibfnamefont
  {T.}~\bibnamefont {Otsuka}}, \bibinfo {author} {\bibfnamefont {T.~R.}\
  \bibnamefont {Rodr\'{\i}guez}}, \bibinfo {author} {\bibfnamefont {J.-Y.}\
  \bibnamefont {Rouss\'e}}, \bibinfo {author} {\bibfnamefont {H.}~\bibnamefont
  {Sakurai}}, \bibinfo {author} {\bibfnamefont {C.}~\bibnamefont {Santamaria}},
  \bibinfo {author} {\bibfnamefont {N.}~\bibnamefont {Shimizu}}, \bibinfo
  {author} {\bibfnamefont {D.}~\bibnamefont {Steppenbeck}}, \bibinfo {author}
  {\bibfnamefont {R.}~\bibnamefont {Taniuchi}}, \bibinfo {author}
  {\bibfnamefont {T.}~\bibnamefont {Togashi}}, \bibinfo {author} {\bibfnamefont
  {Y.}~\bibnamefont {Tsunoda}}, \bibinfo {author} {\bibfnamefont
  {T.}~\bibnamefont {Uesaka}}, \bibinfo {author} {\bibfnamefont
  {T.}~\bibnamefont {Ando}}, \bibinfo {author} {\bibfnamefont {T.}~\bibnamefont
  {Arici}}, \bibinfo {author} {\bibfnamefont {A.}~\bibnamefont {Blazhev}},
  \bibinfo {author} {\bibfnamefont {F.}~\bibnamefont {Browne}}, \bibinfo
  {author} {\bibfnamefont {A.~M.}\ \bibnamefont {Bruce}}, \bibinfo {author}
  {\bibfnamefont {R.}~\bibnamefont {Carroll}}, \bibinfo {author} {\bibfnamefont
  {L.~X.}\ \bibnamefont {Chung}}, \bibinfo {author} {\bibfnamefont {M.~L.}\
  \bibnamefont {Cort\'es}}, \bibinfo {author} {\bibfnamefont {M.}~\bibnamefont
  {Dewald}}, \bibinfo {author} {\bibfnamefont {B.}~\bibnamefont {Ding}},
  \bibinfo {author} {\bibfnamefont {F.}~\bibnamefont {Flavigny}}, \bibinfo
  {author} {\bibfnamefont {S.}~\bibnamefont {Franchoo}}, \bibinfo {author}
  {\bibfnamefont {M.}~\bibnamefont {G\'orska}}, \bibinfo {author}
  {\bibfnamefont {A.}~\bibnamefont {Gottardo}}, \bibinfo {author}
  {\bibfnamefont {A.}~\bibnamefont {Jungclaus}}, \bibinfo {author}
  {\bibfnamefont {J.}~\bibnamefont {Lee}}, \bibinfo {author} {\bibfnamefont
  {M.}~\bibnamefont {Lettmann}}, \bibinfo {author} {\bibfnamefont {B.~D.}\
  \bibnamefont {Linh}}, \bibinfo {author} {\bibfnamefont {J.}~\bibnamefont
  {Liu}}, \bibinfo {author} {\bibfnamefont {Z.}~\bibnamefont {Liu}}, \bibinfo
  {author} {\bibfnamefont {C.}~\bibnamefont {Lizarazo}}, \bibinfo {author}
  {\bibfnamefont {S.}~\bibnamefont {Momiyama}}, \bibinfo {author}
  {\bibfnamefont {K.}~\bibnamefont {Moschner}}, \bibinfo {author}
  {\bibfnamefont {S.}~\bibnamefont {Nagamine}}, \bibinfo {author}
  {\bibfnamefont {N.}~\bibnamefont {Nakatsuka}}, \bibinfo {author}
  {\bibfnamefont {C.}~\bibnamefont {Nita}}, \bibinfo {author} {\bibfnamefont
  {C.~R.}\ \bibnamefont {Nobs}}, \bibinfo {author} {\bibfnamefont
  {L.}~\bibnamefont {Olivier}}, \bibinfo {author} {\bibfnamefont
  {Z.}~\bibnamefont {Patel}}, \bibinfo {author} {\bibfnamefont {Zs.}\
  \bibnamefont {Podoly\'ak}}, \bibinfo {author} {\bibfnamefont
  {M.}~\bibnamefont {Rudigier}}, \bibinfo {author} {\bibfnamefont
  {T.}~\bibnamefont {Saito}}, \bibinfo {author} {\bibfnamefont
  {C.}~\bibnamefont {Shand}}, \bibinfo {author} {\bibfnamefont {P.-A.}\
  \bibnamefont {S\"oderstr\"om}}, \bibinfo {author} {\bibfnamefont
  {I.}~\bibnamefont {Stefan}}, \bibinfo {author} {\bibfnamefont
  {R.}~\bibnamefont {Orlandi}}, \bibinfo {author} {\bibfnamefont
  {V.}~\bibnamefont {Vaquero}}, \bibinfo {author} {\bibfnamefont
  {V.}~\bibnamefont {Werner}}, \bibinfo {author} {\bibfnamefont
  {K.}~\bibnamefont {Wimmer}}, \ and\ \bibinfo {author} {\bibfnamefont
  {Z.}~\bibnamefont {Xu}},\ }\bibfield  {title} {\enquote {\bibinfo {title}
  {{Are There Signatures of Harmonic Oscillator Shells Far from Stability?
  First Spectroscopy of $^{110}\mathrm{Zr}$}},}\ }\href {\doibase
  10.1103/PhysRevLett.118.032501} {\bibfield  {journal} {\bibinfo  {journal}
  {Phys. Rev. Lett.}\ }\textbf {\bibinfo {volume} {118}},\ \bibinfo {pages}
  {032501} (\bibinfo {year} {2017})}\BibitemShut {NoStop}%
\bibitem [{\citenamefont {Wood}\ \emph {et~al.}(1999)\citenamefont {Wood},
  \citenamefont {Zganjar}, \citenamefont {Coster},\ and\ \citenamefont
  {Heyde}}]{Wood99}%
  \BibitemOpen
  \bibfield  {author} {\bibinfo {author} {\bibfnamefont {J.L.}\ \bibnamefont
  {Wood}}, \bibinfo {author} {\bibfnamefont {E.F.}\ \bibnamefont {Zganjar}},
  \bibinfo {author} {\bibfnamefont {C.~De}\ \bibnamefont {Coster}}, \ and\
  \bibinfo {author} {\bibfnamefont {K.}~\bibnamefont {Heyde}},\ }\bibfield
  {title} {\enquote {\bibinfo {title} {{Electric monopole transitions from low
  energy excitations in nuclei}},}\ }\href {\doibase
  https://doi.org/10.1016/S0375-9474(99)00143-8} {\bibfield  {journal}
  {\bibinfo  {journal} {Nucl. Phys. A}\ }\textbf {\bibinfo {volume} {651}},\
  \bibinfo {pages} {323 -- 368} (\bibinfo {year} {1999})}\BibitemShut {NoStop}%
\bibitem [{\citenamefont {Iachello}\ \emph {et~al.}(1998)\citenamefont
  {Iachello}, \citenamefont {Zamfir},\ and\ \citenamefont {Casten}}]{Iach98}%
  \BibitemOpen
  \bibfield  {author} {\bibinfo {author} {\bibfnamefont {F.}~\bibnamefont
  {Iachello}}, \bibinfo {author} {\bibfnamefont {N.~V.}\ \bibnamefont
  {Zamfir}}, \ and\ \bibinfo {author} {\bibfnamefont {R.~F.}\ \bibnamefont
  {Casten}},\ }\bibfield  {title} {\enquote {\bibinfo {title} {Phase
  coexistence in transitional nuclei and the interacting-boson model},}\ }\href
  {\doibase 10.1103/PhysRevLett.81.1191} {\bibfield  {journal} {\bibinfo
  {journal} {Phys. Rev. Lett.}\ }\textbf {\bibinfo {volume} {81}},\ \bibinfo
  {pages} {1191--1194} (\bibinfo {year} {1998})}\BibitemShut {NoStop}%
\bibitem [{\citenamefont {Ising}()}]{Isin25}%
  \BibitemOpen
  \bibfield  {author} {\bibinfo {author} {\bibfnamefont {E.}~\bibnamefont
  {Ising}},\ }\bibfield  {title} {\enquote {\bibinfo {title} {{Beitrag zur
  Theorie des Ferromagnetismus}},}\ }\href@noop {} {\bibfield  {journal}
  {\bibinfo  {journal} {Z. Physik}\ }\textbf {\bibinfo {volume} {31}},\
  \bibinfo {pages} {253}}\BibitemShut {NoStop}%
\bibitem [{\citenamefont {García-Ramos}\ \emph {et~al.}(2001)\citenamefont
  {García-Ramos}, \citenamefont {{De Coster}}, \citenamefont {Fossion},\ and\
  \citenamefont {Heyde}}]{Garc01}%
  \BibitemOpen
  \bibfield  {author} {\bibinfo {author} {\bibfnamefont {J.E.}\ \bibnamefont
  {García-Ramos}}, \bibinfo {author} {\bibfnamefont {C.}~\bibnamefont {{De
  Coster}}}, \bibinfo {author} {\bibfnamefont {R.}~\bibnamefont {Fossion}}, \
  and\ \bibinfo {author} {\bibfnamefont {K.}~\bibnamefont {Heyde}},\ }\bibfield
   {title} {\enquote {\bibinfo {title} {{Two-neutron separation energies,
  binding energies and phase transitions in the interacting boson model}},}\
  }\href {\doibase https://doi.org/10.1016/S0375-9474(00)00592-3} {\bibfield
  {journal} {\bibinfo  {journal} {Nuclear Physics A}\ }\textbf {\bibinfo
  {volume} {688}},\ \bibinfo {pages} {735 -- 754} (\bibinfo {year}
  {2001})}\BibitemShut {NoStop}%
\bibitem [{\citenamefont {Landau}\ and\ \citenamefont
  {Lifshitz}(1969)}]{Land69}%
  \BibitemOpen
  \bibfield  {author} {\bibinfo {author} {\bibfnamefont {L.D.}\ \bibnamefont
  {Landau}}\ and\ \bibinfo {author} {\bibfnamefont {E.M.}\ \bibnamefont
  {Lifshitz}},\ }\href@noop {} {\emph {\bibinfo {title} {Statistical
  Physics}}}\ (\bibinfo  {publisher} {Pergamon Press, Oxford},\ \bibinfo {year}
  {1969})\BibitemShut {NoStop}%
\bibitem [{\citenamefont {Garc\'{\i}a-Ramos}\ and\ \citenamefont
  {Heyde}(2019)}]{Garc19}%
  \BibitemOpen
  \bibfield  {author} {\bibinfo {author} {\bibfnamefont {J.~E.}\ \bibnamefont
  {Garc\'{\i}a-Ramos}}\ and\ \bibinfo {author} {\bibfnamefont {K.}~\bibnamefont
  {Heyde}},\ }\bibfield  {title} {\enquote {\bibinfo {title} {{Quest of shape
  coexistence in Zr isotopes}},}\ }\href {\doibase 10.1103/PhysRevC.100.044315}
  {\bibfield  {journal} {\bibinfo  {journal} {Phys. Rev. C}\ }\textbf {\bibinfo
  {volume} {100}},\ \bibinfo {pages} {044315} (\bibinfo {year}
  {2019})}\BibitemShut {NoStop}%
\bibitem [{\citenamefont {Iachello}(2000)}]{Iach00}%
  \BibitemOpen
  \bibfield  {author} {\bibinfo {author} {\bibfnamefont {F.}~\bibnamefont
  {Iachello}},\ }\bibfield  {title} {\enquote {\bibinfo {title} {Dynamic
  symmetries at the critical point},}\ }\href {\doibase
  10.1103/PhysRevLett.85.3580} {\bibfield  {journal} {\bibinfo  {journal}
  {Phys. Rev. Lett.}\ }\textbf {\bibinfo {volume} {85}},\ \bibinfo {pages}
  {3580--3583} (\bibinfo {year} {2000})}\BibitemShut {NoStop}%
\bibitem [{\citenamefont {Iachello}(2001)}]{Iach01}%
  \BibitemOpen
  \bibfield  {author} {\bibinfo {author} {\bibfnamefont {F.}~\bibnamefont
  {Iachello}},\ }\bibfield  {title} {\enquote {\bibinfo {title} {Analytic
  description of critical point nuclei in a spherical-axially deformed shape
  phase transition},}\ }\href {\doibase 10.1103/PhysRevLett.87.052502}
  {\bibfield  {journal} {\bibinfo  {journal} {Phys. Rev. Lett.}\ }\textbf
  {\bibinfo {volume} {87}},\ \bibinfo {pages} {052502} (\bibinfo {year}
  {2001})}\BibitemShut {NoStop}%
\bibitem [{\citenamefont {Iachello}\ and\ \citenamefont
  {Arima}(1987)}]{iach87}%
  \BibitemOpen
  \bibfield  {author} {\bibinfo {author} {\bibfnamefont {F.}~\bibnamefont
  {Iachello}}\ and\ \bibinfo {author} {\bibfnamefont {A.}~\bibnamefont
  {Arima}},\ }\href@noop {} {\emph {\bibinfo {title} {{The interacting boson
  model}}}}\ (\bibinfo  {publisher} {Cambridge University Press, Cambridge},\
  \bibinfo {year} {1987})\BibitemShut {NoStop}%
\bibitem [{\citenamefont {Duval}\ and\ \citenamefont
  {Barrett}(1981)}]{duval81}%
  \BibitemOpen
  \bibfield  {author} {\bibinfo {author} {\bibfnamefont {Philip~D.}\
  \bibnamefont {Duval}}\ and\ \bibinfo {author} {\bibfnamefont {Bruce~R.}\
  \bibnamefont {Barrett}},\ }\bibfield  {title} {\enquote {\bibinfo {title}
  {{Configuration mixing in the interacting boson model}},}\ }\href {\doibase
  https://doi.org/10.1016/0370-2693(81)90321-X} {\bibfield  {journal} {\bibinfo
   {journal} {Phys. Lett. B}\ }\textbf {\bibinfo {volume} {100}},\ \bibinfo
  {pages} {223 -- 227} (\bibinfo {year} {1981})}\BibitemShut {NoStop}%
\bibitem [{\citenamefont {Duval}\ and\ \citenamefont
  {Barrett}(1982)}]{duval82}%
  \BibitemOpen
  \bibfield  {author} {\bibinfo {author} {\bibfnamefont {Philip~D.}\
  \bibnamefont {Duval}}\ and\ \bibinfo {author} {\bibfnamefont {Bruce~R.}\
  \bibnamefont {Barrett}},\ }\bibfield  {title} {\enquote {\bibinfo {title}
  {{Quantitative description of configuration mixing in the interacting boson
  model}},}\ }\href {\doibase https://doi.org/10.1016/0375-9474(82)90061-6}
  {\bibfield  {journal} {\bibinfo  {journal} {Nucl. Phys. A}\ }\textbf
  {\bibinfo {volume} {376}},\ \bibinfo {pages} {213 -- 228} (\bibinfo {year}
  {1982})}\BibitemShut {NoStop}%
\bibitem [{\citenamefont {Warner}\ and\ \citenamefont
  {Casten}(1983)}]{warner83}%
  \BibitemOpen
  \bibfield  {author} {\bibinfo {author} {\bibfnamefont {D.~D.}\ \bibnamefont
  {Warner}}\ and\ \bibinfo {author} {\bibfnamefont {R.~F.}\ \bibnamefont
  {Casten}},\ }\bibfield  {title} {\enquote {\bibinfo {title} {{Predictions of
  the interacting boson approximation in a consistent $Q$ framework}},}\ }\href
  {\doibase 10.1103/PhysRevC.28.1798} {\bibfield  {journal} {\bibinfo
  {journal} {Phys. Rev. C}\ }\textbf {\bibinfo {volume} {28}},\ \bibinfo
  {pages} {1798--1806} (\bibinfo {year} {1983})}\BibitemShut {NoStop}%
\bibitem [{\citenamefont {Lipas}\ \emph {et~al.}(1985)\citenamefont {Lipas},
  \citenamefont {Toivonen},\ and\ \citenamefont {Warner}}]{lipas85}%
  \BibitemOpen
  \bibfield  {author} {\bibinfo {author} {\bibfnamefont {P.O.}\ \bibnamefont
  {Lipas}}, \bibinfo {author} {\bibfnamefont {P.}~\bibnamefont {Toivonen}}, \
  and\ \bibinfo {author} {\bibfnamefont {D.D.}\ \bibnamefont {Warner}},\
  }\bibfield  {title} {\enquote {\bibinfo {title} {{IBA consistent-Q formalism
  extended to the vibrational region}},}\ }\href {\doibase
  https://doi.org/10.1016/0370-2693(85)91573-4} {\bibfield  {journal} {\bibinfo
   {journal} {Phys. Lett. B}\ }\textbf {\bibinfo {volume} {155}},\ \bibinfo
  {pages} {295 -- 298} (\bibinfo {year} {1985})}\BibitemShut {NoStop}%
\bibitem [{\citenamefont {Heyde}\ \emph {et~al.}(1985)\citenamefont {Heyde},
  \citenamefont {Isacker}, \citenamefont {Casten},\ and\ \citenamefont
  {Wood}}]{Hey85}%
  \BibitemOpen
  \bibfield  {author} {\bibinfo {author} {\bibfnamefont {K.}~\bibnamefont
  {Heyde}}, \bibinfo {author} {\bibfnamefont {P.~Van}\ \bibnamefont {Isacker}},
  \bibinfo {author} {\bibfnamefont {R.F.}\ \bibnamefont {Casten}}, \ and\
  \bibinfo {author} {\bibfnamefont {J.L.}\ \bibnamefont {Wood}},\ }\bibfield
  {title} {\enquote {\bibinfo {title} {{A shell-model interpretation of
  intruder states and the onset of deformation in even-even nuclei}},}\ }\href
  {\doibase https://doi.org/10.1016/0370-2693(85)91575-8} {\bibfield  {journal}
  {\bibinfo  {journal} {Phys. Lett. B}\ }\textbf {\bibinfo {volume} {155}},\
  \bibinfo {pages} {303 -- 308} (\bibinfo {year} {1985})}\BibitemShut {NoStop}%
\bibitem [{\citenamefont {Heyde}\ \emph {et~al.}(1987)\citenamefont {Heyde},
  \citenamefont {Jolie}, \citenamefont {Moreau}, \citenamefont {Ryckebusch},
  \citenamefont {Waroquier}, \citenamefont {Duppen}, \citenamefont {Huyse},\
  and\ \citenamefont {Wood}}]{Hey87}%
  \BibitemOpen
  \bibfield  {author} {\bibinfo {author} {\bibfnamefont {K.}~\bibnamefont
  {Heyde}}, \bibinfo {author} {\bibfnamefont {J.}~\bibnamefont {Jolie}},
  \bibinfo {author} {\bibfnamefont {J.}~\bibnamefont {Moreau}}, \bibinfo
  {author} {\bibfnamefont {J.}~\bibnamefont {Ryckebusch}}, \bibinfo {author}
  {\bibfnamefont {M.}~\bibnamefont {Waroquier}}, \bibinfo {author}
  {\bibfnamefont {P.~Van}\ \bibnamefont {Duppen}}, \bibinfo {author}
  {\bibfnamefont {M.}~\bibnamefont {Huyse}}, \ and\ \bibinfo {author}
  {\bibfnamefont {J.L.}\ \bibnamefont {Wood}},\ }\bibfield  {title} {\enquote
  {\bibinfo {title} {{A shell-model description of 0+ intruder states in
  even-even nuclei}},}\ }\href {\doibase
  https://doi.org/10.1016/0375-9474(87)90439-8} {\bibfield  {journal} {\bibinfo
   {journal} {Nucl. Phys. A}\ }\textbf {\bibinfo {volume} {466}},\ \bibinfo
  {pages} {189 -- 226} (\bibinfo {year} {1987})}\BibitemShut {NoStop}%
\bibitem [{\citenamefont {Garc\'{\i}a-Ramos}\ and\ \citenamefont
  {Heyde}(2009)}]{Garc09}%
  \BibitemOpen
  \bibfield  {author} {\bibinfo {author} {\bibfnamefont {J.E.}\ \bibnamefont
  {Garc\'{\i}a-Ramos}}\ and\ \bibinfo {author} {\bibfnamefont {K.}~\bibnamefont
  {Heyde}},\ }\bibfield  {title} {\enquote {\bibinfo {title} {{The Pt isotopes:
  Comparing the Interacting Boson Model with configuration mixing and the
  extended consistent-Q formalism}},}\ }\href {\doibase
  https://doi.org/10.1016/j.nuclphysa.2009.04.003} {\bibfield  {journal}
  {\bibinfo  {journal} {Nucl. Phys. A}\ }\textbf {\bibinfo {volume} {825}},\
  \bibinfo {pages} {39 -- 70} (\bibinfo {year} {2009})}\BibitemShut {NoStop}%
\bibitem [{\citenamefont {Garc\'{\i}a-Ramos}\ and\ \citenamefont
  {Heyde}(2014)}]{Garc14b}%
  \BibitemOpen
  \bibfield  {author} {\bibinfo {author} {\bibfnamefont {J.~E.}\ \bibnamefont
  {Garc\'{\i}a-Ramos}}\ and\ \bibinfo {author} {\bibfnamefont {K.}~\bibnamefont
  {Heyde}},\ }\bibfield  {title} {\enquote {\bibinfo {title} {{Nuclear shape
  coexistence: A study of the even-even Hg isotopes using the interacting boson
  model with configuration mixing}},}\ }\href {\doibase
  10.1103/PhysRevC.89.014306} {\bibfield  {journal} {\bibinfo  {journal} {Phys.
  Rev. C}\ }\textbf {\bibinfo {volume} {89}},\ \bibinfo {pages} {014306}
  (\bibinfo {year} {2014})}\BibitemShut {NoStop}%
\bibitem [{\citenamefont {Garc\'{\i}a-Ramos}\ and\ \citenamefont
  {Heyde}(2015{\natexlab{c}})}]{Garc15b}%
  \BibitemOpen
  \bibfield  {author} {\bibinfo {author} {\bibfnamefont {J.~E.}\ \bibnamefont
  {Garc\'{\i}a-Ramos}}\ and\ \bibinfo {author} {\bibfnamefont {K.}~\bibnamefont
  {Heyde}},\ }\bibfield  {title} {\enquote {\bibinfo {title} {{Disentangling
  the nuclear shape coexistence in even-even Hg isotopes using the interacting
  boson model}},}\ }\bibfield  {booktitle} {\emph {\bibinfo {booktitle}
  {{Proceedings, 15th International Symposium on Capture Gamma-Ray Spectroscopy
  and Related Topics (CGS15): Dresden, Germany, August 25-29, 2014}}},\ }\href
  {\doibase 10.1051/epjconf/20159301004} {\bibfield  {journal} {\bibinfo
  {journal} {EPJ Web Conf.}\ }\textbf {\bibinfo {volume} {93}},\ \bibinfo
  {pages} {01004} (\bibinfo {year} {2015}{\natexlab{c}})},\ \Eprint
  {http://arxiv.org/abs/1410.2869} {arXiv:1410.2869 [nucl-th]} \BibitemShut
  {NoStop}%
\bibitem [{\citenamefont {Garc\'{\i}a-Ramos}\ and\ \citenamefont
  {Heyde}(2018)}]{Garc18}%
  \BibitemOpen
  \bibfield  {author} {\bibinfo {author} {\bibfnamefont {J.~E.}\ \bibnamefont
  {Garc\'{\i}a-Ramos}}\ and\ \bibinfo {author} {\bibfnamefont {K.}~\bibnamefont
  {Heyde}},\ }\bibfield  {title} {\enquote {\bibinfo {title} {{On the nature of
  the shape coexistence and the quantum phase transition phenomena: lead region
  and Zr isotopes}},}\ }\bibfield  {booktitle} {\emph {\bibinfo {booktitle}
  {{Proceedings, 16th International Symposium on Capture Gamma-Ray Spectroscopy
  and Related Topics (CGS16): Shanghai, China, September 18-22, 2017}}},\
  }\href {\doibase 10.1051/epjconf/201817805005} {\bibfield  {journal}
  {\bibinfo  {journal} {EPJ Web Conf.}\ }\textbf {\bibinfo {volume} {178}},\
  \bibinfo {pages} {05005} (\bibinfo {year} {2018})},\ \Eprint
  {http://arxiv.org/abs/1802.04219} {arXiv:1802.04219 [nucl-th]} \BibitemShut
  {NoStop}%
\bibitem [{\citenamefont {Ginocchio}\ and\ \citenamefont
  {Kirson}(1980)}]{gino80}%
  \BibitemOpen
  \bibfield  {author} {\bibinfo {author} {\bibfnamefont {J.N.}\ \bibnamefont
  {Ginocchio}}\ and\ \bibinfo {author} {\bibfnamefont {M.W.}\ \bibnamefont
  {Kirson}},\ }\bibfield  {title} {\enquote {\bibinfo {title} {{An intrinsic
  state for the interacting boson model and its relationship to the
  Bohr-Mottelson model}},}\ }\href {\doibase
  https://doi.org/10.1016/0375-9474(80)90387-5} {\bibfield  {journal} {\bibinfo
   {journal} {Nucl. Phys. A}\ }\textbf {\bibinfo {volume} {350}},\ \bibinfo
  {pages} {31 -- 60} (\bibinfo {year} {1980})}\BibitemShut {NoStop}%
\bibitem [{\citenamefont {Dieperink}\ and\ \citenamefont
  {Scholten}(1980)}]{diep80a}%
  \BibitemOpen
  \bibfield  {author} {\bibinfo {author} {\bibfnamefont {A.E.L.}\ \bibnamefont
  {Dieperink}}\ and\ \bibinfo {author} {\bibfnamefont {O.}~\bibnamefont
  {Scholten}},\ }\bibfield  {title} {\enquote {\bibinfo {title} {{On shapes and
  shape phase transitions in the interacting boson model}},}\ }\href {\doibase
  https://doi.org/10.1016/0375-9474(80)90492-3} {\bibfield  {journal} {\bibinfo
   {journal} {Nucl. Phys. A}\ }\textbf {\bibinfo {volume} {346}},\ \bibinfo
  {pages} {125 -- 138} (\bibinfo {year} {1980})}\BibitemShut {NoStop}%
\bibitem [{\citenamefont {Dieperink}\ \emph {et~al.}(1980)\citenamefont
  {Dieperink}, \citenamefont {Scholten},\ and\ \citenamefont
  {Iachello}}]{diep80b}%
  \BibitemOpen
  \bibfield  {author} {\bibinfo {author} {\bibfnamefont {A.~E.~L.}\
  \bibnamefont {Dieperink}}, \bibinfo {author} {\bibfnamefont {O.}~\bibnamefont
  {Scholten}}, \ and\ \bibinfo {author} {\bibfnamefont {F.}~\bibnamefont
  {Iachello}},\ }\bibfield  {title} {\enquote {\bibinfo {title} {{Classical
  Limit of the Interacting-Boson Model}},}\ }\href {\doibase
  10.1103/PhysRevLett.44.1747} {\bibfield  {journal} {\bibinfo  {journal}
  {Phys. Rev. Lett.}\ }\textbf {\bibinfo {volume} {44}},\ \bibinfo {pages}
  {1747--1750} (\bibinfo {year} {1980})}\BibitemShut {NoStop}%
\bibitem [{\citenamefont {Gilmore}("1974")}]{Gilm74}%
  \BibitemOpen
  \bibfield  {author} {\bibinfo {author} {\bibfnamefont {R.}~\bibnamefont
  {Gilmore}},\ }\href@noop {} {\emph {\bibinfo {title} {{Lie groups, Lie
  algebras and some applications}}}}\ (\bibinfo  {publisher} {Wiley},\ \bibinfo
  {year} {"1974"})\BibitemShut {NoStop}%
\bibitem [{\citenamefont {Frank}\ \emph {et~al.}(2002)\citenamefont {Frank},
  \citenamefont {Casta\~nos}, \citenamefont {Isacker},\ and\ \citenamefont
  {Padilla}}]{Frank02}%
  \BibitemOpen
  \bibfield  {author} {\bibinfo {author} {\bibfnamefont {A.}~\bibnamefont
  {Frank}}, \bibinfo {author} {\bibfnamefont {O.}~\bibnamefont {Casta\~nos}},
  \bibinfo {author} {\bibfnamefont {P.~Van}\ \bibnamefont {Isacker}}, \ and\
  \bibinfo {author} {\bibfnamefont {E.}~\bibnamefont {Padilla}},\ }\bibfield
  {title} {\enquote {\bibinfo {title} {{The Geometry of the IBM with
  Configuration Mixing}},}\ }\bibfield  {booktitle} {\emph {\bibinfo
  {booktitle} {{Proceedings, International Conference on Nuclear Structure:
  Mapping the Triangle: Grand Teton National Park, Wyoming, May 22-25,
  2002}}},\ }\href {\doibase 10.1063/1.1517933} {\bibfield  {journal} {\bibinfo
   {journal} {AIP Conf. Proc.}\ }\textbf {\bibinfo {volume} {638}},\ \bibinfo
  {pages} {23} (\bibinfo {year} {2002})}\BibitemShut {NoStop}%
\bibitem [{\citenamefont {Frank}\ \emph {et~al.}(2004)\citenamefont {Frank},
  \citenamefont {Van~Isacker},\ and\ \citenamefont {Vargas}}]{Frank04}%
  \BibitemOpen
  \bibfield  {author} {\bibinfo {author} {\bibfnamefont {Alejandro}\
  \bibnamefont {Frank}}, \bibinfo {author} {\bibfnamefont {Piet}\ \bibnamefont
  {Van~Isacker}}, \ and\ \bibinfo {author} {\bibfnamefont {Carlos~E.}\
  \bibnamefont {Vargas}},\ }\bibfield  {title} {\enquote {\bibinfo {title}
  {{Evolving shape coexistence in the lead isotopes: The geometry of
  configuration mixing in nuclei}},}\ }\href {\doibase
  10.1103/PhysRevC.69.034323} {\bibfield  {journal} {\bibinfo  {journal} {Phys.
  Rev. C}\ }\textbf {\bibinfo {volume} {69}},\ \bibinfo {pages} {034323}
  (\bibinfo {year} {2004})}\BibitemShut {NoStop}%
\bibitem [{\citenamefont {Frank}\ \emph {et~al.}(2006)\citenamefont {Frank},
  \citenamefont {Isacker},\ and\ \citenamefont {Iachello}}]{Frank06}%
  \BibitemOpen
  \bibfield  {author} {\bibinfo {author} {\bibfnamefont {A.}~\bibnamefont
  {Frank}}, \bibinfo {author} {\bibfnamefont {P.~Van}\ \bibnamefont {Isacker}},
  \ and\ \bibinfo {author} {\bibfnamefont {F.}~\bibnamefont {Iachello}},\
  }\bibfield  {title} {\enquote {\bibinfo {title} {{Phase transitions in
  configuration mixed models}},}\ }\href {\doibase 10.1103/PhysRevC.73.061302}
  {\bibfield  {journal} {\bibinfo  {journal} {Phys. Rev. C}\ }\textbf {\bibinfo
  {volume} {73}},\ \bibinfo {pages} {061302} (\bibinfo {year}
  {2006})}\BibitemShut {NoStop}%
\bibitem [{\citenamefont {Morales}\ \emph {et~al.}(2008)\citenamefont
  {Morales}, \citenamefont {Frank}, \citenamefont {Vargas},\ and\ \citenamefont
  {Isacker}}]{Mora08}%
  \BibitemOpen
  \bibfield  {author} {\bibinfo {author} {\bibfnamefont {Irving~O.}\
  \bibnamefont {Morales}}, \bibinfo {author} {\bibfnamefont {Alejandro}\
  \bibnamefont {Frank}}, \bibinfo {author} {\bibfnamefont {Carlos~E.}\
  \bibnamefont {Vargas}}, \ and\ \bibinfo {author} {\bibfnamefont {P.~Van}\
  \bibnamefont {Isacker}},\ }\bibfield  {title} {\enquote {\bibinfo {title}
  {{Shape coexistence and phase transitions in the platinum isotopes}},}\
  }\href {\doibase 10.1103/PhysRevC.78.024303} {\bibfield  {journal} {\bibinfo
  {journal} {Phys. Rev. C}\ }\textbf {\bibinfo {volume} {78}},\ \bibinfo
  {pages} {024303} (\bibinfo {year} {2008})}\BibitemShut {NoStop}%
\bibitem [{\citenamefont {Garc\'{\i}a-Ramos}\ \emph {et~al.}(2014)\citenamefont
  {Garc\'{\i}a-Ramos}, \citenamefont {Heyde}, \citenamefont {Robledo},\ and\
  \citenamefont {Rodr\'{\i}guez-Guzm\'an}}]{Garc14a}%
  \BibitemOpen
  \bibfield  {author} {\bibinfo {author} {\bibfnamefont {J.~E.}\ \bibnamefont
  {Garc\'{\i}a-Ramos}}, \bibinfo {author} {\bibfnamefont {K.}~\bibnamefont
  {Heyde}}, \bibinfo {author} {\bibfnamefont {L.~M.}\ \bibnamefont {Robledo}},
  \ and\ \bibinfo {author} {\bibfnamefont {R.}~\bibnamefont
  {Rodr\'{\i}guez-Guzm\'an}},\ }\bibfield  {title} {\enquote {\bibinfo {title}
  {{Shape evolution and shape coexistence in Pt isotopes: Comparing interacting
  boson model configuration mixing and Gogny mean-field energy surfaces}},}\
  }\href {\doibase 10.1103/PhysRevC.89.034313} {\bibfield  {journal} {\bibinfo
  {journal} {Phys. Rev. C}\ }\textbf {\bibinfo {volume} {89}},\ \bibinfo
  {pages} {034313} (\bibinfo {year} {2014})}\BibitemShut {NoStop}%
\bibitem [{\citenamefont {Kumar}(1972)}]{kumar72}%
  \BibitemOpen
  \bibfield  {author} {\bibinfo {author} {\bibfnamefont {Krishna}\ \bibnamefont
  {Kumar}},\ }\bibfield  {title} {\enquote {\bibinfo {title} {{Intrinsic
  Quadrupole Moments and Shapes of Nuclear Ground States and Excited
  States}},}\ }\href {\doibase 10.1103/PhysRevLett.28.249} {\bibfield
  {journal} {\bibinfo  {journal} {Phys. Rev. Lett.}\ }\textbf {\bibinfo
  {volume} {28}},\ \bibinfo {pages} {249--253} (\bibinfo {year}
  {1972})}\BibitemShut {NoStop}%
\bibitem [{\citenamefont {Cline}(1986)}]{Cline86}%
  \BibitemOpen
  \bibfield  {author} {\bibinfo {author} {\bibfnamefont {Douglas}\ \bibnamefont
  {Cline}},\ }\bibfield  {title} {\enquote {\bibinfo {title} {{Nuclear shapes
  studied by coulomb excitation}},}\ }\href {\doibase
  10.1146/annurev.ns.36.120186.003343} {\bibfield  {journal} {\bibinfo
  {journal} {Ann. Rev. Nucl. Part. Sci.}\ }\textbf {\bibinfo {volume} {36}},\
  \bibinfo {pages} {683--716} (\bibinfo {year} {1986})}\BibitemShut {NoStop}%
\bibitem [{\citenamefont {Hilairea}\ and\ \citenamefont
  {Girod}(2008)}]{Hila08}%
  \BibitemOpen
  \bibfield  {author} {\bibinfo {author} {\bibfnamefont {S.}~\bibnamefont
  {Hilairea}}\ and\ \bibinfo {author} {\bibfnamefont {M.}~\bibnamefont
  {Girod}},\ }\bibfield  {title} {\enquote {\bibinfo {title} {The amedee
  nuclear structure database},}\ }\bibfield  {booktitle} {\emph {\bibinfo
  {booktitle} {ND 2007, International Conference on Nuclear Data for Science
  and Technology 2007}},\ }\href {\doibase
  10.1051/ndata:077090.1051/ndata:07709} {\bibfield  {journal} {\bibinfo
  {journal} {EDP Sciences}\ ,\ \bibinfo {pages} {107--110}} (\bibinfo {year}
  {2008})}\BibitemShut {NoStop}%
\bibitem [{\citenamefont {CEA}()}]{Bruyere-surfaces}%
  \BibitemOpen
  \bibfield  {author} {\bibinfo {author} {\bibnamefont {CEA}},\ }\href@noop {}
  {}\bibinfo {howpublished} {http://www-phynu.cea.fr/HFB-Gogny.htm}\BibitemShut
  {NoStop}%
\bibitem [{\citenamefont {Nomura}\ \emph {et~al.}(2016)\citenamefont {Nomura},
  \citenamefont {Rodr\'{\i}guez-Guzm\'an},\ and\ \citenamefont
  {Robledo}}]{Nomu16}%
  \BibitemOpen
  \bibfield  {author} {\bibinfo {author} {\bibfnamefont {K.}~\bibnamefont
  {Nomura}}, \bibinfo {author} {\bibfnamefont {R.}~\bibnamefont
  {Rodr\'{\i}guez-Guzm\'an}}, \ and\ \bibinfo {author} {\bibfnamefont {L.~M.}\
  \bibnamefont {Robledo}},\ }\bibfield  {title} {\enquote {\bibinfo {title}
  {{Structural evolution in $A\ensuremath{\approx}100$ nuclei within the mapped
  interacting boson model based on the Gogny energy density functional}},}\
  }\href {\doibase 10.1103/PhysRevC.94.044314} {\bibfield  {journal} {\bibinfo
  {journal} {Phys. Rev. C}\ }\textbf {\bibinfo {volume} {94}},\ \bibinfo
  {pages} {044314} (\bibinfo {year} {2016})}\BibitemShut {NoStop}%
\bibitem [{\citenamefont {Togashi}\ \emph {et~al.}(2016)\citenamefont
  {Togashi}, \citenamefont {Tsunoda}, \citenamefont {Otsuka},\ and\
  \citenamefont {Shimizu}}]{Togashi16}%
  \BibitemOpen
  \bibfield  {author} {\bibinfo {author} {\bibfnamefont {Tomoaki}\ \bibnamefont
  {Togashi}}, \bibinfo {author} {\bibfnamefont {Yusuke}\ \bibnamefont
  {Tsunoda}}, \bibinfo {author} {\bibfnamefont {Takaharu}\ \bibnamefont
  {Otsuka}}, \ and\ \bibinfo {author} {\bibfnamefont {Noritaka}\ \bibnamefont
  {Shimizu}},\ }\bibfield  {title} {\enquote {\bibinfo {title} {{Quantum Phase
  Transition in the Shape of Zr isotopes}},}\ }\href {\doibase
  10.1103/PhysRevLett.117.172502} {\bibfield  {journal} {\bibinfo  {journal}
  {Phys. Rev. Lett.}\ }\textbf {\bibinfo {volume} {117}},\ \bibinfo {pages}
  {172502} (\bibinfo {year} {2016})}\BibitemShut {NoStop}%
\bibitem [{\citenamefont {Rodr\'{\i}guez-Guzm\'an}\ \emph
  {et~al.}(2010)\citenamefont {Rodr\'{\i}guez-Guzm\'an}, \citenamefont
  {Sarriguren}, \citenamefont {Robledo},\ and\ \citenamefont
  {Perez-Martin}}]{Rodr10}%
  \BibitemOpen
  \bibfield  {author} {\bibinfo {author} {\bibfnamefont {R.}~\bibnamefont
  {Rodr\'{\i}guez-Guzm\'an}}, \bibinfo {author} {\bibfnamefont
  {P.}~\bibnamefont {Sarriguren}}, \bibinfo {author} {\bibfnamefont {L.M.}\
  \bibnamefont {Robledo}}, \ and\ \bibinfo {author} {\bibfnamefont
  {S.}~\bibnamefont {Perez-Martin}},\ }\bibfield  {title} {\enquote {\bibinfo
  {title} {Charge radii and structural evolution in sr, zr, and mo isotopes},}\
  }\href {\doibase https://doi.org/10.1016/j.physletb.2010.06.035} {\bibfield
  {journal} {\bibinfo  {journal} {Physics Letters B}\ }\textbf {\bibinfo
  {volume} {691}},\ \bibinfo {pages} {202 -- 207} (\bibinfo {year}
  {2010})}\BibitemShut {NoStop}%
\bibitem [{\citenamefont {Abusara}\ and\ \citenamefont {Ahmad}(2017)}]{Abus17}%
  \BibitemOpen
  \bibfield  {author} {\bibinfo {author} {\bibfnamefont {H.}~\bibnamefont
  {Abusara}}\ and\ \bibinfo {author} {\bibfnamefont {Shakeb}\ \bibnamefont
  {Ahmad}},\ }\bibfield  {title} {\enquote {\bibinfo {title} {{Shape evolution
  in Kr, Zr, and Sr isotopic chains in covariant density functional theory}},}\
  }\href {\doibase 10.1103/PhysRevC.96.064303} {\bibfield  {journal} {\bibinfo
  {journal} {Phys. Rev. C}\ }\textbf {\bibinfo {volume} {96}},\ \bibinfo
  {pages} {064303} (\bibinfo {year} {2017})}\BibitemShut {NoStop}%
\bibitem [{\citenamefont {Bucurescu}\ and\ \citenamefont
  {Zamfir}(2018)}]{Buca18}%
  \BibitemOpen
  \bibfield  {author} {\bibinfo {author} {\bibfnamefont {D.}~\bibnamefont
  {Bucurescu}}\ and\ \bibinfo {author} {\bibfnamefont {N.~V.}\ \bibnamefont
  {Zamfir}},\ }\bibfield  {title} {\enquote {\bibinfo {title} {Empirical
  signatures of shape phase transitions in nuclei with odd nucleon numbers},}\
  }\href {\doibase 10.1103/PhysRevC.98.024301} {\bibfield  {journal} {\bibinfo
  {journal} {Phys. Rev. C}\ }\textbf {\bibinfo {volume} {98}},\ \bibinfo
  {pages} {024301} (\bibinfo {year} {2018})}\BibitemShut {NoStop}%
\bibitem [{\citenamefont {Garc\'{\i}a-Ramos}\ \emph {et~al.}(2005)\citenamefont
  {Garc\'{\i}a-Ramos}, \citenamefont {Heyde}, \citenamefont {Fossion},
  \citenamefont {Hellemans},\ and\ \citenamefont {De~Baerdemacker}}]{Garc05}%
  \BibitemOpen
  \bibfield  {author} {\bibinfo {author} {\bibfnamefont {J.~E.}\ \bibnamefont
  {Garc\'{\i}a-Ramos}}, \bibinfo {author} {\bibfnamefont {K.}~\bibnamefont
  {Heyde}}, \bibinfo {author} {\bibfnamefont {R.}~\bibnamefont {Fossion}},
  \bibinfo {author} {\bibfnamefont {V.}~\bibnamefont {Hellemans}}, \ and\
  \bibinfo {author} {\bibfnamefont {S.}~\bibnamefont {De~Baerdemacker}},\
  }\bibfield  {title} {\enquote {\bibinfo {title} {{A theoretical description
  of energy spectra and two-neutron separation energies for neutron-rich
  zirconium isotopes}},}\ }\href {\doibase 10.1140/epja/i2005-10176-1}
  {\bibfield  {journal} {\bibinfo  {journal} {Eur. Phys. J. A}\ }\textbf
  {\bibinfo {volume} {26}},\ \bibinfo {pages} {221--225} (\bibinfo {year}
  {2005})}\BibitemShut {NoStop}%
\bibitem [{\citenamefont {Arias}\ \emph {et~al.}(2003)\citenamefont {Arias},
  \citenamefont {Dukelsky},\ and\ \citenamefont {Garc\'{\i}a-Ramos}}]{Aria03}%
  \BibitemOpen
  \bibfield  {author} {\bibinfo {author} {\bibfnamefont {J.~M.}\ \bibnamefont
  {Arias}}, \bibinfo {author} {\bibfnamefont {J.}~\bibnamefont {Dukelsky}}, \
  and\ \bibinfo {author} {\bibfnamefont {J.~E.}\ \bibnamefont
  {Garc\'{\i}a-Ramos}},\ }\bibfield  {title} {\enquote {\bibinfo {title}
  {{Quantum Phase Transitions in the Interacting Boson Model: Integrability,
  Level Repulsion, and Level Crossing}},}\ }\href {\doibase
  10.1103/PhysRevLett.91.162502} {\bibfield  {journal} {\bibinfo  {journal}
  {Phys. Rev. Lett.}\ }\textbf {\bibinfo {volume} {91}},\ \bibinfo {pages}
  {162502} (\bibinfo {year} {2003})}\BibitemShut {NoStop}%
\bibitem [{\citenamefont {Poves}(2018)}]{Poves18}%
  \BibitemOpen
  \bibfield  {author} {\bibinfo {author} {\bibfnamefont {Alfredo}\ \bibnamefont
  {Poves}},\ }\bibfield  {title} {\enquote {\bibinfo {title} {{Shape
  Coexistence and Islands of Inversion Monopole vs Multipole}},}\ }\bibfield
  {booktitle} {\emph {\bibinfo {booktitle} {{Proceedings, IIRC Symposium on
  Perspectives of the Physics of Nuclear Structure: Tokyo, Japan, November 1-4,
  2017}}},\ }\href {\doibase 10.7566/JPSCP.23.012015} {\bibfield  {journal}
  {\bibinfo  {journal} {JPS Conf. Proc.}\ }\textbf {\bibinfo {volume} {23}},\
  \bibinfo {pages} {012015} (\bibinfo {year} {2018})}\BibitemShut {NoStop}%
\bibitem [{\citenamefont {Gavrielov}\ \emph
  {et~al.}(2019{\natexlab{a}})\citenamefont {Gavrielov}, \citenamefont
  {Leviatan},\ and\ \citenamefont {Iachello}}]{Gavr19}%
  \BibitemOpen
  \bibfield  {author} {\bibinfo {author} {\bibfnamefont {N.}~\bibnamefont
  {Gavrielov}}, \bibinfo {author} {\bibfnamefont {A.}~\bibnamefont {Leviatan}},
  \ and\ \bibinfo {author} {\bibfnamefont {F.}~\bibnamefont {Iachello}},\
  }\bibfield  {title} {\enquote {\bibinfo {title} {Intertwined quantum phase
  transitions in the zr isotopes},}\ }\href {\doibase
  10.1103/PhysRevC.99.064324} {\bibfield  {journal} {\bibinfo  {journal} {Phys.
  Rev. C}\ }\textbf {\bibinfo {volume} {99}},\ \bibinfo {pages} {064324}
  (\bibinfo {year} {2019}{\natexlab{a}})}\BibitemShut {NoStop}%
\bibitem [{\citenamefont {Gavrielov}\ \emph
  {et~al.}(2019{\natexlab{b}})\citenamefont {Gavrielov}, \citenamefont
  {Leviatan},\ and\ \citenamefont {Iachello}}]{Gavr19b}%
  \BibitemOpen
  \bibfield  {author} {\bibinfo {author} {\bibfnamefont {N}~\bibnamefont
  {Gavrielov}}, \bibinfo {author} {\bibfnamefont {A}~\bibnamefont {Leviatan}},
  \ and\ \bibinfo {author} {\bibfnamefont {F}~\bibnamefont {Iachello}},\
  }\bibfield  {title} {\enquote {\bibinfo {title} {{Interplay between
  shape-phase transitions and shape coexistence in the Zr isotopes}},}\ }\href
  {\doibase 10.1088/1402-4896/ab456b} {\bibfield  {journal} {\bibinfo
  {journal} {Physica Scripta}\ }\textbf {\bibinfo {volume} {95}},\ \bibinfo
  {pages} {024001} (\bibinfo {year} {2019}{\natexlab{b}})}\BibitemShut
  {NoStop}%
\end{thebibliography}%
\end{document}